\documentclass[%
 aip,
 jcp,
 amsmath,amssymb,
 reprint,%
]{revtex4-1}

\usepackage{graphicx}% Include figure files
\usepackage{dcolumn}% Align table columns on decimal point
\usepackage{physics}
\usepackage[utf8]{inputenc}
\usepackage[T1]{fontenc}
\usepackage{mathptmx}
\usepackage{etoolbox}
\usepackage{booktabs}
\usepackage{bbding}
\usepackage{hyperref}
\usepackage{textgreek}
\usepackage{float}
\usepackage{pifont}
\usepackage{siunitx}  % SI-units, table formatting
\usepackage{bm}% bold math

\newcommand{\ve}[1]{\bm{#1}}

\newcommand{\sv}[1]{\mathbf{#1}}
\newcommand{\vc}[1]{\mathbf{#1}}
\newcommand{\one}{\hat{I}}
\newcommand{\xs}{\times_{\sigma}}
\newcommand{\pk}{\hat{\Theta}}

\makeatletter
\def\@email#1#2{%
 \endgroup
 \patchcmd{\titleblock@produce}
  {\frontmatter@RRAPformat}
  {\frontmatter@RRAPformat{\produce@RRAP{*#1\href{mailto:#2}{#2}}}\frontmatter@RRAPformat}
  {}{}
}%
\makeatother
\begin{document}

%%%%%%%%%%%%%%%%%%%%%%%%%%%%%%%%%%%%%%%%%%%%%%%%%%%%%%%%%%%%%%%%%%%%%
% Authors
%%%%%%%%%%%%%%%%%%%%%%%%%%%%%%%%%%%%%%%%%%%%%%%%%%%%%%%%%%%%%%%%%%%%%

\title{Current-Dependent Meta-GGA and Local Hybrid Functionals for Relativistic NMR Shifts}

\author{Yannick J. Franzke$^*$}
\affiliation{Institute of Nanotechnology, Karlsruhe
Institute of Technology (KIT), Kaiserstra\ss{}e 12, 76131 Karlsruhe, Germany}
\email{yannick.franzke@kit.edu}
%ORCID: 0000-0002-8344-113X

\author{Christof Holzer$^*$}
\affiliation{Institute of Quantum Materials and Technologies, Karlsruhe
Institute of Technology (KIT), Kaiserstra\ss{}e 12, 76131 Karlsruhe,
Germany}
\email{christof.holzer@kit.edu}
%ORCID: 0000-0001-8234-260X

%%%%%%%%%%%%%%%%%%%%%%%%%%%%%%%%%%%%%%%%%%%%%%%%%%%%%%%%%%%%%%%%%%%%%
%%% Title and Abstract
%%%%%%%%%%%%%%%%%%%%%%%%%%%%%%%%%%%%%%%%%%%%%%%%%%%%%%%%%%%%%%%%%%%%%
\begin{abstract}
NMR chemical shifts are key for the characterization of novel compounds.
Especially for molecules with heavy elements, the experimental data
is often corroborated with computational studies in density functional
theory (DFT). However, modern meta-generalized gradient approximations
(meta-GGAs) and local hybrid functionals are not widely supported in
relativistic calculations with quantum chemistry program suites.
So far, their current-dependent generalization in the spirit of Dobson,
Becke, and Tao was not presented at all. This generalization ensures
gauge invariance and does not introduce artifacts--unlike the more
popular field-dependent generalization of the kinetic-energy density.
In non-relativistic and scalar-relativistic calculations, the first was
shown to be superior but the current density only arises due to the
magnetic field perturbation. 
In contrast, spin--orbit coupling also induces current densities
in the preceding ground-state calculation and the DFT framework 
needs to be generalized to a current density functional theory (CDFT)
for consistency with the current-dependent response equations.
Herein, we present the two-component relativistic generalization
of the current-dependent kinetic-energy density of a Kramers-restricted
Kohn--Sham reference for meta-GGAs and local hybrid functionals in NMR
calculations. The extension is invariant under a position-dependent phase
and a position-dependent spin rotation of the spinors, or a
local U(1)$\times$SU(2) transformation.
Relativistic effects are considered with exact two-component theory
using a superposition of atomic potentials to include the two-electron
picture-change correction (SAP-X2C). 
Results show that the more common field-dependent generalization for the
kinetic-energy density leads to outliers for some functionals.
In this regard, the current-dependent generalization reduces the spread
of the results and increases the predicative capabilities of
\textit{in silico} NMR studies. The accuracy is demonstrated for
Cd NMR spectra highlighting the importance of Boltzmann isomer sampling
and NMR simulations of Zintl pseudo-tetrahedrons in solution. Here,
inclusion of the current density results in small changes only for the
chemical shifts due to cancellation of the impact on the shielding constants.
\end{abstract}

\maketitle

%%%%%%%%%%%%%%%%%%%%%%%%%%%%%%%%%%%%%%%%%%%%%%%%%%%%%%%%%%%%%%%%%%%%%
%%% Introduction
%%%%%%%%%%%%%%%%%%%%%%%%%%%%%%%%%%%%%%%%%%%%%%%%%%%%%%%%%%%%%%%%%%%%%
\section{Introduction}
\label{sec:introduction}
The accurate prediction of nuclear magnetic resonance (NMR) properties
with density functional theory (DFT) is a prime target in theoretical
chemistry and physics. NMR spectra not only allow to elucidate the molecular
structure but also to gain a deeper understanding of the electronic
structure. \cite{Helgaker.Jaszunski.ea:Ab.1999, Kaupp.Buhl.ea:Calculation.2004,
Vaara:Theory.2007, Autschbach:Relativistic.2014, InC-Repisky.Komorovsky.ea:2016, 
Grimme.Bannwarth.ea:Fully.2017, Sundholm.Dimitrova.ea:Current.2021, Aucar:2025}
Here, generalized gradient approximations (GGAs) and their hybrids are
still the workhorse of quantum chemistry, \cite{Aucar:2025} despite the
fact that meta-GGAs have witnessed tremendous improvement in the last
two decades and often perform superior for thermochemistry.
\cite{Mardirossian.Head-Gordon:Thirty.2017, Teale.Helgaker.ea:DFT.2022, InB-Perdew:2024}
This is mainly due to the necessary additional generalizations to ensure
gauge invariance of the kinetic energy density in the presence of a magnetic
perturbation,\cite{Dobson:Alternative.1993, Becke:Current.2002, Tao:Explicit.2005,
Maximoff.Scuseria:Nuclear.2004, Bates.Furche:Harnessing.2012}
which still significantly limits the number of quantum chemical program
suites supporting meta-GGAs for \textit{in silico} studies in NMR spectroscopy.
We refer to the reviews and perspectives in recent special issues on quantum chemical
codes concerning this limitation.\cite{Sherrill.Manolopoulos.ea:Electronic.2020, Gagliardi:Sustainable.2023}
First, Maximoff and Scuseria implemented a generalization based on the
magnetic field and its vector potential itself, \cite{Maximoff.Scuseria:Nuclear.2004}
which allowed for the first gauge-origin invariant NMR calculations with
meta-GGAs. It was later shown, however, that this comes at the price
of violating the iso-orbital condition, which is at the heart of
the theoretical construction of meta-GGAs. \cite{Bates.Furche:Harnessing.2012}
In contrast, a current-dependent generalization in the spirit of Dobson,
Becke, and Tao ensures both the iso-orbital condition and gauge invariance.
\cite{Dobson:Alternative.1993, Becke:Current.2002, Tao:Explicit.2005}
Moreover, it removes spurious artifacts and leads to more accurate
results in theoretical benchmark calculations. \cite{Schattenberg.Kaupp:Effect.2021}
Crucially, only the actual NMR or generally electrochemical
response part of the computational workflow requires a generalization due
to the perturbative approach for the magnetic induction.
\cite{Lee.Handy.ea:density.1995, Bates.Furche:Harnessing.2012, 
Reimann.Ekstrom.ea:importance.2015, Liang.Feng.ea:Revisiting.2022,
Grotjahn.Furche:Gauge-Invariant.2023,
Schattenberg.Kaupp:Effect.2021, Franzke:Reducing.2023,
Holzer.Franzke.ea:Assessing.2021,
Franzke.Holzer:Impact.2022, Bruder.Franzke.ea:Paramagnetic.2022,
Bruder.Franzke.ea:Zero-Field.2023, Franzke.Bruder.ea:Paramagnetic.2024}
That is, the self-consistent field (SCF) procedure is not
affected with either generalization.
Matters are different in finite field calculations, where
the induced current densities need to be considered already at the
SCF level and DFT is always generalized to the current density
functional (CDFT) framework.
\cite{Vignale.Rasolt:Current-.1988, Tellgren.Kvaal.ea:Choice.2012,
Tellgren.Teale.ea:Non-perturbative.2014,
Tellgren.Kvaal.ea:Fermion.2014, Lieb.Schrader:Current.2013,
Furness.Verbeke.ea:Current.2015, Irons.Spence.ea:Analyzing.2020,
Sen.Tellgren:Benchmarking.2021, Irons.David.ea:Optimizing.2021}

This issue is also of importance in relativistic calculations with
spin--orbit coupling (SOC) such as two-component (2c) and four-component (4c)
calculations. Here, SOC directly induces spin current densities in the
ground-state SCF for a closed-shell system.
\cite{Brack.Jennings.ea:On.1976} Unlike in an external magnetic field,
all spin directions generally feature non-zero spin current densities.
This also calls for a generalization of the kinetic energy density
in the exchange-correlation potential and its derivatives.
\cite{Holzer.Franzke.ea:Current.2022, Franzke.Holzer:Current.2024, 
Maier.Ikabata.ea:Restoring.2019, Desmarais.Maul.ea:Spin.2024, 
Desmarais.Vignale.ea:Electron.2024, Franzke.Pausch.ea:Application.2025,
Huebsch.Tran.ea:Capturing.2026, Franzke.Holzer:Application.2026}
Therefore, the current-dependent generalization is the more
natural choice for relativistic calculations, as it treats both
SOC and the external magnetic field on an equal footing and is also able
to describe the cross/coupling terms of the two types of induced currents.
\cite{Holzer.Franzke.ea:Current.2022, Pausch.Holzer:Linear.2022}
To the best of our knowledge, only our previously work described
a meta-GGA implementation with gauge-including atomic orbitals (GIAOs) 
for 2c NMR shifts in the literature at all. \cite{Franzke.Holzer:Exact.2023} 
The given work, however, uses the Maximoff--Scuseria generalization
and neglects the SOC-induced current densities for the kinetic energy
density in the preceding SCF step.
Further considering the known artifacts of this generalization in
the NMR response part, \cite{Schattenberg.Kaupp:Effect.2021}
this means that functionals such as the strongly constrained and
appropriately Normed (SCAN) family or the Minnesota functionals should
only be applied with great caution. \cite{Franzke.Holzer:Exact.2023}

Furthermore, the kinetic energy density is also a key variable of other
functionals such as local hybrid functionals (LHFs).
\cite{Jaramillo.Scuseria.ea:Local.2003}
LHFs use a position-dependent admixture of exact exchange
controlled with a local mixing function, which often depends on the
iso-orbital indicator using the kinetic energy density.
\cite{Jaramillo.Scuseria.ea:Local.2003, Maier.Arbuznikov.ea:Local.2019}
Local hybrid functionals have recently shown promising results for
NMR shifts in both benchmark studies and practical applications.
\cite{Schattenberg.Kaupp:Extended.2021, Holzer.Franzke.ea:Assessing.2021,
Schattenberg.Kaupp:Implementation.2021, Holzer.Franzke:Local.2022, 
Schattenberg.Lehmann.ea:Systematic.2022, Holzer.Franzke:General.2025,
Ketzel.Li.ea:Benchmark.2025, Steffen.Szych.ea:Di.2025}
However, implementations are currently restricted to the non-relativistic
or scalar-relativistic framework, which limits their applicability to
light elements.

In the present work, we derive current-dependent meta-GGA functionals
incorporating both the SOC-induced current density and that of the
external magnetic field. Additionally, we present the implementation
of spin--orbit 2c NMR shifts with local hybrids using both the
Maximoff--Scuseria and the current-dependent generalization.
Throughout this work, gauge-origin invariance is ensured with GIAOs,
as done in the non-relativistic limit.

%%%%%%%%%%%%%%%%%%%%%%%%%%%%%%%%%%%%%%%%%%%%%%%%%%%%%%%%%%%%%%%%%%%%%
%%% Theory
%%%%%%%%%%%%%%%%%%%%%%%%%%%%%%%%%%%%%%%%%%%%%%%%%%%%%%%%%%%%%%%%%%%%%
\section{Theory}
\label{sec:2c-nmr}

\subsection{Preliminaries: Ground-State Density and DFT Variables}
\label{subsec:preliminaries}

We consider a complex two-component (2c) Kramers-restricted reference
with an even number of total electrons, so that
the one-electron particle density reads
\begin{equation}
  \hat{P} = \sum_i n_i\,|\phi_i\rangle\langle\phi_j| =
  \begin{pmatrix}
    P^{\alpha \alpha} & P^{\alpha \beta}\\
    P^{\beta \alpha} & P^{\beta \beta}
  \end{pmatrix}
  \quad \text{with} \quad
  \phi_i =
  \begin{pmatrix}
    \phi_{i}^{\alpha} \\ \phi_{i}^{\beta}
  \end{pmatrix}
\end{equation}
with the Kohn--Sham spinors $|\phi_i\rangle$ and their occupations $n_i$. Here,
$n_i$ is either zero or one. This can be rewritten in the Pauli form as
\begin{equation}
  \hat{P} =\frac12\left(\,{P}_0 \one + \bm{P}\circ\ve\hat{\boldsymbol{\sigma}}\right), \qquad {\bm{P}}\circ\ve\hat{\boldsymbol{\sigma}}=\sum_{u=x,y,z}{{P}}_{m, u}\hat{\sigma}_{u}\
  \label{eq:pauli}
\end{equation}
with the particle density ${P}_0$ and the spin vector $\bm{P}$ reading
\begin{equation}
\begin{split}
  P_0 = & P_{\alpha \alpha} + P_{\beta \beta}, \quad
  P_z =  P_{\alpha \alpha} - P_{\beta \beta}, \\
  P_x = & P_{\alpha\beta} + P_{\beta\alpha}, \quad
  P_y = \text{i}\,(P_{\alpha\beta} - P_{\beta\alpha}).
\end{split}
\end{equation}
Here, $\one$ is the $2 \times 2$ identity $\one$ and $\hat{\ve\sigma}=(\hat{\sigma}_x,\hat{\sigma}_y,\hat{\sigma}_z)$ is the vector of the Pauli spin matrices,
\begin{equation}
  \one=\begin{pmatrix}1&0\\0&1\end{pmatrix}, 
  \hat{\sigma}_x=\begin{pmatrix}0&1\\1&0\end{pmatrix},
  \hat{\sigma}_y=\begin{pmatrix}0&-\text{i}\\\text{i}&0\end{pmatrix},
  \hat{\sigma}_z=\begin{pmatrix}1&0\\0&-1\end{pmatrix}.
\end{equation}
We use the indices $u, v, w$ for the spin components, $k$ for the combined index of the
particle and spin components, $i, j, \dots $ for occupied
spinors, $a, b, \dots$ for virtual spinors, and $p, q, \dots $ for general spinors.
Bold letters indicate vectors in spatial or spin space, while a hat indicates a matrix
in the $2 \times 2$ spin space. The indices clarify the
corresponding space. Further, $\alpha, \beta$ denote the Cartesian or spatial indices. 
As usual, bold letters also denote matrices in the spinor or atomic orbital (AO) basis.
In the atomic orbital basis, only the one-component (1c) linear combinations
$\Re(P_{\mu \nu}^{\alpha \alpha}) + \Re(P_{\mu \nu}^{\beta \beta})$
corresponding to the particle density, and 
$\Im(P_{\mu \nu}^{\alpha \beta}) + \Im(P_{\mu \nu}^{\beta \alpha})$,
$\Re(P_{\mu \nu}^{\alpha \beta}) - \Re(P_{\mu \nu}^{\beta \alpha})$, as well as
$\Im(P_{\mu \nu}^{\alpha \alpha}) - \Im(P_{\mu \nu}^{\beta \beta})$ corresponding
to the spin current densities are non-zero for a Hermitian ground-state reference.
\cite{Armbruster.Weigend.ea:Self-consistent.2008, Peralta.Scuseria.ea:Noncollinear.2007,
Bulik.Scalmani.ea:Noncollinear.2013} The first one is symmetric and the three spin
current density linear combinations are antisymmetric. Note that we distinguish this
notation with $P_{\mu \nu} = \sum_i n_i C_{\mu, i} C_{\nu, i}^{\dagger}$ from the common
chemist's notation $D_{\mu \nu} = \sum_i n_i C_{\mu, i}^{\dagger} C_{\nu, i}$, where
$C_{\mu, i}$ is the 2c AO coefficient of a spinor.

In 2c Kohn--Sham DFT, the electron density is obtained as
\begin{equation}
  \rho_0  = \sum_i \phi_i^{\dagger} \phi_i = \sum_{\mu \nu} \left[\Re(P_{\mu \nu}^{\alpha \alpha}) + \Re(P_{\mu \nu}^{\beta \beta})\right] \chi_{\mu} \chi_{\nu}^*
\end{equation}
with the symmetric particle linear combination of the AO density matrix and the AO basis functions
$\chi$. For the KS reference, the basis functions are real-valued, however, we use explicitly 
complex notation to derive the NMR equations more straightforwardly given that complex gauge-including
atomic orbitals (GIAOs) will be used for NMR.
The spin density magnetization, $m_u  = \sum_i \phi_i^{\dagger} \hat{\sigma}_u \phi_i$,
of the ground state vanishes. So that the spin-up and spin-down variables follows as
\begin{align}
    \rho_{\uparrow, \downarrow} = & \frac{1}{2} (\rho_0 \pm m_{\text{s}}) = \rho_{\downarrow, \uparrow} = \frac{1}{2} \rho_0 \text{,} \\
    m_{\text{s}} = & \sqrt{ m_x^2 + m_y^2 + m_z^2 } \text{,}
\end{align}
due to $m_{\text{s}} = 0$. Note that the notation $\rho_k$ implies that we consider
the particle and spin densities. The spatial gradient of the density $\boldsymbol{\nabla} \rho_{\uparrow, \downarrow}$
follows accordingly and the particle contribution of the kinetic energy density reads
\begin{equation}
\begin{split}
    \tau_0 = & \frac{1}{2} \sum_i \left(\boldsymbol{\nabla} \phi_i\right)^{\dagger} \cdot \left(\boldsymbol{\nabla} \phi_i\right) \\
    = & \frac{1}{2} \sum_{\mu \nu} \left[\Re(P_{\mu \nu}^{\alpha \alpha}) + \Re(P_{\mu \nu}^{\beta \beta})\right] \left[ \left( \boldsymbol{\nabla} \chi_{\mu} \right) \cdot \left( \boldsymbol{\nabla} \chi_{\nu}^* \right) \right] \text{.}
\end{split}
\end{equation}
The spin kinetic energy density is defined accordingly, i.e.\ $ u_u = \tfrac{1}{2} \sum_{i}(\boldsymbol{\nabla}\phi_i)^\dagger\hat{\sigma}_u(\boldsymbol{\nabla}\phi_i)$
with $\boldsymbol{\nabla}$ acting on the spatial directions. Here, the spin kinetic energy density is zero for the SCF ground-state density.
In the presence of spin--orbit coupling, the paramagnetic spin current densities $\mathbf{Y}_u$ are non-zero,
while the paramagnetic particle current density $\mathbf{j}$ vanishes.
In the 2c SOC treatment, the three spin current densities read
\begin{equation}
\begin{split}
  Y_{\alpha, u} = & -\frac{\text{i}}{2} \sum_i \left[ \phi_i^{\dagger} \hat{\sigma}_u (\nabla_{\alpha} \phi_i) - (\nabla_{\alpha} \phi_i)^{\dagger} \hat{\sigma}_u \phi_i \right] \\
  = & - \frac{\text{i}}{2} \sum_{\mu \nu} P_{\mu \nu, u} \left[ \left(\nabla_{\alpha} \chi_{\mu} \right) \chi_{\nu}^* - \chi_{\mu} \left(\nabla_{\alpha} \chi_{\nu}^*\right) \right]
\end{split}
\end{equation}
and are available from the antisymmetric linear combinations of the AO density matrix.
The particle current density is defined accordingly.

Magnetic perturbations and spin--orbit coupling, which is a form of magnetic induction,
can be treated on the same footing with the current-dependent generalization in the spirit
of Dobson, Becke, and Tao. \cite{Dobson:Alternative.1993, Becke:Current.2002, Tao:Explicit.2005}
For the 2c Kramers-restricted ground state SCF calculation, the generalized kinetic energy density reads
\begin{align}
    \tilde{\tau}_0 & = t_0 = \tau_0 - \frac{1}{2} \frac{\sum_{u} {\bf{Y}}_u \cdot {\bf{Y}}_u}{\rho_0} = \tau_0 - \frac{1}{2} \frac{\mathbf{Y} \odot \mathbf{Y}}{\rho_0}
    \label{eq:gen-tau}
\end{align}
using all three spin current densities. Scalar products in spatial and spin space are denoted with
$\cdot$ and $\circ$, respectively. Accordingly, we define the cross product for the spatial space
with $\crossproduct$ and that of the spin space with $\crossproduct_{\sigma}$.
This $\tilde{\tau}_0$ is invariant under a position-dependent phase and a position-dependent
spin rotation of the spinors, or a local U(1)$\times$SU(2) transformation.
\cite{Vignale.Rasolt:Current-.1988, Bencheikh:Spin-orbit.2003,
Pittalis.Vignale.ea:U1.2017, Desmarais.Maul.ea:Spin.2024}

For the Kramers-restricted reference, the spin-resolved variables of the XC functional reduce to
$\rho_{\uparrow} = \rho_{\downarrow} = \rho_0/2$, $\gamma_{\uparrow\uparrow} = \gamma_{\uparrow\downarrow} = \gamma_{\downarrow\downarrow} = \gamma_0/4$,
and $\tilde{\tau}_{\uparrow} = \tilde{\tau}_{\downarrow} = \tilde{\tau}_0/2 = t_0/2$,
as the spin current density enters $\tilde{\tau}_{\uparrow, \downarrow}$ through
$|\mathbf{j}_{\uparrow, \downarrow}|^2 = \tfrac{1}{4} \sum_{\alpha} \sum_u Y_{\alpha, u}^2 = \tfrac{1}{4} \mathbf{Y} \odot \mathbf{Y}$.
Thus, the XC energy is a functional $E^{\text{XC}} = \int e^{\text{XC}} (\rho_0, \gamma_0, \tilde{\tau}_0) \, \text{d}\mathbf{r}$
with $\gamma_0 = \boldsymbol{\nabla} \rho_0 \cdot \boldsymbol{\nabla} \rho_0$.
With $V^{\text{XC}}_{\mu \nu, k} = \partial E^{\text{XC}} / \partial P_{\mu \nu, k}$ and
\begin{equation}
    \frac{\partial \tilde{\tau}_0}{\partial \rho_0} = \frac{\mathbf{Y} \odot \mathbf{Y}}{2 \rho_0^2} \text{,} \qquad
    \frac{\partial \tilde{\tau}_0}{\partial Y_{\alpha, u}} = - \frac{Y_{\alpha, u}}{\rho_0} \text{,}
\end{equation}
the particle part of the XC potential reads
\begin{equation}
\begin{split}
V^{\text{XC}}_{\mu \nu, 0}
= & \int \left[ v^{\text{XC,s}}_{\text{LDA}}
    + v^{\text{XC,s}}_{\tau} \frac{\mathbf{Y} \odot \mathbf{Y}}{2 \rho_0^2} \right]
    \chi_{\mu} \chi_{\nu}^* \, \text{d}\mathbf{r} \\
& + \int v^{\text{XC,s}}_{\text{GGA}} \, \boldsymbol{\nabla} \rho_0 \cdot
    \left[ (\boldsymbol{\nabla} \chi_{\mu}) \chi_{\nu}^*
    + \chi_{\mu} \boldsymbol{\nabla} \chi_{\nu}^* \right] \text{d}\mathbf{r} \\
& + \frac{1}{2} \int v^{\text{XC,s}}_{\tau} \,
    (\boldsymbol{\nabla} \chi_{\mu}) \cdot (\boldsymbol{\nabla} \chi_{\nu}^*) \, \text{d}\mathbf{r} \text{,}
\end{split}
\label{eq:vxc-particle}
\end{equation}
where the second term in the first bracket is the diamagnetic-like contribution arising from the
$\rho_0$-dependence of $\tilde{\tau}_0$. The spin parts, $u = x, y, z$, only arise from the spin current densities,
\begin{equation}
\begin{split}
    V^{\text{XC}}_{\mu \nu, u}
    = &  \frac{\text{i}}{2} \int \frac{v^{\text{XC,s}}_{\tau}}{\rho_0} \,
        \mathbf{Y}_u 
    \cdot \left[ (\boldsymbol{\nabla} \chi_{\mu}) \chi_{\nu}^*
        - \chi_{\mu} \boldsymbol{\nabla} \chi_{\nu}^* \right] \text{d}\mathbf{r} \text{,}
\end{split}
\label{eq:vxc-spin}
\end{equation}
and enter the antisymmetric part of the KS matrix. All terms depending on $m_u$, $\boldsymbol{\nabla} m_u$, $u_u$, and
the particle current density $\mathbf{j}_0$ vanish for the Kramers-restricted reference.
All three spin current densities contribute to Eqs.~\eqref{eq:gen-tau} and \eqref{eq:vxc-spin}, 
which retains the invariance under global spin rotations.
The terms such as $v^{\text{XC,s}}_{\text{LDA}}$ are the functional ingredients itself.
The local density approximation (LDA), GGA, and meta-GGA quantities are defined as
\begin{align}
    v^{\text{XC,s}}_{\text{LDA}} = & \frac{\partial e^{\text{XC}}}{\partial \rho_{\uparrow}}
        =  \frac{\partial e^{\text{XC}}}{\partial \rho_{0}} \text{,} \\
    v^{\text{XC,s}}_{\text{GGA}} = & \frac{\partial e^{\text{XC}}}{\partial \gamma_{\uparrow \uparrow}} + \frac{1}{2} \frac{\partial e^{\text{XC}}}{\partial \gamma{\uparrow \downarrow}} = 2 \frac{\partial e^{\text{XC}}}{\partial \gamma_{0}} \text{,} \\
    v^{\text{XC,s}}_{\tau} = & \frac{\partial e^{\text{XC}}}{\partial \tilde{\tau}_{\uparrow}}
        = \frac{\partial e^{\text{XC}}}{\partial \tilde{\tau}_{0}} \text{,}
\end{align}
where we again used that spin density variables vanish for the ground state, i.e.\ $m_u = u_u = \boldsymbol{\nabla} m_u = 0$.
The GGA singlet potential $v^{\text{XC,s}}_{\text{GGA}}$ follows from the chain rule in the Kramers-restricted
limit, c.f.\ Refs.~\citenum{Peralta.Scuseria.ea:Noncollinear.2007, Armbruster.Weigend.ea:Self-consistent.2008,
Desmarais.Komorovsky.ea:Spin-orbit.2021} for details. We refer to the literature for details on
the chain rule of current-dependent meta-GGAs. \cite{Holzer.Franzke.ea:Current.2022, Grotjahn.Furche.ea:Importance.2022,
Irons.David.ea:Optimizing.2021, Pausch.Holzer:Linear.2022} In practice, the
generalized kinetic energy density is formed and then the specific density functional
approximation, e.g. TPSS or r$^2$SCAN, is evaluated by replacing $\tau$ with $\tilde{\tau}$
to evaluate $\frac{\partial e^{\text{XC}}}{\partial \tilde{\tau}_{0}}$.

\subsection{NMR Shielding Tensor}
\label{subsec:nmr-shielding}
The NMR shielding tensor $\boldsymbol{\sigma}_{\beta, \alpha}^N$ is defined as the mixed second-order
derivative of the external magnetic field $\mathbf{B}$ and the nuclear magnetic moments
$\boldsymbol{\mu}$ according to
\begin{equation}
\small
\sigma_{\beta \alpha}^N = \left. \frac{\text{d}^2 E}{\text{d} B_{\alpha} \text{d} \mu_{I, \beta}} \right|_{\mathbf{B},\boldsymbol{\mu} = 0} =
\tr \left( \textbf{P} \frac{\partial^2 \textbf{h}}{\partial B_{\alpha} \partial \mu_{N, \beta}} \right)+
\tr \left( \frac{\partial \textbf{P}}{\partial B_{\alpha}} \frac{\partial \textbf{h}}{\partial \mu_{N, \beta}} \right)
\label{eq:shielding}
\end{equation}
in the limit of vanishing magnetic perturbations. Here, we also use the Greek indices $\alpha, \beta$ for the
Cartesian components of the perturbations. $\mathbf{h}$ is the one-electron Hamiltonian, for which we consider
exact two-component (X2C) theory \cite{Dyall:exact.1994, Kutzelnigg.Liu:Quasirelativistic.2005,
Ilias.Saue:infinite-order.2007} with a superposition of atomic potentials (SAP) for the two-electron part.
\cite{Surjuse.Valeev:SAP-X2C.2026, Franzke.Holzer:Application.2026}
The first term of Eq.~\eqref{eq:shielding} describes the unperturbed or SCF density contribution,
whereas the second term describes the response of the density with respect to the external
magnetic field. Here, complex GIAOs
\begin{equation}
  \chi_\mu(\mathbf B)=
  \exp\!\left[-\frac{\mathrm i}{2c}
  (\mathbf B\!\times\!\mathbf R_\mu)\!\cdot\!\mathbf r\right]\chi_\mu,
  \label{eq:giao}
\end{equation}
introduce additional derivatives of the two-electron integrals and the exchange-correlation
potential. \cite{Ditchfield:Self-consistent.1974} $\chi$ without explicit $\mathbf{B}$-dependence
marks the real-valued field-free AO basis function.
Capital letters refer to nuclear coordinates and minor letters to electronic coordinates.

The occupied-occupied contribution to the perturbed density matrix of Eq.~\eqref{eq:shielding}
is obtained from the orbital rotation matrix
\begin{equation}
O_{ij}^{B_{\alpha}} = - \frac{1}{2} S_{ij}^{B_{\alpha}} \text{,}
\end{equation}
where $\mathbf{S}^B$ indicates the derivative of the overlap matrix with respect to the magnetic
field. This contribution does not require to iteratively solve response equations and is independent
of the XC functional rung. However, the occupied-virtual part necessitates to iteratively solve the
coupled perturbed Kohn--Sham (CPKS) equations
\cite{Franzke.Holzer:Exact.2023, Olejniczak.Bast.ea:simple.2012, Komorovsky.Repisky.ea:Fully.2010,
Autschbach:role.2013, Holzer.Franzke:Relativistic.2026}
\begin{equation}
\label{eq:cpks}
    \begin{pmatrix}
        \mathbf{A}^{\phantom{*}} & \mathbf{B}^{\phantom{*}} \\
        \mathbf{B}^{*} & \mathbf{A}^{*}
    \end{pmatrix}
    \begin{pmatrix}
        \mathbf{U}^B \\
        \mathbf{V}^B
    \end{pmatrix}
    = 
    \begin{pmatrix}
        \mathbf{b}{\phantom{^*}} \\
        \mathbf{b}^*
    \end{pmatrix}
\end{equation}
with the right-hand side (RHS) contribution $\mathbf{b}$ and the solution
vectors $(\mathbf{U}^B, \mathbf{V}^B)$.
The electronic Hessian matrix on the left hand side (LHS) reads
\begin{align}
\small
    A_{ai,bj} = & (\epsilon_a - \epsilon_i) \delta_{ij} \delta_{ab} + C_{ai,jb} \\
    B_{ai,bj} = & C_{ai,bj} 
\end{align}
where $\mathbf{C}$ denotes the interaction kernel, 
\begin{equation}
  C_{pq,rs} = (pq|rs) + f^{\text{XC}}_{pq,rs} - c_{\text{X}}(ps|qr)
  \text{.}
\end{equation}
Here, $(pq|rs)$ is a complex electron repulsion integral in Mulliken notation,
$\mathbf{f}^{\text{XC}}$ the non-collinear exchange-correlation (XC)
kernel, and $c_{\text{X}}$ the static admixture of exact exchange.
\cite{Gao.Zou.ea:Time-dependent.2005, Wang.Ziegler.ea:calculation.2005, Egidi.Sun.ea:Two-Component.2017,
Kehry.Franzke.ea:Quasirelativistic.2020, Komorovsky.Cherry.ea:Four-component.2019,
Franzke.Holzer:Exact.2023, Franzke.Holzer:Application.2026}
The RHS consists of a one-electron (1e) and a two-electron (2e) contribution,
\begin{equation}
    {b}^B_{ai} = {b}^{B, \text{1e}}_{ai} + {b}^{B, \text{2e}}_{ai} =
    \left({h}^B_{ai} - {\bf\epsilon}_i {S}^B_{ai} \right) +
    \left({b}^{B, {\text{pot}}}_{ai} + {b}^{B, \text{kernel}}_{ai} \right) \text{.}
\end{equation}
${\bf\epsilon}$ refers to the KS spinor eigenvalues. The two-electron part is partitioned
into a potential and a kernel contribution.
\cite{Olejniczak.Bast.ea:simple.2012, Autschbach:role.2013, Franzke.Holzer:Exact.2023}
We refer to Ref.~\citenum{Franzke.Holzer:Exact.2023} for the derivation up to meta-GGAs and their hybrids
with the Maximoff--Scuseria generalization of $\tau$. Crucially, the two-component spin--orbit formalism
always requires to solve the CPKS equations iteratively---also with pure or semilocal meta-GGAs
in the Maximoff--Scuseria generalization. This is due to the kernel being of triplet character and
the spin current densities give rise to response spin densities during the CPKS procedure.
\cite{Olejniczak.Bast.ea:simple.2012, Autschbach:role.2013, Franzke.Holzer:Exact.2023}
As the ground-state spin density variables vanish, the kernel is still a substantial
simplification of the complete non-collinear XC kernel.
In a non-relativistic or scalar-relativistic framework, only the current-dependent generalization
requires an iterative procedure, \cite{Schattenberg.Kaupp:Effect.2021, Bates.Furche:Harnessing.2012}
while other pure semilocal density functional rungs can be solved in one shot.
\cite{Malkin.Malkina.ea:Calculations.1993, Cheeseman.Trucks.ea:comparison.1996, Maximoff.Scuseria:Nuclear.2004}
Therefore, the computational overhead due to the current-dependent generalization will be
basically vanishing in the presence of spin--orbit coupling.

\subsection{Derivative of the meta-GGA XC Potential with Respect to the External Magnetic Field}
\label{susec:xc-derivative}

The two-electron part of the RHS consists of the potential and the kernel contribution.
The XC potential contribution, ${b}^{B, \text{pot, XC}}$, is the derivative of the XC potential
of Eqs.~\eqref{eq:vxc-particle} and \eqref{eq:vxc-spin} with respect to the magnetic field
for a fixed SCF density matrix, i.e.\ only the GIAOs of Eq.~\eqref{eq:giao} are differentiated.
The kernel contribution, ${b}^{B, \text{kernel, XC}}$, collects the response of the XC potential
to the field-perturbed density variables.
For a Kramers-restricted KS reference, the particle part ($k = 0$) of the XC potential contribution reads
\begin{equation}
\begin{split}
    \left[ {b}^{B, {\text{pot, XC}}}_{\mu \nu} \right]_{0, \alpha}
   & =  \frac{\partial {V}^{\text{XC}}_{\mu \nu, 0}}{\partial B_{\alpha}} \\
   & =  - \frac{\text{i}}{2c} \int v^{\text{XC,s}}_{\text{LDA}}
        (\mathbf{R}_{\mu \nu} \crossproduct \mathbf{r})_{\alpha} \,
        \chi_{\mu} \chi_{\nu} \, \text{d}\mathbf{r} \\
    & - \frac{\text{i}}{2c} \int v^{\text{XC,s}}_{\text{GGA}} \,
        (\mathbf{R}_{\mu \nu} \crossproduct \mathbf{r})_{\alpha} \\
    & \qquad \times \boldsymbol{\nabla}\rho_0 \cdot
        \left[ (\boldsymbol{\nabla} \chi_{\mu}) \chi_{\nu}
        + \chi_{\mu} \boldsymbol{\nabla} \chi_{\nu} \right] \text{d}\mathbf{r} \\
    & - \frac{\text{i}}{2c} \int v^{\text{XC,s}}_{\text{GGA}} \,
        \boldsymbol{\nabla}\rho_0 \cdot
        \left( \mathbf{B} \crossproduct \mathbf{R}_{\mu \nu} \right)^{B_{\alpha}}
        \chi_{\mu} \chi_{\nu} \, \text{d}\mathbf{r} \\
    & - \frac{\text{i}}{2c} \int v^{\text{XC,s}}_{\tau}
        \frac{\mathbf{Y} \odot \mathbf{Y}}{2 \rho_0^2}
        (\mathbf{R}_{\mu \nu} \crossproduct \mathbf{r})_{\alpha} \,
        \chi_{\mu} \chi_{\nu} \, \text{d}\mathbf{r} \\
    & - \frac{\text{i}}{4c} \int v^{\text{XC,s}}_{\tau} \Big[
        (\mathbf{R}_{\mu \nu} \crossproduct \mathbf{r})_{\alpha}
        (\boldsymbol{\nabla} \chi_{\mu}) \cdot (\boldsymbol{\nabla} \chi_{\nu}) \\
    & \qquad + \chi_{\mu} ( \mathbf{R}_{\mu} \crossproduct \boldsymbol{\nabla} \chi_{\nu} )_{\alpha}
        - \chi_{\nu} ( \mathbf{R}_{\nu} \crossproduct \boldsymbol{\nabla} \chi_{\mu} )_{\alpha}
        \Big] \text{d}\mathbf{r} \\
    & + \frac{\text{i}}{2} \int \frac{v^{\text{XC,s}}_{\tau}}{\rho_0} \,
        \mathbf{j}^{B_{\alpha}}_{D, 0} \cdot
        \left[ (\boldsymbol{\nabla} \chi_{\mu}) \chi_{\nu}
        - \chi_{\mu} \boldsymbol{\nabla} \chi_{\nu} \right] \text{d}\mathbf{r} \text{,}
\end{split}
\label{eq:xc-b}
\end{equation}
with $\mathbf{R}_{\mu \nu} = \mathbf{R}_{\mu} - \mathbf{R}_{\nu}$ and $\mathbf{j}^{B_{\alpha}}_{D, 0}$ is given below.
Here, the first meta-GGA term arises from the $\rho_0$-dependence of $\tilde{\tau}_0$, the second term
from the GIAO derivative of $\tfrac{1}{2} (\boldsymbol{\nabla} \chi_{\mu}^*(\mathbf B)) \cdot (\boldsymbol{\nabla} \chi_{\nu}(\mathbf B))$
or $\tfrac{1}{2} (\boldsymbol{\nabla} \chi_{\mu}(\mathbf B)) \cdot (\boldsymbol{\nabla} \chi_{\nu}^*(\mathbf B))$
including the gradient acting on the GIAO phase, and the third from the field derivative of the
paramagnetic particle current density at fixed density matrix, $\mathbf{j}^{B_{\alpha}}_{D, 0}$, which is
non-zero although $\mathbf{j}_0$ vanishes for the Kramers-restricted reference. \cite{Schattenberg.Kaupp:Effect.2021}
In contrast to the non-relativistic limit, the non-zero spin current densities of the SCF reference
lead to additional spin contributions of the potential derivative, cf.\ Ref.~\citenum{Schattenberg.Kaupp:Effect.2021}.
Differentiating the spin-dependent part of the XC potential in Eq.~\eqref{eq:vxc-spin} with GIAOs yields
\begin{equation}
\begin{split}
    \left[ {b}^{B, {\text{pot, XC}}}_{\mu \nu} \right]_{u, \alpha}
    = & \frac{\partial {V}^{\text{XC}}_{\mu \nu, u}}{\partial B_{\alpha}} = \frac{1}{4c} \int \frac{v^{\text{XC,s}}_{\tau}}{\rho_0} \Big\{
        (\mathbf{R}_{\mu \nu} \crossproduct \mathbf{r})_{\alpha} \\
    & \qquad \times \mathbf{Y}_u \cdot
        \left[ (\boldsymbol{\nabla} \chi_{\mu}) \chi_{\nu}
        - \chi_{\mu} \boldsymbol{\nabla} \chi_{\nu} \right] \\
    & + \left[ ( \mathbf{R}_{\mu} + \mathbf{R}_{\nu} ) \crossproduct \mathbf{Y}_u \right]_{\alpha}
        \chi_{\mu} \chi_{\nu} \Big\} \text{d}\mathbf{r} \text{,}
\end{split}
\label{eq:xc-b-spin}
\end{equation}
where the first term stems from the GIAO phase of the basis function pair and the second one from the
gradient acting on the phase. The latter also survives for one-center pairs, $\mathbf{R}_{\mu \nu} = 0$.
Up to the GGA rung, the XC potential and its derivative are only non-zero for the particle
contribution, as we only consider closed-shell systems herein and $m_{\text{s}} = 0$ for the
SCF density matrix. For the current-dependent meta-GGAs, the spin contributions of
Eq.~\eqref{eq:xc-b-spin} are non-zero due to the spin current densities $\mathbf{Y}_u$
of the SCF reference. The RHS kernel always gives rise to non-zero spin contributions.
The magnetic field perturbed density variables $\rho^B$ are given as
\begin{align}
    \rho^{B_{\alpha}}_{k} = & \rho^{B_{\alpha}}_{R, k} + \rho^{B_{\alpha}}_{D, k} \text{,}
\end{align}
and its gradient, the kinetic energy density, and current density follow the same way.
The reorthonormalization part, utilizing the modified density matrix $O_{\mu \nu}^{B_{\alpha}}$, reads
\begin{align}
    \rho^{B_{\alpha}}_{R, k} = & \sum_{\mu \nu} O_{\mu \nu, k}^{B_{\alpha}} \,
        \chi_{\mu} \chi_{\nu} \text{,} \\
    \boldsymbol{\nabla} \rho^{B_{\alpha}}_{R, k} = & \sum_{\mu \nu} O_{\mu \nu, k}^{B_{\alpha}}
        \left[ (\boldsymbol{\nabla} \chi_{\mu}) \chi_{\nu} + \chi_{\mu} \boldsymbol{\nabla} \chi_{\nu} \right] \text{,} \\
    \tau^{B_{\alpha}}_{R, k} = & \frac{1}{2} \sum_{\mu \nu} O_{\mu \nu, k}^{B_{\alpha}}
        (\boldsymbol{\nabla} \chi_{\mu}) \cdot (\boldsymbol{\nabla} \chi_{\nu}) \text{,} \\
    \mathbf{j}^{B_{\alpha}}_{R, k} = & - \frac{\text{i}}{2} \sum_{\mu \nu} O_{\mu \nu, k}^{B_{\alpha}}
        \left[ (\boldsymbol{\nabla} \chi_{\mu}) \chi_{\nu} - \chi_{\mu} \boldsymbol{\nabla} \chi_{\nu} \right] \text{.}
\end{align}
For the density variables, only the symmetric part of $O_{\mu \nu}^{B_{\alpha}}$ yields a non-vanishing
response, whereas the antisymmetric part gives rise to the response of the current density variables.
The contributions from the direct GIAO terms read
\begin{align}
    \rho^{B_{\alpha}}_{D, k} = & - \frac{\text{i}}{2c} \sum_{\mu \nu} P_{\mu \nu, k} \,
        (\mathbf{R}_{\mu \nu} \crossproduct \mathbf{r})_{\alpha} \, \chi_{\mu} \chi_{\nu} \text{,}
        \label{eq:rho-direct} \\
    \begin{split}
    \boldsymbol{\nabla} \rho^{B_{\alpha}}_{D, k} = & - \frac{\text{i}}{2c} \sum_{\mu \nu} P_{\mu \nu, k} \Big[
        (\mathbf{R}_{\mu \nu} \crossproduct \mathbf{r})_{\alpha} \\
    & \qquad \times \left( (\boldsymbol{\nabla} \chi_{\mu}) \chi_{\nu} + \chi_{\mu} \boldsymbol{\nabla} \chi_{\nu} \right) \\
    & + \left( \mathbf{B} \crossproduct \mathbf{R}_{\mu \nu} \right)^{B_{\alpha}} \chi_{\mu} \chi_{\nu} \Big] \text{,}
    \end{split} \\
    \begin{split}
    \tau^{B_{\alpha}}_{D, k} = & - \frac{\text{i}}{4c} \sum_{\mu \nu} P_{\mu \nu, k} \Big[
        (\mathbf{R}_{\mu \nu} \crossproduct \mathbf{r})_{\alpha}
        (\boldsymbol{\nabla} \chi_{\mu}) \cdot (\boldsymbol{\nabla} \chi_{\nu}) \\
    & + \chi_{\mu} ( \mathbf{R}_{\mu} \crossproduct \boldsymbol{\nabla} \chi_{\nu} )_{\alpha}
      - \chi_{\nu} ( \mathbf{R}_{\nu} \crossproduct \boldsymbol{\nabla} \chi_{\mu} )_{\alpha} \Big] \text{,}
    \end{split} \label{eq:tau-direct} \\
    \begin{split}
    \mathbf{j}^{B_{\alpha}}_{D, k} = & - \frac{1}{4c} \sum_{\mu \nu} P_{\mu \nu, k} \Big\{
        (\mathbf{R}_{\mu \nu} \crossproduct \mathbf{r})_{\alpha} \\
    & \qquad \times \left[ (\boldsymbol{\nabla} \chi_{\mu}) \chi_{\nu} - \chi_{\mu} \boldsymbol{\nabla} \chi_{\nu} \right] \\
    & + \left[ \mathbf{B} \crossproduct ( \mathbf{R}_{\mu} + \mathbf{R}_{\nu} ) \right]^{B_{\alpha}}
        \chi_{\mu} \chi_{\nu} \Big\} \text{.}
    \end{split} \label{eq:j-direct}
\end{align}
The prefactor of Eq.~\eqref{eq:j-direct} is real, since the imaginary unit of the GIAO phase
and that of the current operator combine.
The brackets of Eqs.~\eqref{eq:rho-direct} and \eqref{eq:tau-direct} are antisymmetric with respect
to the interchange of $\mu$ and $\nu$, whereas that of Eq.~\eqref{eq:j-direct} is symmetric.
Hence, only the spin contributions, $k=u$, are non-zero for the density variables
and only the particle contribution, $k=0$, is non-zero for the current density.
The first-derivative term of the latter, $\mathbf{j}^{B_{\alpha}}_{D, 0}$, i.e.\ its contraction with
$\partial^2 e^{\text{XC}} / \partial \mathbf{j}_0 \partial \mathbf{j}_0 = - v^{\text{XC,s}}_{\tau} / \rho_0$,
is already accounted for in the potential contribution of Eq.~\eqref{eq:xc-b}. Therefore, only its coupling
to the spin channels via the spin current densities $\mathbf{Y}_u$ of the reference enters the kernel contribution.

\subsection{Second-Order Variation and XC Kernel}
\label{subsec:variation-tau}
Due to the non-zero response of the spin magnetization, a generalization of the ground-state
$\tilde{\tau}_0$ only is not sufficient for NMR. The second-order variation and the XC kernel
also have to meet the transformation requirements for the exact exchange correlation
energy. \cite{Vignale.Rasolt:Current-.1988, Bencheikh:Spin-orbit.2003, Pittalis.Vignale.ea:U1.2017}
Here, we stress that these transformation requirements already necessitate changes
in the GGA kernel, as outlined in the appendix~\ref{sec:trafo}. In short, the ordinary
gradient of the magnetization component by component \cite{Autschbach:role.2013,
Kuehn.Weigend:Implementation.2013, Franzke.Holzer:Exact.2023} breaks invariance for
the GGA kernel in NMR and time-dependent DFT calculations with a closed-shell
ground-state reference. Therefore, a covariant GGA kernel is derived in appendix~\ref{sec:gga-covariant}.
In the main text, we focus on the second-order variation of the generalized $\tau$-dependent
part and present the main findings for the XC kernel of meta-GGAs with a closed-shell ground-state
reference. Further details are given in appendix~\ref{sec:detailed-derivation} and we
compare our covariant formulation to the kernel which would arise from minimal
substitution proposed previously \cite{Pittalis.Vignale.ea:U1.2017, Desmarais.Maul.ea:Spin.2024}
in appendix~\ref{bec:comparison-sub}.

We define the $2 \times 2$ spin space matrices $\hat{R} = \rho_0\,\one + \sv m\circ\hat{\boldsymbol{\sigma}}$,
$\hat{Q} = \tau_0\,\one + \sv u\circ\hat{\boldsymbol{\sigma}}$ and $\hat{J}_{\alpha} = j_{0,\alpha}\,\one + \sv Y_{\alpha}\circ \hat{\boldsymbol{\sigma}}$.
The kinetic energy density variable for the functional becomes
\begin{equation}
  \hat{T} = \pk + \hat{W},
\end{equation}
with the Pauli kinetic energy density matrix $\pk$ given by
\begin{align}
  \pk &= \hat{Q} - \frac12\sum_{\alpha} \hat{K}_{\alpha} \hat{R}^{-1} \hat{K}_{\alpha}^\dagger, \label{eq:P}
\end{align} 
using the auxiliary variables $\hat{K}_{\alpha}$, $\hat{R}^{-1}$, and $D$,
\begin{align}
  \hat{K}_{\alpha} = & \tfrac12\,\nabla_{\alpha} \hat{R} + \text{i} \hat{J}_{\alpha}, \\
  \hat{R}^{-1} = & \frac{\rho_0\,\one - \sv m\circ \hat{\boldsymbol{\sigma}}}{D}, \\
  D = & \rho_0^2 - \sv m\circ\sv m \text{,}
\end{align}
as well as the von-Weizs\"acker matrix $\hat{W}$,
\begin{align}
  \hat{W} &= \frac{1}{8D}\sum_{\alpha} \Big[S\big(\nabla_{\alpha} \rho_0,\mathcal D_{\alpha} \sv m\big)\,\one
  + \sv V\big(\nabla_{\alpha} \rho_0,\mathcal D_{\alpha} \sv m\big)\circ\hat{\bm\sigma}\Big], \label{eq:W}
\end{align}
which is formed with the auxiliary variables $S$, $\sv V$, and the covariant gradient variable for
GGAs discussed in appendix \ref{sec:gga-covariant},
\begin{align}
  S(x,\sv y) &= \rho_0(x^2 + \sv y\circ\sv y) - 2x\,(\sv y\circ\sv m), \\
  \sv V(x,\sv y) & = 2\rho_0x\,\sv y - (x^2 - \sv y\circ\sv y)\,\sv m - 2(\sv m\circ\sv y)\,\sv y, \\
  \mathcal D_{\alpha}\sv m & = \nabla_{\alpha}\sv m + \frac{2}{\rho_0}\,\sv Y_{\alpha}\xs\sv m \text{.}
\end{align}
Details are given in appendix \ref{sec:detailed-derivation}. Just like the other matrices $\hat{R}$,
$\hat{Q}$, and $\hat{J}_{\alpha}$, the kinetic energy variable $T$ consists of a charge and a spin contribution,
i.e.\ $\hat{T} = t_0\,\one + \sv t\circ\hat{\boldsymbol{\sigma}}$. For the ground-state SCF solution,
$\sv t = 0$ holds and we obtain
\begin{equation}
  t_0 = \tau_0 - \frac{1}{2\rho_0}\sum_{\alpha}\sv Y_{\alpha}\circ\sv Y_{\alpha}
  \ge \frac{|\boldsymbol{\nabla}\rho_0|^2}{8\rho_0},
\end{equation}
with equality for one Kramers pair. So, the iso-orbital condition and the von-Weizs\"acker
inequality hold as required and the generalized kinetic energy density $\tilde{\tau}$ is
recovered.

For the response of the kinetic energy variable, we further define the spin current density
variables
\begin{equation}
  \kappa = \frac{1}{2\rho_0^2}\sum_{\alpha}\sv Y_{\alpha}\circ\sv Y_{\alpha}, \qquad
  G_{uv} = \sum_{\alpha} Y_{\alpha,u}Y_{\alpha,v}
\end{equation}
and use the covariant derivative of the gradient variable, $X_{\alpha, u}$, as defined in Eq.~\eqref{eq:scX}.
This way, the first variation of the spin part and the second variation of the charge part
for a time-odd response such as a magnetic field read
\begin{align}
  \delta\sv t &= \delta\sv u - \frac{1}{\rho_0}\sum_{\alpha}\sv Y_{\alpha}\,\delta j_{0,\alpha}
  + \kappa\,\delta\sv m
  - \frac{1}{2\rho_0}\sum_{\alpha}\sv X_{\alpha}\xs\sv Y_{\alpha} , \label{eq:scdt}\\
  \begin{split}
  \tfrac12\,\delta^2t_0 &= -\frac{1}{2\rho_0}\sum_{\alpha}(\delta j_{0,\alpha})^2
  + \frac{1}{\rho_0^2}\sum_{\alpha}\delta j_{0,\alpha}\,(\sv Y_{\alpha}\circ\delta\sv m) \\
  & - \frac{1}{2\rho_0^3}\sum_{uv}\delta m_u\,G_{uv}\,\delta m_v . \label{eq:scd2t}
  \end{split}
\end{align}

Based on the response, the second-order XC energy can be derived. Here, the spin part of the
second-order XC energy is of quadratic form in $\delta\sv m$, $X_{\alpha, u}$, and $\delta\sv t$.
For the latter, the first derivative regarding the kinetic variable adds the second variation of $t_0$.
Hence,
\begin{align}
  E^{\text{XC}, (2)}_{\text{spin}} = & \int\Big\{\sum_u \tfrac12 f_1\,\delta m_u^2  \nonumber\\
  & + \sum_u \Big[  f_2\,\delta m_u\,(\vc g\cdot\vc X_u)
  + \tfrac12 f_3\,(\vc g\cdot\vc X_u)^2
  + \tfrac12 f_4\sum_{\alpha} X_{\alpha,u}^2 \Big] \nonumber\\
  & + \sum_u \Big[  f_5\,\delta m_u\,\delta t_u + \tfrac12 f_6\,\delta t_u^2
  + f_7\,(\vc g\cdot\vc X_u)\,\delta t_u\Big] \nonumber\\
  & + f_\tau\,\tfrac12\,\delta^2t_0\Big\}\,\text{d} \mathbf{r} .
  \label{eq:scE2m}
\end{align}
The GGA weights $f_1$ to $f_4$ are given in the appendix~\ref{sec:gga-covariant}. The meta-GGA weights
read
\begin{align}
  f_5 &= \frac12\Big(\frac{\partial^2 e^{\text{XC}}}{\partial\rho_{\uparrow}\partial\tilde{\tau}_{\uparrow}}
  - \frac{\partial^2 e^{\text{XC}}}{\partial\rho_{\uparrow}\partial\tilde{\tau}_{\downarrow}}\Big), &
  f_6 &= \frac12\Big(\frac{\partial^2 e^{\text{XC}}}{\partial\tilde{\tau}_{\uparrow}\partial\tilde{\tau}_{\uparrow}}
  - \frac{\partial^2 e^{\text{XC}}}{\partial\tilde{\tau}_{\uparrow}\partial\tilde{\tau}_{\downarrow}}\Big),
  \nonumber\\
  f_7 &= \frac12\Big(\frac{\partial^2 e^{\text{XC}}}{\partial\gamma_{\uparrow\uparrow}\partial\tilde{\tau}_{\uparrow}}
  - \frac{\partial^2 e^{\text{XC}}}{\partial\gamma_{\uparrow\uparrow}\partial\tilde{\tau}_{\downarrow}}\Big), &
  f_\tau &= \frac{\partial e^{\text{XC}}}{\partial\tilde{\tau}_{\uparrow}} ,
\end{align}
and are evaluated at the Kramers-restricted reference so that $\tilde{\tau}_{\uparrow} = \tilde{\tau}_{\downarrow} = t_0/2 = \tilde{\tau}_0/2$.
Just like for the GGA weights, the derivatives remain unchanged upon exchange of two spin labels.
The weights refer to one spin component $u$ at one time. That is, the kinetic variables of the two
channels change by $+\delta t_u/2$ and $-\delta t_u/2$ for the spin channels along the axis $u$. 

For the first-order potential, the derivatives of the curly bracket of Eq.~\eqref{eq:scE2m} with
respect to the spin variables $\delta m_u$, $X_{\alpha,u}$, and $\delta t_u$ read
\begin{align}
  h_u &= f_1\,\delta m_l + f_2\,(\vc g\cdot\vc X_l) + f_5\,\delta t_l , \\
  v_{\alpha,u} &= \big(f_2\,\delta m_u + f_3\,(\vc g\cdot\vc X_u) + f_7\,\delta t_u\big)\,g_{\alpha}
  + f_4\,X_{\alpha,u} , \\
  \eta_u &= f_5\,\delta m_u + f_7\,(\vc g\cdot\vc X_u) + f_6\,\delta t_u .
\end{align}
Note that each derivative is formed with the other two variables held fixed.
Therefore, the so-called rows of the first-order potential follow with the chain rule
as usual by considering Eqs.~\eqref{eq:scX}, \eqref{eq:scdt}, and \eqref{eq:scd2t}.
See also appendix \ref{sec:gga-covariant} for a detailed explanation of the rows and the
first-order potential of the GGA part. This finally yields the expressions
\begin{align}
  \text{kinetic row, spin:} \quad & \bm\eta , \\
  \text{gradient row, spin:} \quad & \sv v_{\alpha}' = \sv v_{\alpha}
  - \frac{1}{2\rho_0}\,\sv Y_{\alpha}\xs\bm\eta , \\
  \begin{split}
  \text{density row, spin:} \quad & \sv h + \kappa\,\bm\eta
  + \frac{2}{\rho_0}\sum_{\alpha}\sv v_{\alpha}'\xs\sv Y_{\alpha} \\
  & + f_\tau\Big[\frac{1}{\rho_0^2}\sum_{\alpha}\delta j_{0,\alpha}\,\sv Y_{\alpha}
  - \frac{1}{\rho_0^3}\,G\,\delta\sv m\Big] , 
  \end{split} \\
 \begin{split}
  \text{current row, particle:} \quad & -\frac{1}{\rho_0}\,\sv Y_{\alpha}\circ\bm\eta
  + \frac{f_\tau}{\rho_0^2}\,\sv Y_{\alpha}\circ\delta\sv m \\
  & - \frac{f_\tau}{\rho_0}\,\delta j_{0,\alpha} .
  \end{split} 
\end{align}
Here, the kinetic row is multiplied with the kinetic energy density term
$\tfrac12\sum_{\alpha}(\nabla_{\alpha}\chi_\mu)(\nabla_{\alpha}\chi_\nu)$
and $\hat{\sigma}_u$. Accordingly, the current row is multiplied with the
paramagnetic current density term
$\frac{\text{i}}{2} \left[ \left(\nabla_{\alpha} \chi_{\mu} \right) \chi_{\nu} - \chi_{\mu} \left(\nabla_{\alpha} \chi_{\nu}\right) \right]$
and the identity matrix $\one$.
A time-odd perturbation such as a magnetic field does not give rise to a
charge channel and to the rows of the spin-current density operators.
Overall, the kernel is invariant under a local phase transformation and
a local spin rotation of the reference and the response.

A few comments are in order. First, the last term of the current row
is omitted for the direct GIAO RHS kernel, as it is included in the field
derivative of the XC potential. Second, the given kernel reduces to the
triplet kernel and the magnetic kernel of the one-component current-dependent
meta-GGA formulation. \cite{Bates.Furche:Harnessing.2012, Bates.Heiche.ea:Erratum.2022}
Third, the Pauli kinetic energy matrix $\pk$ vanishes along the perturbation
for a closed-shell two-electron system. Then, the kernel becomes that of a
GGA, as the kinetic variable is replaced by its von-Weizs\"acker value
in both the non-relativistic limit and in the presence of SOC.
This allows for a further validation. \cite{Bates.Heiche.ea:Erratum.2022}

For Se NMR shieldings of H$_2$Se with a double-zeta basis set and a large grid,
the changes from the NMR-limit of the kernel of Ref.~\citenum{Franzke.Holzer:Application.2026},
which does not guarantee local U(1)$\times$SU(2) invariance for the response,
to the covariant form herein amounts to only about 0.04\,ppm. To compare with,
the absolute shielding constant is about 1980\,ppm. For GGAs, the change is
even smaller. This shows that the impact of the invariance on $\tilde{\tau}$
is small for the NMR shieldings from a practical point of view. In contrast,
the impact of generalizing $\tau$ with the currents at all is much larger
(see below in Sec.~\ref{sec:results}).
Same holds for NMR coupling constants and excitation energies in time-dependent DFT,
c.f.\ Ref.~\citenum{Holzer.Franzke.ea:Current.2022}. The presented kernel
can be directly used for NMR coupling constants. \cite{Franzke.Mack.ea:NMR.2021}
The kernel for excitation energies in time-dependent DFT is discussed in
appendix \ref{sec:tddft} and also needs the time-even part for the response.

\subsection{Extension to Local Hybrid Functionals}
\label{suseb:lhf}

Local hybrid functionals replace the constant admixture $c_{\text{X}}$ of exact exchange
by a position-dependent local mixing function (LMF) $a(\mathbf{r})$,
\cite{Jaramillo.Scuseria.ea:Local.2003, Maier.Arbuznikov.ea:Local.2019}
\begin{equation}
  E^{\text{XC}}_{\text{LH}} = \int \Big\{ a \, e^{\text{X}}_{\text{exact}}
  + (1 - a) \, e^{\text{X}}_{\text{sl}} + e^{\text{C}}_{\text{sl}} \Big\} \, \text{d}\mathbf{r} \text{,}
  \label{eq:lhf}
\end{equation}
where $e^{\text{X}}_{\text{sl}}$ and $e^{\text{C}}_{\text{sl}}$ are the semilocal exchange and
correlation energy densities. The exact-exchange energy density is formed with the 2c
one-particle density matrix $\hat{P}(\mathbf{r}, \mathbf{r}') = \sum_i \phi_i(\mathbf{r}) \phi_i^{\dagger}(\mathbf{r}')
= \sum_{\mu \nu} \hat{P}_{\mu \nu} \chi_{\mu}(\mathbf{r}) \chi_{\nu}^*(\mathbf{r}')$, which is a
$2 \times 2$ matrix in spin space, as
\begin{equation}
\begin{split}
  e^{\text{X}}_{\text{exact}}(\mathbf{r})
  = & - \frac{1}{2} \int \frac{\operatorname{tr} \left[ \hat{P}(\mathbf{r}, \mathbf{r}')
      \hat{P}(\mathbf{r}', \mathbf{r}) \right]}{|\mathbf{r} - \mathbf{r}'|} \, \text{d}\mathbf{r}' \\
  = & - \frac{1}{2} \sum_{\mu \nu \kappa \lambda} \operatorname{tr} \left[ \hat{P}_{\mu \nu} \hat{P}_{\kappa \lambda} \right]
      \chi_{\mu}(\mathbf{r}) \chi_{\lambda}^*(\mathbf{r}) \, A_{\nu \kappa}(\mathbf{r}) \text{,}
\end{split}
\label{eq:exx}
\end{equation}
with the trace over the spin indices and the three-center integrals
$A_{\nu \kappa}(\mathbf{r}) = \int \chi_{\nu}^*(\mathbf{r}') \chi_{\kappa}(\mathbf{r}') / |\mathbf{r} - \mathbf{r}'| \, \text{d}\mathbf{r}'$.
The latter are evaluated analytically, while the integration over $\mathbf{r}$ is carried out
using the seminumerical scheme.
\cite{Plessow.Weigend:Seminumerical.2012, Holzer:improved.2020}
Inserting the Pauli form of Eq.~\eqref{eq:pauli}, the trace yields the sum of a particle
and three spin contributions, so all four components of the density matrix enter.

The XC potential consists of a local and a non-local part. The local part
has the form of Eqs.~\eqref{eq:vxc-particle} and \eqref{eq:vxc-spin}, where the functional
derivatives with respect to the semilocal variables $q \in \{\rho_0, \gamma_0, \tilde{\tau}_0\}$ are
replaced by
\begin{equation}
  \frac{\partial e^{\text{XC}}_{\text{LH}}}{\partial q} =
  (1 - a) \frac{\partial e^{\text{X}}_{\text{sl}}}{\partial q}
  + \frac{\partial e^{\text{C}}_{\text{sl}}}{\partial q}
  + \frac{\partial a}{\partial q} \left( e^{\text{X}}_{\text{exact}} - e^{\text{X}}_{\text{sl}} \right) \text{.}
  \label{eq:lh-deriv}
\end{equation}
Thus, the LMF behaves as an additional meta-GGA ingredient, and its dependence on $\tilde{\tau}_0$
gives rise to spin-current contributions to the potential in addition to the local
exchange and correlation parts. The non-local part is an exchange
operator with the symmetrized LMF,
\cite{Bahmann.Kaupp:Efficient.2015, Schattenberg.Reiter.ea:Efficient.2020}
\begin{equation}
  \hat{K}^{\text{LH}}_{\mu \nu} = - \frac{1}{2} \int \!\! \int \chi_{\mu}^*(\mathbf{r})
  \frac{\left[ a(\mathbf{r}) + a(\mathbf{r}') \right] \hat{P}(\mathbf{r}, \mathbf{r}')}{|\mathbf{r} - \mathbf{r}'|}
  \chi_{\nu}(\mathbf{r}') \, \text{d}\mathbf{r} \, \text{d}\mathbf{r}' \text{,}
  \label{eq:klh}
\end{equation}
which is again a $2 \times 2$ matrix in spin space. The SCF implementation for 
current-dependent complex 2c spinors was presented in Ref.~\citenum{Holzer.Franzke.ea:Current.2022}.

The NMR response requires a dedicated 2c implementation. While the GIAO derivative of the
LH Kohn--Sham matrix structurally follows the 1c implementation,
\cite{Schattenberg.Reiter.ea:Efficient.2020, Holzer.Franzke.ea:Assessing.2021}
the 2c reference carries spin current densities and the field induces a spin magnetization,
which gives rise to terms without a counterpart for a 1c closed-shell reference.
Three contributions are needed. First, the potential contribution ${b}^{B, \text{pot}}$,
i.e.\ the GIAO derivative of the LH Kohn--Sham matrix at fixed density. The local part follows
from Eqs.~\eqref{eq:xc-b} and \eqref{eq:xc-b-spin} with the derivatives of Eq.~\eqref{eq:lh-deriv}.
For the non-local part, the GIAO phases of all four basis functions in Eq.~\eqref{eq:klh}
need to be differentiated,
\begin{equation}
\begin{split}
  \frac{\partial \hat{K}^{\text{LH}}_{\mu \nu}}{\partial B_{\alpha}}
  = & \frac{\text{i}}{4c} \sum_{\kappa \lambda} \int \!\! \int
      \chi_{\mu}^*(\mathbf{r}) \chi_{\kappa}(\mathbf{r})
      \frac{\left[ a(\mathbf{r}) + a(\mathbf{r}') \right] \hat{P}_{\kappa \lambda}}{|\mathbf{r} - \mathbf{r}'|}
      \chi_{\lambda}^*(\mathbf{r}') \chi_{\nu}(\mathbf{r}') \\
    & \qquad \times \left[ \mathbf{R}_{\kappa \mu} \crossproduct \mathbf{r}
      + \mathbf{R}_{\nu \lambda} \crossproduct \mathbf{r}' \right]_{\alpha}
      \text{d}\mathbf{r} \, \text{d}\mathbf{r}' \text{,}
\end{split}
\label{eq:klh-b}
\end{equation}
with $\mathbf{R}_{\kappa \mu} = \mathbf{R}_{\kappa} - \mathbf{R}_{\mu}$. This again requires
the particle and all three spin components of the density matrix, where the latter are
non-zero due to SOC. Second, the kernel contribution ${b}^{B, \text{kernel}}$, i.e.\ the
second variation of the semilocal part contracted with the GIAO-induced density variables of
Eqs.~\eqref{eq:rho-direct} to \eqref{eq:j-direct}. Third, the LH kernel of the CPKS equations.
Here, the exchange integral of the interaction kernel is replaced by its LMF-weighted counterpart,
\begin{equation}
  c_{\text{X}} (ps|qr) \; \rightarrow \;
  \int \!\! \int \phi_p^{\dagger}(\mathbf{r}) \phi_s(\mathbf{r})
  \frac{a(\mathbf{r}) + a(\mathbf{r}')}{2 \, |\mathbf{r} - \mathbf{r}'|}
  \phi_q^{\dagger}(\mathbf{r}') \phi_r(\mathbf{r}') \, \text{d}\mathbf{r} \, \text{d}\mathbf{r}' \text{,}
  \label{eq:lh-kernel}
\end{equation}
and the semilocal XC kernel is that of Sec.~\ref{subsec:variation-tau}, with the weights
evaluated for the LH energy density of Eq.~\eqref{eq:lhf} at fixed $e^{\text{X}}_{\text{exact}}$.
In particular, $f_\tau$ includes $(\partial a / \partial \tilde{\tau}_{\uparrow})
(e^{\text{X}}_{\text{exact}} - e^{\text{X}}_{\text{sl}})$. In general, the LH kernel further
contains the coupling of the LMF response with the response of the exact-exchange energy
density and potential. \cite{Holzer:improved.2020}

For a Kramers-restricted reference and the time-odd magnetic field, these coupling terms
simplify but do not vanish. The particle contributions to $\delta a$ and to the
response of $e^{\text{X}}_{\text{exact}}$ are zero, as both are time-even. This also
holds for the GIAO-induced variation of $e^{\text{X}}_{\text{exact}}$, which is not the
response to a density-matrix perturbation, as the GIAO phases also act on the basis
functions at the second coordinate $\mathbf{r}'$ of Eq.~\eqref{eq:exx}. In contrast, the spin
channels of $\delta a$, i.e.\ the response to $\delta \mathbf{m}$, to the covariant gradient
variable, and to $\delta \mathbf{t}$, survive whenever the LMF depends on spin-resolved
variables. Therefore, the coupling of $\delta a$ with $e^{\text{X}}_{\text{sl}}$ and with the
exact-exchange operator is retained in both ${b}^{B, \text{kernel}}$ and the CPKS kernel,
and only vanishes for a common LMF.

%%%%%%%%%%%%%%%%%%%%%%%%%%%%%%%%%%%%%%%%%%%%%%%%%%%%%%%%%%%%%%%%%%%%%
%%% Computational Methods
%%%%%%%%%%%%%%%%%%%%%%%%%%%%%%%%%%%%%%%%%%%%%%%%%%%%%%%%%%%%%%%%%%%%%
\section{Computational Methods}
\label{sec:comput}
The theoretical framework outlined above was implemented into
TURBOMOLE \cite{Franzke.Holzer.ea:TURBOMOLE.2023, Franzke.Sharma.ea:Recent.2026}
by extending our previous work.\cite{Franzke.Holzer:Exact.2023, Holzer.Franzke:Relativistic.2026,
Holzer.Franzke.ea:Current.2022, Franzke.Pausch.ea:Application.2025, InB-Franzke:2025}
CPKS equations are solved as outlined in Ref.~\citenum{Kehry.Franzke.ea:Quasirelativistic.2020}
and in case of stalling convergence a minium residual subspace solver of
the complex response equations is used as a fallback.
Expressions for the response were validated numerically. Here, the variations as well
as the covariance of $\hat{T}$ were confirmed by finite differences with random spinors at
a time-reversal symmetric reference point.

First, initial studies on heavy atoms and molecules as mainly collected previously
\cite{Franzke.Holzer:Application.2026} were performed with the SAP-X2C Hamiltonian
\cite{Surjuse.Valeev:SAP-X2C.2026, Franzke.Holzer:Application.2026}
in the diagonal local approximation to the unitary decoupling transformation
\cite{Peng.Middendorf.ea:efficient.2013, Franzke.Middendorf.ea:Efficient.2018} (DLU)
and with a finite nuclear Gaussian charge model \cite{Visscher.Dyall:Dirac-Fock.1997, Franzke.Weigend:NMR.2019}
for both the scalar and the vector potential. The uncontracted Dyall-VDZ basis
\cite{Dyall:Relativistic.2004, Dyall:Relativistic.2006, Dyall:Relativistic.2007A,
Dyall:Relativistic.2009, Gomes.Dyall.ea:Relativistic.2010, Dyall:Relativistic.2011,
Dyall:Relativistic.2012, Dyall:Relativistic.2016, Dyall-Repo}
is employed for the atomic calculations, while molecular calculations used the
uncontracted Dyall-VDZ basis set for the heavy elements \cite{Dyall:Relativistic.2006, Dyall-Repo}
and the decontracted cc-pVDZ basis set was employed for H.
\cite{Dunning:Gaussian.1989, Woon.Dunning:Gaussian.1993}
The SCF procedures was solved with a two-component augmented Roothaan--Hall solver
\cite{Holzer.Franzke:Relativistic.2026}
and criteria of $10^{-9}$\,$E_{\text{h}}$ for the total energy,
$10^{-4}$\,$E_{\text{h}}$ for the one-electron energy,
$10^{-9}$ for the root-mean square of the density matrix change,
and $3.2 \cdot 10^{-6}$ for the orbital rotation gradient
(square 1/100 of the total energy threshold).
Here, the TPSS, \cite{Tao.Perdew.ea:Climbing.2003}
TPSSh, \cite{Staroverov.Scuseria.ea:Comparative.2003}
TASK, \cite{Aschebrock.Kummel:Ultranonlocality.2019}
r$^2$SCAN, \cite{Furness.Kaplan.ea:Accurate.2020}
M06-2X, \cite{Zhao.Truhlar:M06.2008}
and CHYF-B95~\cite{Holzer.Franzke:General.2025} density
functional approximations were selected with a medium-sized
grid tailored for X2C (grid size 3a).
\cite{Treutler.Ahlrichs:Efficient.1995, Franzke.Tress.ea:Error-consistent.2019}
Inclusion of the current density is denoted with the prefix ``c'' or with the acronym CDFT.
The two-electron Coulomb integrals were evaluated analytically, i.e.\ without
the resolution of the identity approximation (RI-J).
Gauge-origin invariance is demonstrated for H$_2$Po with the
x2c-SVPall basis set \cite{Pollak.Weigend:Segmented.2017}
and a very large grid (grid size 7a with reference settings).
\cite{Treutler.Ahlrichs:Efficient.1995, Franzke.Tress.ea:Error-consistent.2019, TURBOMOLE-manual}
Local hybrids always use a seminumerical integration scheme.
\cite{Plessow.Weigend:Seminumerical.2012, Holzer:improved.2020}

Second, the Cd and P NMR shifts of the tertiary cluster [Cd$_3$(Ge$_3$P)$_3$]$^{3-}$
were calculated. This cluster was originally prepared in Ref.~\citenum{Mitzinger.Bandemehr.ea:Ge2P22-.2018}
and its NMR properties were studied computationally in Ref.~\citenum{Spang.Beuthert.ea:From.2026}.
Herein, we complement this study with the more rigorous theoretical approach.
Molecular structures of the three isomers were taken from Ref.~\citenum{Spang.Beuthert.ea:From.2026}.
Modified x2c-TZVPall-2c-s basis sets were employed for the nuclei of
interest as given in the Supporting Information and x2c-TZVPall-s basis for
all other elements. \cite{Franzke.Tress.ea:Error-consistent.2019}
The SAP-DLU-X2C Hamiltonian was applied with the TPSS, \cite{Tao.Perdew.ea:Climbing.2003}
TPSSh, \cite{Staroverov.Scuseria.ea:Comparative.2003}
r$^2$SCAN, \cite{Furness.Kaplan.ea:Accurate.2020}
TASK, \cite{Aschebrock.Kummel:Ultranonlocality.2019}
M06-2X, \cite{Zhao.Truhlar:M06.2008}
MN15-L, \cite{Yu.He.ea:MN15-L.2016}
MN15, \cite{Yu.He.ea:MN15.2016}
LH12ct-SsirPW92 (LH12ct), \cite{Arbuznikov.Kaupp:Importance.2012}
LH20t$^*$ (LH20t without calibration function), \cite{Haasler.Maier.ea:Local.2020}
and CHYF-B95~\cite{Holzer.Franzke:General.2025}
functionals using the same grid as above. The D4 correction was included
for the energy sampling except for TASK, MN15-L, and MN15. \cite{Caldeweyher.Bannwarth.ea:Extension.2017, Caldeweyher.Ehlert.ea:Generally.2019}
The conductor-like screening model (COSMO) was employed with the settings
for N,N-dimethylformamide (DMF) to implicitly account for solvent effects.
\cite{Schafer.Klamt.ea:COSMO.2000, Reiter.Mack.ea:Calculation.2018, Pausch:Consistent.2024}
For efficiency, the multipole-accelerated resolution of the identity approximation
\cite{Sierka.Hogekamp.ea:Fast.2003, Reiter.Mack.ea:Calculation.2018} (MARI-J)
with tailored auxiliary basis sets \cite{Pollak.Weigend:Segmented.2017, Franzke.Spiske.ea:Segmented.2020}
and the seminumerical exchange (snK) approximation \cite{Holzer:improved.2020, Holzer.Franzke.ea:Assessing.2021}
were turned on. SCF densities were converged with criteria of
$10^{-7}$\,$E_{\text{h}}$ for the total energy,
$10^{-4}$\,$E_{\text{h}}$ for the one-electron energy,
$10^{-6}$ for the root-mean square of the density matrix change,
and $3.2 \cdot 10^{-5}$ for the orbital rotation gradient of the
augmented Roothaan--Hall solver. \cite{Holzer.Franzke:Relativistic.2026}
NMR response equations were converged with a threshold of $10^{-6}$
for the norm of the residuum. \cite{Kehry.Franzke.ea:Quasirelativistic.2020}
NMR shifts are reported with respect to the standard reference of
85\% aqueous H$_3$PO$_4$ and neat dimethylcadmium (CdMe$_2$), respectively.
Calculations employed PH$_3$ and CdMe$_2$ in the gas phase as
secondary reference. Molecular structures of the references were optimized
herein for each functional and are given in the Supporting Information.
The shift to the standard scale is $-266.1$\,ppm \cite{Jameson.De-Dios.ea:Absolute.1990}
and $+62$\,ppm. \cite{Ellis.Odom.ea:Comments.1993} This allows for
comparison with the experimental results. \cite{Mitzinger.Bandemehr.ea:Ge2P22-.2018}

Third, we study the pseudo-tetrahedral Zintl anions (HGe$_3$P)$^{2-}$
and (HSn$_3$As)$^{2-}$. \cite{Spang.Beuthert.ea:From.2026, Kubis.Peerless.ea:SnAs}
According to previous studies, these systems are subject to pronounced
solvation effects and require the explicit inclusion of a solvent shell,
as do even heavier anions such as (Sn$_2$Sb$_2$)$^{2-}$ and (Sn$_2$Bi$_2$)$^{2-}$.
\cite{Spang.Beuthert.ea:From.2026, Kubis.Peerless.ea:SnAs}
In line with the given studies, we consider 12 and 25 DMF
molecules and a COSMO shell around the solute-solvent system.
Thus, the structures were taken from the literature and computational
settings were generally chosen accordingly. That is, we applied
the MARI-J~\cite{Sierka.Hogekamp.ea:Fast.2003, Reiter.Mack.ea:Calculation.2018}
and snK approximations. \cite{Holzer:improved.2020, Holzer.Franzke.ea:Assessing.2021}
The SAP-DLU-X2C Hamiltonian is again applied with modified x2c-TZVPall-2c-s
basis sets for H, P, and Sn as given in the Supporting Information, while the
x2c-TZVPall-s basis is used otherwise. \cite{Franzke.Tress.ea:Error-consistent.2019}
In this regard, the computational parameters are theoretically more accurate
than those of the initial study.
Based on the study of the tertiary cluster [Cd$_3$(Ge$_3$P)$_3$]$^{3-}$,
we considered the TPSS, \cite{Tao.Perdew.ea:Climbing.2003}
TPSSh, \cite{Staroverov.Scuseria.ea:Comparative.2003}
r$^2$SCAN, \cite{Furness.Kaplan.ea:Accurate.2020}
TASK, \cite{Aschebrock.Kummel:Ultranonlocality.2019}
MN15-L, \cite{Yu.He.ea:MN15-L.2016}
MN15, \cite{Yu.He.ea:MN15.2016}
LH12ct, \cite{Arbuznikov.Kaupp:Importance.2012}
and CHYF-B95~\cite{Holzer.Franzke:General.2025} functionals
to represent the common meta-GGAs and two local hybrids.
Thresholds for the SCF and CPKS equations and the DFT grid are
the same as for the [Cd$_3$(Ge$_3$P)$_3$]$^{3-}$.
Sn NMR shifts were calculated relative to SnMe$_4$, also with
an explicit DMF solvent shell and a COSMO cavity around the
solute-solvent system. \cite{Kubis.Peerless.ea:SnAs}
P NMR shifts were calculated using PH$_3$ as secondary reference as
done for [Cd$_3$(Ge$_3$P)$_3$]$^{3-}$. Due to the large SOC
contribution to the NMR shieldings and shifts of H and Sn for the
given clusters, we list the scalar-relativistic results in the
Supporting Information but not in the main text.

%%%%%%%%%%%%%%%%%%%%%%%%%%%%%%%%%%%%%%%%%%%%%%%%%%%%%%%%%%%%%%%%%%%%%
%%% Results
%%%%%%%%%%%%%%%%%%%%%%%%%%%%%%%%%%%%%%%%%%%%%%%%%%%%%%%%%%%%%%%%%%%%%
\begin{table*}[tbp]
    \centering
    \caption{NMR isotropic shielding constants in ppm at the DFT and CDFT level with the uncontracted Dyall-VDZ (heavy elements)
    and decontracted cc-pVDZ (hydrogen) basis. GIAOs and the SAP-X2C Hamiltonian in a magnetically balanced basis set are used throughout all calculations.
    For the molecules, the nucleus of the isotropic NMR shielding constant is listed in parenthesis.
    DFT denotes that the Maximoff--Scuseria generalization is applied and the current densities are not considered in the SCF calculation,
    while CDFT refers to the current-dependent generalization for the SCF and the NMR part.
    Both formalisms include the GIAO terms originally arising from the spin-current density contributions for the RHS kernel and the spin
    density contributions to the LHS kernel, see Refs.~\citenum{Franzke.Holzer:Exact.2023} and \citenum{Olejniczak.Bast.ea:simple.2012} for details and derivation.
    CHYF-B95 represents the class of local hybrids, while TPSS, TASK, r$^2$SCAN represent pure semilocal meta-GGAs,
    and TPSSh, M06-2X are examples for global hybrid meta-GGA functionals.
    Molecular geometries are the same as in Ref.~\citenum{Franzke.Holzer:Application.2026} and the atoms are placed at the origin of the coordinate system.}
    \label{tab:nmr-shieldings}
    \begin{tabular}{@{\extracolsep{4pt}}
    l
    S[table-format = 5.1]
    S[table-format = 5.1]
    S[table-format = 5.1]
    S[table-format = 5.1]
    S[table-format = 5.1]
    S[table-format = 5.1]
    S[table-format = 5.1]
    S[table-format = 5.1]
    S[table-format = 5.1]
    S[table-format = 5.1]
    S[table-format = 5.1]
    S[table-format = 5.1]@{}
    }
    \toprule
     & \multicolumn{2}{S}{{\text{TPSS}}} & \multicolumn{2}{S}{{\text{TPSSh}}} & \multicolumn{2}{S}{{\text{TASK}}} & \multicolumn{2}{S}{{\text{r$^2$SCAN}}} & \multicolumn{2}{S}{{\text{M06-2X}}} & \multicolumn{2}{S}{{\text{CHYF-B95}}} \\
    \cmidrule{2-3} \cmidrule{4-5} \cmidrule{6-7} \cmidrule{8-9} \cmidrule{10-11} \cmidrule{12-13}
    System & {\text{DFT}} & {\text{CDFT}} & {\text{DFT}} & {\text{CDFT}} & {\text{DFT}} & {\text{CDFT}} & {\text{DFT}} & {\text{CDFT}} & {\text{DFT}} & {\text{CDFT}} & {\text{DFT}} & {\text{CDFT}} \\
    \midrule
    Sr & 3936.2  & 3905.4  & 3933.3  & 3905.4  & 4169.7  & 3903.7  & 3989.9  & 3905.6  & 3836.0  & 3906.9  & 3930.0  & 3903.9  \\
    Cd & 5778.2  & 5723.1  & 5773.1  & 5723.1  & 6243.1  & 5720.8  & 5877.2  & 5722.8  & 5584.6  & 5724.5  & 5726.8  & 5721.0\\
    Xe & 7068.1  & 7016.6  & 7063.9  & 7016.7  & 7791.3  & 7016.8  & 7306.5  & 7018.1  & 6972.4  & 7020.9  & 7136.4  & 7013.1 \\
    Ba & 7546.7  & 7481.7  & 7540.7  & 7481.6  & 7985.4  & 7470.7  & 7650.4  & 7478.1  & 7351.2  & 7485.6  & 7522.3  & 7477.4 \\
    Lu$^+$ & 12157.7 & 12044.9 & 12004.4 & 11969.2 & 12792.6 & 12029.7 & 12305.9 & 12045.2 & 11899.8 & 12065.8 & 12049.9 & 12042.7 \\
    Hg & 16083.6 & 15953.3 & 16070.3 & 15952.5 & 17164.1 & 15941.6 & 16304.1 & 15944.8 & 15636.9 & 15937.5 & 15952.9 & 15935.2 \\
    Rn & 19487.2 & 19351.8 & 19481.9 & 19358.0 & 21195.8 & 19373.2 & 20045.3 & 19370.1 & 19264.4 & 19412.4 & 19607.7 & 19318.9 \\
    Ra & 20745.9 & 20619.3 & 20739.2 & 20624.1 & 22017.3 & 20635.9 & 21237.6 & 20633.4 & 20536.7 & 20661.9 & 20839.5 & 20591.0 \\
    Lr$^+$ & 35016.9 & 34850.3 & 34999.5 & 34850.8 & 35731.5 & 34379.4 & 35206.4 & 34797.4 & 34509.3 & 34976.6 & 34731.8 & 34755.5 \\
    Cn & 50150.7 & 49827.8 & 50096.3 & 49805.3 & 53810.0 & 49863.1 & 51165.7 & 49797.1 & 49949.8 & 49719.2 & 49701.8 & 49654.3 \\
    Og & 65968.4 & 65592.4 & 66070.7 & 65725.1 & 70489.6 & 65156.0 &66238.0 & 65324.0 & 66038.6 & 66763.9 & 65763.1 & 65304.0 \\
    SnH$_4$ (H) & 27.7 & 27.7 & 27.7 & 27.7 & 28.0 & 28.1 & 27.7 & 27.8 & 27.9 & 27.8 & 27.7 & 27.7 \\
	SnH$_4$ (Sn) & 3936.1 & 3917.4 & 3962.3 & 3945.1 & 3974.6 & 3879.9 & 4016.4 & 3980.5 & 4003.4 & 4002.6 & 4041.3 & 4034.3 \\
    PbH$_4$ (H) & 25.9 & 26.0 & 26.0 & 26.0 & 26.3 & 26.1 & 26.5 & 26.4 & 24.8 & 24.6 & 26.4 & 26.2 \\
	PbH$_4$ (Pb) & 11970.1 & 11929.8 & 12080.4 & 12043.1 & 12058.0 & 11831.1 & 12237.8 & 12141.9 & 12338.9 & 12347.1 & 12354.2 & 12322.3 \\
    SbH$_3$ (H) & 30.5 & 30.5 & 30.5 & 30.5 & 32.0 & 31.6 & 31.2 & 31.1 & 31.0 & 31.0 & 30.5 & 30.5 \\
	SbH$_3$ (Sb) & 4221.0 & 4205.5 & 4246.9 & 4232.7 & 4338.0 & 4192.2 & 4373.8 & 4329.7 & 4271.2 & 4266.1 & 4346.8 & 4319.6 \\
    BiH$_3$ (H) & 26.5 & 26.4 & 26.5 & 26.4 & 30.5 & 28.3 & 28.9 & 28.2 & 22.5 & 23.0 & 26.8 & 26.3 \\
	BiH$_3$ (Bi) & 12067.7 & 12037.9 & 12159.9 & 12132.1 & 12422.1 & 11999.1 & 12660.3 & 12538.3 & 11762.5 & 11829.8 & 12452.1 & 12340.3 \\
    H$_2$Te (H) & 37.8 & 37.6 & 37.8 & 37.6 & 39.7 & 38.5 & 38.7 & 38.1 & 38.4 & 38.6 & 37.6 & 37.4 \\
	H$_2$Te (Te) & 4608.2 & 4584.6 & 4636.1 & 4614.4 & 4978.3 & 4652.3 & 4850.5 & 4745.0 & 4604.9 & 4622.5 & 4754.6 & 4693.0 \\
    H$_2$Po	(H) & 42.1 & 41.5 & 42.3 & 41.8 & 53.3 & 47.0 & 46.7 & 43.8 & 42.8 & 44.2 & 43.1 & 42.3 \\
    H$_2$Po	(Po) & 13765.2 & 13716.1 & 13902.8 & 13855.4 & 14850.3 & 13968.4 & 14473.1 & 14192.0 & 13762.9 & 13936.0 & 14295.7 & 14148.5 \\
    HI (H) & 45.1 & 44.7 & 44.9 & 44.6 & 45.8 & 43.5 & 45.5 & 44.3 & 44.4 & 44.9 & 44.1 & 43.8 \\
    HI (I) & 5714.0 & 5678.6 & 5737.2 & 5704.5 & 6321.6 & 5776.3 & 5985.6 & 5795.4 & 5696.5 & 5732.6 & 5851.9 & 5761.4 \\
    HAt (H) & 62.7 & 61.6 & 63.0 & 62.0 & 73.7 & 63.0 & 68.1 & 63.1 & 64.6 & 66.6 & 63.0 & 62.0 \\
    HAt (At) & 16571.5 & 16479.0 & 16701.7 & 16615.0 & 18148.9 & 16701.7 & 17362.8 & 16860.6 & 16636.5 & 16830.4 & 17055.0 & 16841.6 \\
    \bottomrule
    \end{tabular}
\end{table*}

\begin{table*}[tbp]
    \centering
    \caption{Po and H NMR isotropic shielding constants in ppm of H$_2$Po at CDFT TASK and CHYF-B95 levels with
    the segmented-contracted x2c-SVPall basis set and the decontracted Dyall-VDZ/cc-pVDZ basis sets using various spatial alignments.
    GIAOs and the SAP-X2C Hamiltonian in a magnetically balanced basis set are used throughout all calculations.
    CMS denotes that the origin is the center of mass and the molecule is aligned so that the Cartesian axes correspond to the principal axes of inertia.
    Then the molecule is rotated by 100$^{\circ}$ along each Cartesian axes (rot) and further moved along 100\,$a_0$ along each axes (rot+move).
    The Po--H bond lengths amount to 174.70\,pm and the H--Po--H angle is 90.52$^{\circ}$. 
    The largest available grid was employed for the numerical integration of the XC terms and thus the shieldings differ from those listed in Table~\ref{tab:nmr-shieldings}.
    The x2c-SVPall basis set is not optimized for 2c NMR calculations and only applied for illustrative purposes of the gauge-origin independence.
    Thus, the order of magnitude with this basis set and the Dyall basis differs notably for Po.}
    \label{tab:nmr-gauge}
    \begin{tabular}{@{\extracolsep{15pt}}
    l
    S[table-format = 4.3]
    S[table-format = 2.3]
    S[table-format = 2.3]
    S[table-format = 5.3]
    S[table-format = 2.3]
    S[table-format = 2.3]
    S[table-format = 4.3]
    S[table-format = 2.3]
    S[table-format = 2.3]@{}
    }
    \toprule
     & \multicolumn{3}{S}{{\text{x2c-SVPall cTASK}}} & \multicolumn{3}{S}{{\text{Dyall-VDZ/cc-pVDZ cTASK}}} & \multicolumn{3}{S}{{\text{x2c-SVPall cCHYF-B95}}} \\
    \cmidrule{2-4} \cmidrule{5-7} \cmidrule{8-10} 
    Alignment & {\text{Po}} & {\text{H1}} & {\text{H2}} & {\text{Po}} & {\text{H1}} & {\text{H2}} & {\text{Po}} & {\text{H1}} & {\text{H2}} \\
    \midrule
    yz plane (CMS) & 5543.989 & 48.635 & 48.635 & 13972.026 & 48.586 & 48.586 & 5519.289 & 43.814 & 43.814 \\
    xz plane (CMS) & 5543.989 & 48.635 & 48.635 & 13972.026 & 48.586 & 48.586 & 5519.289 & 43.814 & 43.814 \\
    xy plane (CMS) & 5543.989 & 48.635 & 48.635 & 13972.026 & 48.586 & 48.586 & 5519.289 & 43.814 & 43.814 \\
    yz plane (rot) & 5543.992 & 48.634 & 48.634 & 13972.030 & 48.586 & 48.586 & 5519.286 & 43.812 & 43.813 \\
    xz plane (rot) & 5543.992 & 48.634 & 48.634 & 13972.031 & 48.586 & 48.586 & 5519.325 & 43.814 & 43.809 \\
    xy plane (rot) & 5543.992 & 48.634 & 48.634 & 13972.032 & 48.586 & 48.586 & 5519.279 & 43.812 & 43.813 \\
    yz plane (rot+move) & 5543.992 & 48.634 & 48.634 & 13972.030 & 48.586 & 48.586 & 5519.286 & 43.812 & 43.813 \\
    xz plane (rot+move) & 5543.992 & 48.634 & 48.634 & 13972.030 & 48.586 & 48.586 & 5519.324 & 43.814 & 43.809 \\
    xy plane (rot+move) & 5543.992 & 48.634 & 48.634 & 13972.030 & 48.586 & 48.586 & 5519.278 & 43.812 & 43.813 \\
    \bottomrule
    \end{tabular}
\end{table*}

\section{Results \& Discussion}
\label{sec:results}

\subsection{Shieldings of Atoms and Small Molecules}
\label{subsec:shieldigns}

As an initial test of the impact of the current density on NMR shielding constants,
we performed calculations on closed-shell atoms and heavy-element hydrides, as listed in
Table~\ref{tab:nmr-shieldings}. As observed previously for NMR shieldings and
excitation energies, \cite{Schattenberg.Kaupp:Effect.2021, Grotjahn.Furche.ea:Importance.2022,
Holzer.Franzke.ea:Assessing.2021, Holzer.Franzke.ea:Current.2022}
the impact of the current density strongly depends on the density functional approximation.
For the heavy nuclei, the change from the Maximoff--Scuseria generalization (DFT) to the
current-dependent generalization (CDFT) is smallest for TPSS and TPSSh with about 0.6\,\%
and 0.5\,\% on average and at most 1\%, i.e.\ some tens of ppm---except for superheavy atoms.
This is in line with the rather weak dependence of TPSS on the kinetic energy density or,
more precisely, the shallow slope of its exchange enhancement factor with respect to the
iso-orbital indicator $\alpha$. \cite{Tao.Perdew.ea:Climbing.2003}
In contrast, TASK was specifically designed to feature a steep slope of the enhancement
factor with respect to $\alpha$, \cite{Aschebrock.Kummel:Ultranonlocality.2019} and thus
shows the largest dependence on the current density. Here, the shieldings change by 6\%
on average and by up to 10\% for the heavy nuclei, i.e.\ by up to about 1700\,ppm for Rn
and Ra. r$^2$SCAN is in between with an average change of 2\%. Clearly, a crucial part
of the physics included in the functional design is missing if the current density is
neglected for such functionals.
Admixture of exact exchange only slightly reduces the dependence for TPSSh (10\,\% exact
exchange) compared to TPSS. The local hybrid CHYF-B95 shows an average change of 0.9\%,
which is notably smaller than for TASK and r$^2$SCAN, even though both its LMF and the
B95 correlation depend on the kinetic energy density. This reflects that the LMF switches
to large amounts of exact exchange in critical regions such as the high-density core and
iso-orbital regions. Except for M06-2X, the inclusion of the current density
consistently lowers the shielding constants of the heavy nuclei. Here, the shieldings
increase for 16 of the 19 heavy nuclei.

\begin{figure*}[ht!]
  \centering
    \begin{tabular}[b]{c}
        \includegraphics[width=0.3\linewidth]{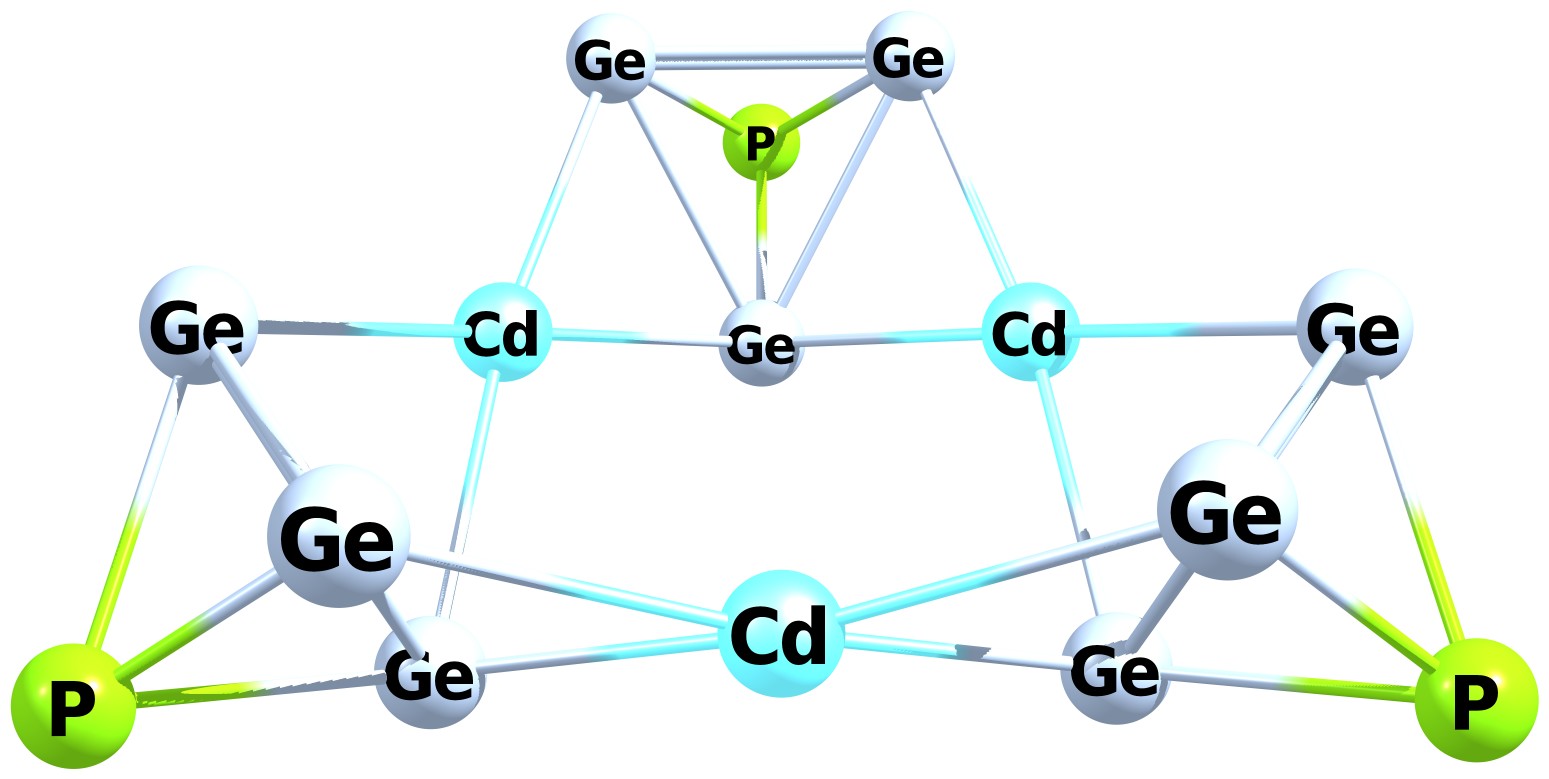} \\
        \small Isomer A
    \end{tabular}
    \hfill
    \begin{tabular}[b]{c}
        \includegraphics[width=0.3\linewidth]{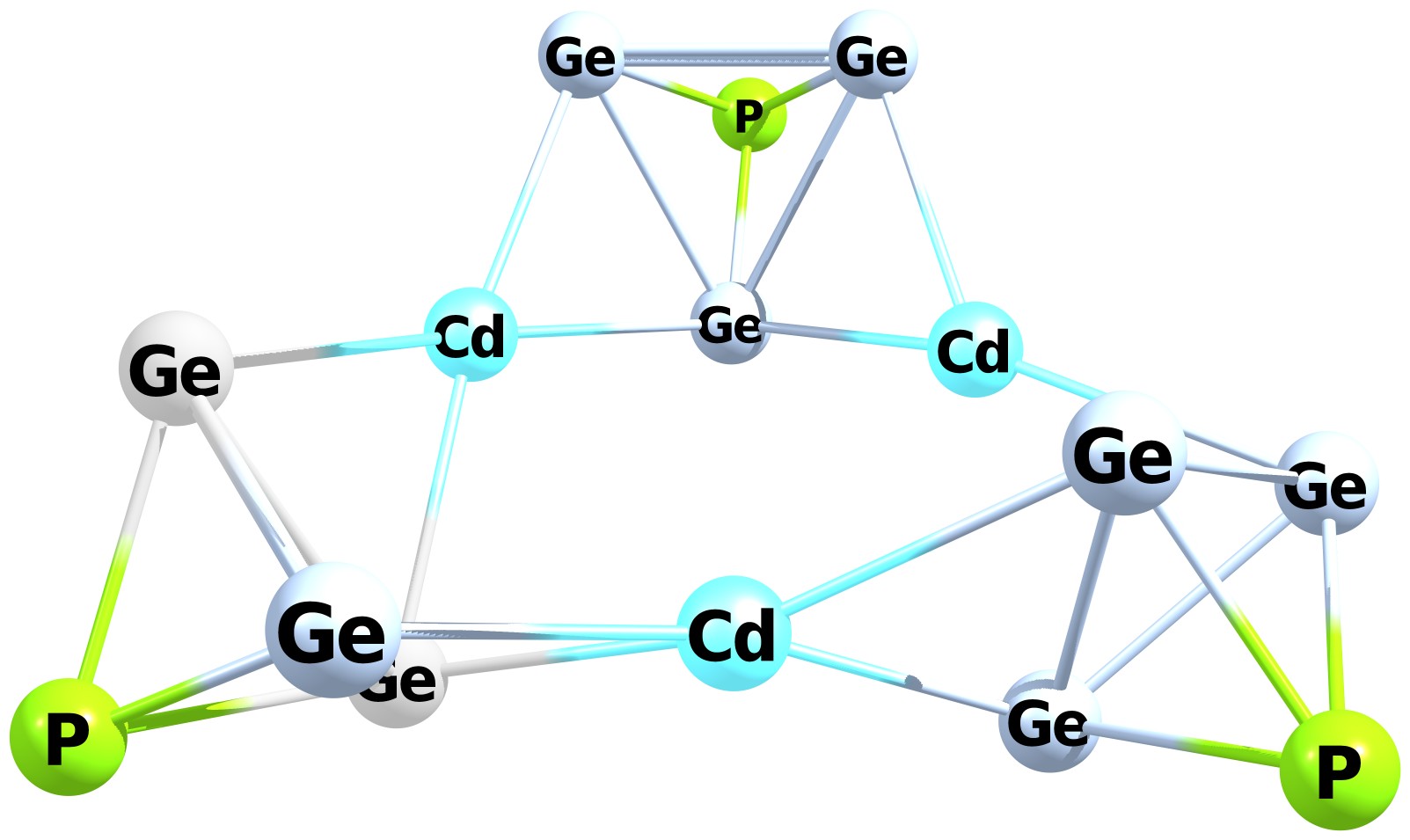} \\
        \small Isomer B
    \end{tabular}
    \hfill
    \begin{tabular}[b]{c}
        \includegraphics[width=0.3\linewidth]{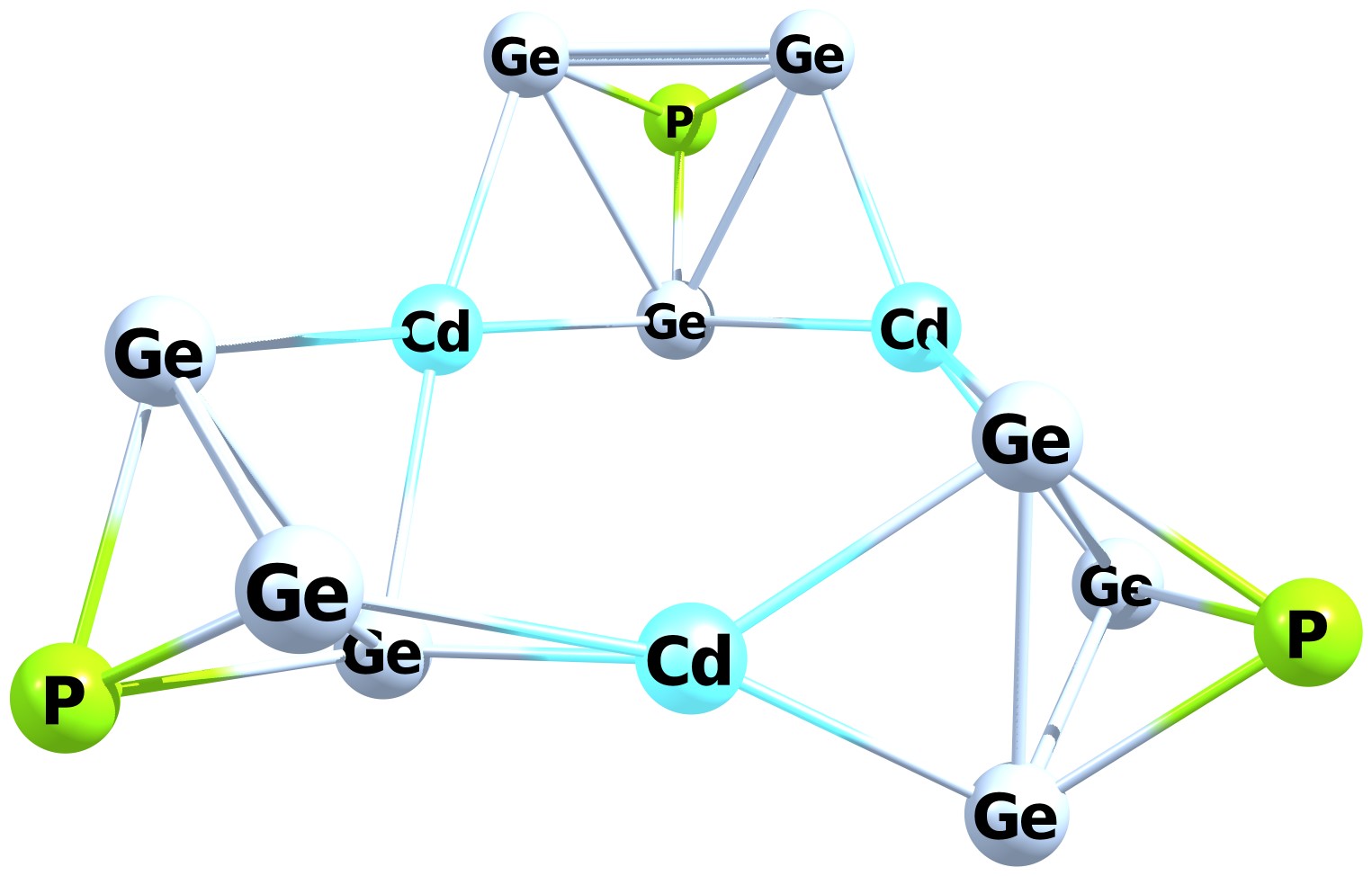} \\
        \small Isomer C
    \end{tabular} 
	\caption{Molecular structure of the three energetically lowest isomers of [Cd$_3$(Ge$_3$P)$_3$]$^{3-}$,
    which are related by the lowest vibrational mode as discussed in Ref.~\citenum{Mitzinger.Bandemehr.ea:Ge2P22-.2018}.
    Relative energies are listed in Table~\ref{tab:cd-energies}.}
	\label{fig:cdgep-structures}
\end{figure*}

A further remarkable finding concerns the closed-shell atoms. With the Maximoff--Scuseria
generalization, the shieldings of the six functionals spread by 300 to 1900\,ppm up to Ra,
whereas the spread is reduced to 3 to 90\,ppm with the current-dependent generalization.
This is even more pronounced for the superheavy atoms Cn and Og, where the spread
reaches 5000\, ppm and is reduced to about 200\,ppm. This confirms the atomic
results observed in the non-relativistic limit. \cite{Schattenberg.Kaupp:Effect.2021}
Therein, it was shown explicitly that the Maximoff--Scuseria generalization introduces
artifacts. Herein, it also does not include the spin current densities, as a relativistic
generalization of the Maximoff--Scuseria generalization is not straightforward.
Overall, all functionals yield almost the same atomic shieldings within CDFT, as expected
for the diamagnetically dominated shielding of a closed-shell atom.

For the heavy nuclei of the molecules, the spread is reduced as well, for instance from
1580 to 380\,ppm for At and from 630 to 120\,ppm for I. Only for SnH$_4$ and PbH$_4$ is
the spread slightly increased, as TASK yields the lowest and CHYF-B95 or M06-2X yield the
highest shieldings.
Notably, the effect on the light hydrogen nucleus depends on the neighboring atom.
While the H shieldings of SnH$_4$ and PbH$_4$ are virtually unaffected, the heavy halogen and
chalcogen atoms induce sizable changes of up to 11\,ppm for HAt and 6\,ppm for H$_2$Po
with TASK, and of up to 5\,ppm with r$^2$SCAN. These are large changes on the scale of
typical H shifts and follow the heavy-atom effect on the light atom (HALA) due to the
spin--orbit coupling of the neighboring atom. \cite{Vicha.Novotny.ea:Relativistic.2020}
This is important as it indicates that there might only be limited error cancellation
for NMR chemical shifts, which are usually computed relative to a reference compound such as
tetramethylsilane (TMS) for hydrogen and carbon nuclei. Such light-element references barely
feature spin current densities due to their weak spin--orbit coupling interaction.

The gauge-origin, translational, and rotational invariance were validated explicitly
for TASK and CHYF-B95 and H$_2$Po with small basis sets as shown in Table~\ref{tab:nmr-gauge}.
With TASK, the Po shieldings change by at most 0.006\,ppm while the absolute shielding
constant amounts to 13972\,ppm. A similar picture is revealed for the H shieldings,
which change by at most 0.001\,ppm for an absolute shielding of about 48\,ppm.
With CHYF-B95, the deviations between the different spatial placements are somewhat
larger, i.e.\ Po shieldings differ by 0.047\,ppm and H shieldings by 0.005\,ppm,
respectively. We attribute this to the increased numerical requirements of local hybrids.
In this regard, CHYF-B95 is one of the more challenging local hybrid functionals without
a calibration function, as its exchange uses the $\alpha$ variable and similar interpolation
functions as the SCAN functional. \cite{Sun.Ruzsinszky.ea:Strongly.2015}

\subsection{P and Cd NMR Shifts of the Cluster [Cd$_3$(Ge$_3$P)$_3$]$^{3-}$}
\label{subsec:cdp}

As NMR is at the forefront of elucidating the structure of new chemical compounds,
accurate predictions of chemical shifts can be a valuable addition in the process
of assigning the correct structure. This is especially true if isomers are energetically
close, as for the [Cd$_3$(Ge$_3$P)$_3$]$^{3-}$ clusters in Fig.~\ref{fig:cdgep-structures}
and Table~\ref{tab:cd-energies}. Here, the shifts are Boltzmann-averaged over the three
isomers at room temperature and compared to experiment
($-129$\,ppm for $^{31}$P and $637$\,ppm for $^{113}$Cd).
\cite{Mitzinger.Bandemehr.ea:Ge2P22-.2018}
The three isomers are connected by the lowest vibrational modes of about 10\,cm$^{-1}$
and the [Cd$_3$(Ge$_3$P)$_3$]$^{3-}$ is thus expected to be highly flexible.\cite{Mitzinger.Bandemehr.ea:Ge2P22-.2018}
Moreover, dispersion effects are important for the energetic order. \cite{Spang.Beuthert.ea:From.2026}

\begin{table}[tbp]
    \centering
    \caption{Relative energies in kJ/mol for the three considered isomers of [Cd$_3$(Ge$_3$P)$_3$]$^{3-}$.
    A value of zero indicates the most stable isomer and this is isomer B in all cases.
    Spin--orbit coupling and inclusion of the spin-current densities in the SCF do not notably
    affect the relative energies, as these change by less than 0.1\,kJ/mol.
    So, we only report the 2c CDFT relative energies and the scalar-relativistic results as well as the
    2c DFT results are given in the Supporting Information. D4 dispersion correction is included for all functionals but TASK, MN15-L, and MN15.
    Without dispersion effects, isomer B is significantly destabilized. \cite{Spang.Beuthert.ea:From.2026}
    However, a large Boltzmann population of isomer B is crucial for reasonable agreement with the
    experimental P NMR shifts, whereas the Cd shift is less affected by the different isomers.}
    \label{tab:cd-energies}
    \begin{tabular}{@{\extracolsep{28pt}}
    l
    S[table-format = 2.2]
    S[table-format = 2.2]
    S[table-format = 2.2]@{}
    }
    \toprule
     Method & {\text{Isomer A}} & {\text{Isomer B}} &  {\text{Isomer C}} \\
    \midrule
    cTPSS & 15.18 & 0.00 & 10.96 \\
    cTPSSh & 13.43 & 0.00 & 8.66 \\
    cr$^2$SCAN & 5.86 & 0.00 & 0.05 \\
    cTASK & 7.28 & 12.28 & 0.00 \\
    cM06-2X & 8.06 & 0.00 & 1.43 \\
    cMN15-L & 15.25 & 0.00 & 9.56 \\
    cMN15 & 12.41 & 0.00 & 6.05 \\
    cLH12ct & 28.22 & 0.00 & 24.71 \\
    cLH20t$^*$ & 11.73 & 0.00 & 4.56 \\
    cCHYF-B95 & 31.79 & 0.00 & 25.50 \\
    \bottomrule
    \end{tabular}
\end{table}

\begin{table}[tbp]
    \centering
    \caption{Calculated and experimental $^{31}$P and $^{113}$Cd NMR shifts of [Cd$_3$(Ge$_3$P)$_3$]$^{3-}$
    in ppm. Computational results for the three isomers are averaged according to the Boltzmann population
    at room temperature. We report 1c scalar-relativistic DFT (SR), 2c spin--orbit DFT (DFT), and 2c spin--orbit CDFT (CDFT) results.
    Here, scalar-relativistic calculations use the current-density response due to the external magnetic field.
    Experimental results in DMF are taken from Ref.~\citenum{Mitzinger.Bandemehr.ea:Ge2P22-.2018}.
    Due to the missing dispersion correction for TASK, we directly report the results of isomer B for this functional.
    See Supporting Information for shieldings and shifts of the different isomers.}
    \label{tab:cd-nmr}
    \begin{tabular}{@{\extracolsep{10pt}}
    l
    S[table-format = -3.0]
    S[table-format = -3.0]
    S[table-format = -3.0]
    S[table-format = 3.0]
    S[table-format = 3.0]
    S[table-format = 3.0]@{}
    }
    \toprule
     & \multicolumn{3}{S}{{\text{$^{31}$P NMR Shift}}} & \multicolumn{3}{S}{{\text{$^{113}$Cd NMR Shift}}} \\
    \cmidrule{2-4} \cmidrule{5-7}
     & {\text{SR}} & {\text{DFT}} & {\text{CDFT}} & {\text{SR}} & {\text{DFT}} & {\text{CDFT}} \\
    \midrule
    TPSS      & -151 & -158 & -163 & 773 & 646 & 640 \\
    TPSSh     & -177 & -185 & -190 & 766 & 630 & 624 \\
    r$^2$SCAN & -212 & -211 & -222 & 684 & 564 & 538 \\
    TASK      & -281 & -272 & -290 & 774 & 673 & 580 \\
    M06-2X    & -270 & -293 & -286 & 708 & 588 & 597 \\
    MN15-L    & -249 & -236 & -257 & 778 & 701 & 668 \\
    MN15      & -122 & -136 & -139 & 947 & 733 & 718 \\
    LH12ct    & -190 & -199 & -203 & 729 & 590 & 582 \\
    LH20t$^*$ & -216 & -227 & -228 & 772 & 641 & 641 \\
    CHYF-B95  & -160 & -160 & -174 & 749 & 606 & 606 \\
    \midrule
    Experiment & \multicolumn{3}{S}{{\text{$-$129}}} & \multicolumn{3}{S}{{\text{637}}} \\
    \bottomrule
    \end{tabular}
\end{table}

Isomer B is the global minimum for all functionals except cTASK, which lacks a dispersion
correction, and cr$^2$SCAN, where B and C are energetically degenerate.
Spin--orbit coupling and the current density change the relative energies by less than 0.1\,kJ/mol.
The $^{31}$P shifts of isomers A and C are 45 to 75\,ppm lower than those of isomer B, so that
the isomer population matters (e.g., 28\,ppm for r$^2$SCAN), whereas the $^{113}$Cd shifts
vary by at most about 55\,ppm between the three isomers.

As shown in Table~\ref{tab:cd-nmr}, spin--orbit coupling is mandatory for $^{113}$Cd, where it
lowers the shifts by 77 to 214\,ppm and reduces the mean absolute error from 131\,ppm (SR) to
about 41\,ppm (DFT and CDFT). For $^{31}$P, the scalar-relativistic one-component (SR) and
the 2c results differ by at most 23\,ppm.
The current density does not change the average accuracy, but its impact depends on the functional
approximation. For TPSS, TPSSh, LH12ct, LH20t$^*$, and CHYF-B95, the $^{113}$Cd shifts change by 8\,ppm
or less, presumably due to error cancellation with the reference compound. In contrast, TASK, r$^2$SCAN,
and MN15-L, which depend strongly on the kinetic energy density, change by 93, 26, and 33\,ppm,
respectively. For TASK and r$^2$SCAN, this moves the Cd shift away from experiment
(673 $\rightarrow$ 580\,ppm and 564 $\rightarrow$ 538\,ppm), so that the better DFT agreement
is likely fortuitous. Moreover, the spread over the functionals is not reduced for the shifts
(151\,ppm for $^{31}$P and 180\,ppm for $^{113}$Cd with CDFT).

At the CDFT level, TPSS, TPSSh, and LH20t$^*$ reproduce the Cd shift within 13\,ppm, while the P shifts
are generally too negative. MN15 has the smallest P error but overestimates Cd by 81\,ppm.
Overall, for this example TPSS and CHYF-B95 offer the most balanced description of both
NMR-active nuclei. The greater accuracy for the Cd shifts also suggests that dynamics plays
a more critical role for the P shifts, which is also confirmed by the individual shieldings
for the three isomers. The mean shieldings and shifts of the three P nuclei show larger absolute
changes than that of Cd, even though the shielding constants of P are much smaller than for Cd.
Especially the isomers A and C reveal way too negative P shifts.

\subsection{NMR Shifts of Zintl Pseudo-Tetrahedrons}
\label{subsec:zintl}

\begin{figure*}[tbp]
  \centering
    \begin{tabular}[b]{c}
        \includegraphics[width=0.38\linewidth]{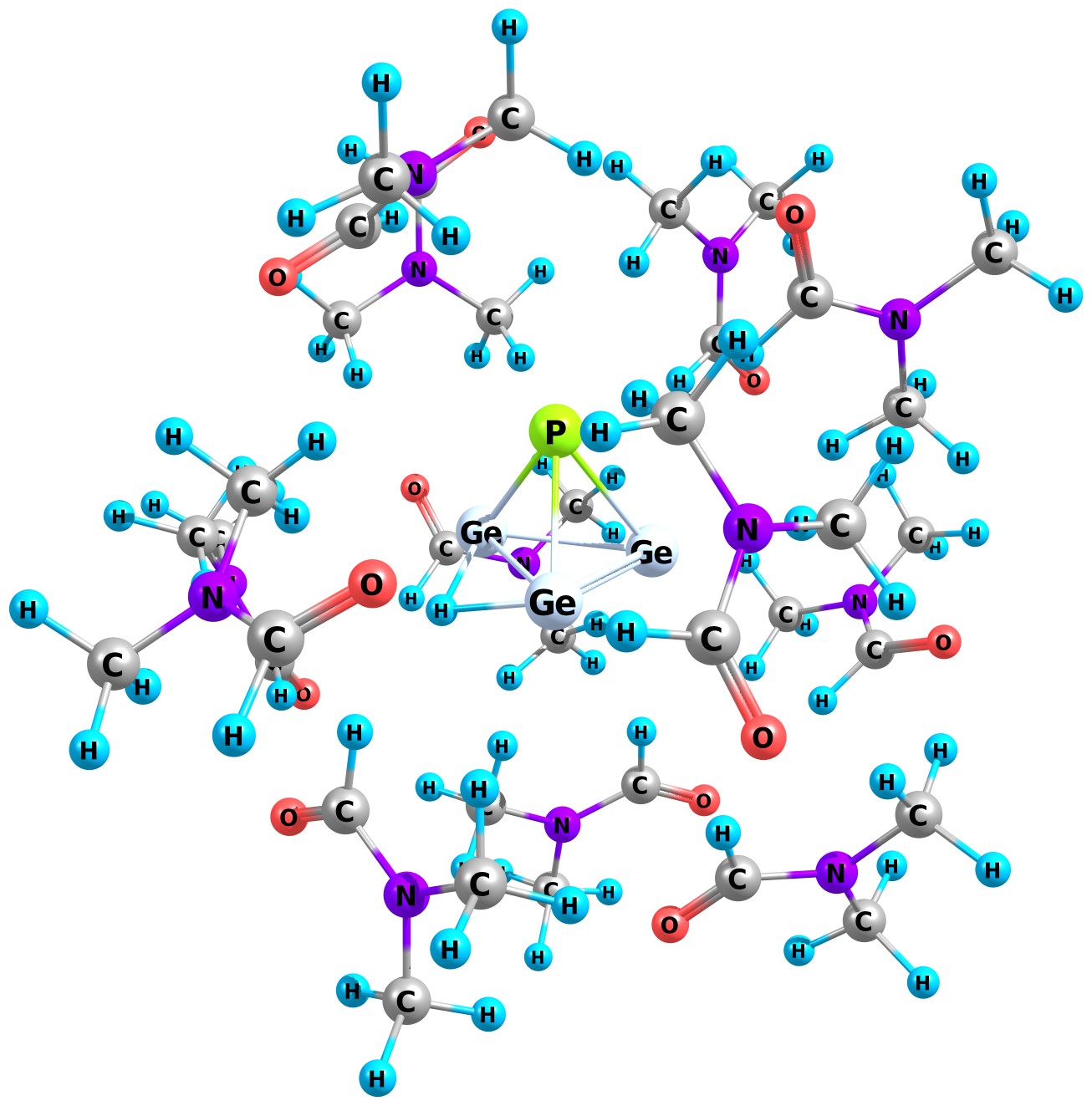}\\
        \small (HGe$_3$P)$^{2-}$ in DMF
    \end{tabular}
    \hfill
    \begin{tabular}[b]{c}
        \includegraphics[width=0.5\linewidth]{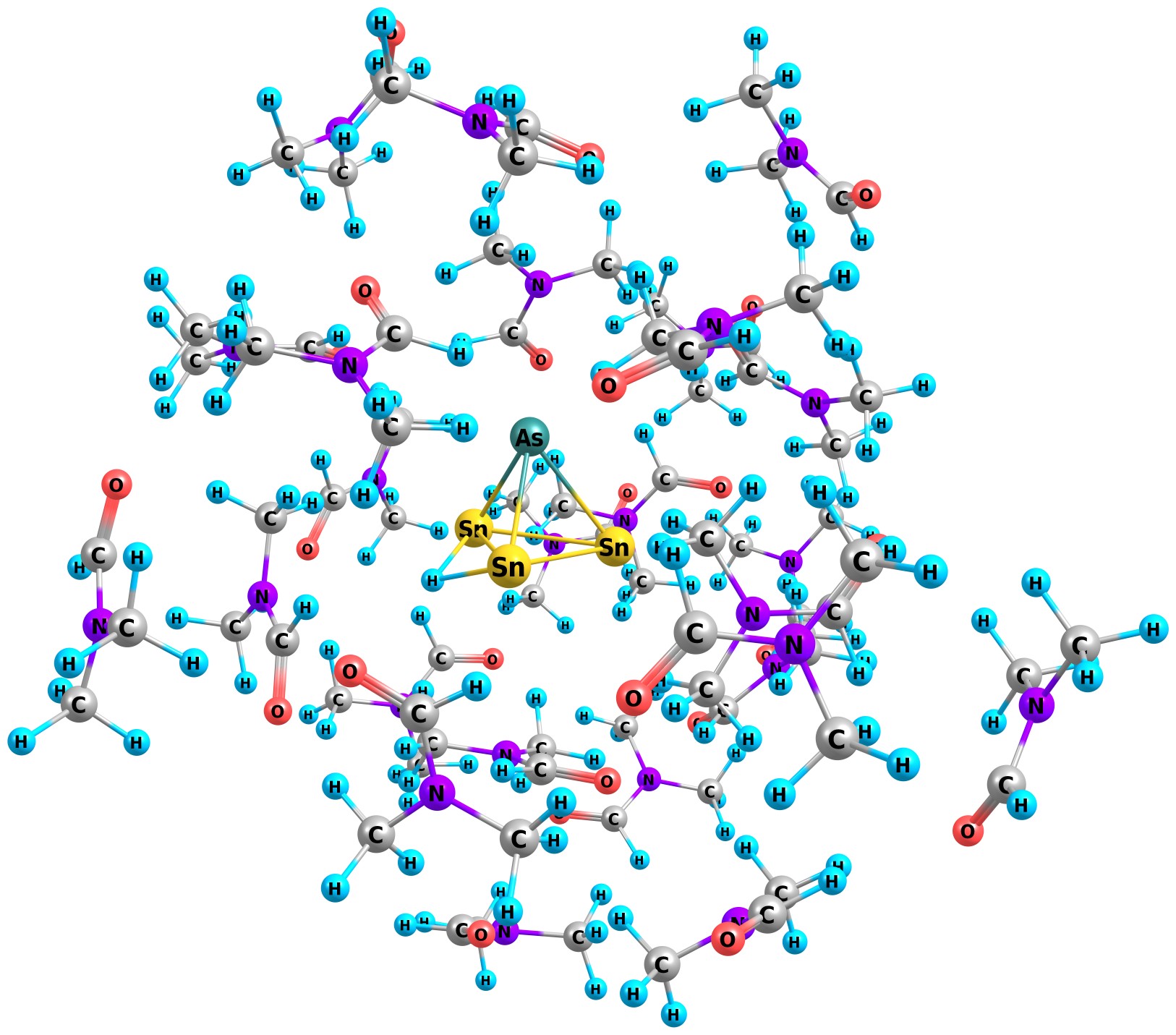} \\
        \small (HSn$_3$As)$^{2-}$ in DMF
    \end{tabular}
	\caption{Solute-solvent structure of (HGe$_3$P)$^{2-}$ with 12 DMF (left) and (HSn$_3$As)$^{2-}$
    with 25 DMF molecules (right) to explicitly  consider solvation effects.
    The whole systems consist of 149 and 305 atoms, respectively.
    For (HGe$_3$P)$^{2-}$, only 12 DMF molecules are considered due to the smaller atomic radii
    of Ge/P compared to Sn/As. See Refs.~\citenum{Spang.Beuthert.ea:From.2026} and \citenum{Kubis.Peerless.ea:SnAs}
    for details on the structures and their optimization.
    Therein, the impact of different solute-solvent structures on the NMR shifts was rather small compared to
    the number of DMF molecules to close the shell around the central cluster is more important.}
	\label{fig:tetrahedrons}
\end{figure*}

Finally, we turn to the protonated pseudo-tetrahedral Zintl anions (HGe$_3$P)$^{2-}$ and (HSn$_3$As)$^{2-}$
(Fig.~\ref{fig:tetrahedrons}, Table~\ref{tab:tetrahedrons-nmr}), for which solvation is the dominant
effect. Describing the solvent with COSMO alone is however insufficient, and shifts the $^{31}$P 
signal of (HGe$_3$P)$^{2-}$ by about 145\,ppm and the $^{119}$Sn signals of (HSn$_3$As)$^{2-}$ by
about 250\,ppm away from the experimental values, and the $^1$H shifts by 1--1.5\,ppm.
Explicit DMF molecules in the first solvation shell are thus mandatory, in line with previous studies.
\cite{Spang.Beuthert.ea:From.2026, Kubis.Peerless.ea:SnAs}
Spin--orbit coupling is of importance for the $^1$H and $^{119}$Sn shifts. For (HGe$_3$P)$^{2-}$,
it changes the $^1$H shift by about 2.5\,ppm, which corresponds to a small but notable HALA effect.
With about 9\,ppm, the HALA effect is much strong for (HSn$_3$As)$^{2-}$, as both systems would
yield similar $^1$H shifts of about 7 to 8\,ppm with the given functionals. For the $^{119}$Sn shifts,
spin--orbit coupling is of utmost importance and may change shifts by up to around 200\,ppm.
In contrast, the impact of the current density is small, especially for $^1$H. For the latter, it
changes the shifts by 0.3\,ppm or less for all functionals, while the $^{31}$P and $^{119}$Sn shifts
of cTPSS, cTPSSh, and cLH12ct differ by less than 5 and 10\,ppm from the DFT results, respectively.
Larger changes occur again for the functionals with a pronounced dependence on the kinetic energy
density, i.e.\ up to 30\,ppm for $^{31}$P (MN15-L) and up to about 55\,ppm for $^{119}$Sn (TASK, MN15).
This is, however, small compared to the deviations from experiment discussed below and resembles the
findings for the Cd cluster, where the effect on shifts largely cancels between the shielding constants
of the cluster and the reference compound.

\begin{table}[tbp]
    \centering
    \caption{Calculated and experimental $^1$H, $^{31}$P and $^{119}$Sn NMR shifts of (HGe$_3$P)$^{2-}$ and (HSn$_3$As)$^{2-}$ 
    in ppm. The two different $^{119}$Sn shifts for (HSn$_3$As)$^{2-}$ refer to the \textit{Sn}--H--\textit{Sn} and Sn--\textit{Sn}--Sn moieties.
    Experimental results in DMF are taken from Refs.~\citenum{Spang.Beuthert.ea:From.2026} and \citenum{Kubis.Peerless.ea:SnAs}.
    Due to strong spin--orbit effects on the $^1$H and $^{119}$Sn shifts, we only list 2c results and refer to the Supporting Information
    for the scalar-relativistic results.}
    \label{tab:tetrahedrons-nmr}
    \begin{tabular}{@{\extracolsep{4pt}}
    l
    S[table-format = -2.2]
    S[table-format = -3.1]
    S[table-format = -2.2]
    S[table-format = -4.0]
    S[table-format = -4.0]@{}
    }
    \toprule
     & \multicolumn{2}{S}{{\text{(HGe$_3$P)$^{2-}$}}} & \multicolumn{3}{S}{{\text{(HSn$_3$As)$^{2-}$ }}} \\
    \cmidrule{2-3} \cmidrule{4-6}
     & {\text{$^1$H}} & {\text{$^{31}$P}} & {\text{$^1$H}} & {\text{$^{119}$Sn}} & {\text{$^{119}$Sn}} \\
    \midrule
    TPSS COSMO-only & -10.97 & -575.1 & -16.30 & -1664 & -2108 \\
    cTPSS COSMO-only & -11.02 & -578.4 & -16.23 & -1881 & -2101 \\
    TPSS & -9.66 & -429.6 & -15.33 & -1410 & -1855 \\
    cTPSS & -9.73 & -433.6 & -15.32 & -1407 & -1853 \\
    TPSSh & -9.76 & -442.1 & -15.61 & -1404 & -1861 \\
    cTPSSh & -9.86 & -445.8 & -15.63 & -1402 & -1860 \\
    r$^2$SCAN & -9.69 & -414.7 & -16.06 & -1450 & -1895 \\
    cr$^2$SCAN & -9.81 & -426.5 & -15.96 & -1451 & -1896 \\
    TASK & -9.94 & -461.2 & -17.51 & -1486 & -2004 \\
    cTASK & -10.25 & -475.2 & -17.28 & -1466 & -1960 \\
    MN15-L & -9.26 & -419.1 & -15.53 & -1475 & -1907 \\
    cMN15-L & -9.31& -448.9 & -15.61 & -1459 & -1891 \\
    MN15 & -10.93 & -439.3 & -19.10 & -1462 & -2075 \\
    cMN15 & -10.76 & -431.4 & -19.08 & -1428 & -2019 \\
    LH12ct & -9.57 & -446.6 & -15.61 & -1474 & -1934 \\
    cLH12ct & -9.54 & -450.1 & -15.54 & -1467 & -1925 \\
    CHYF-B95 & -10.13 & -426.6 & -16.41 & -1406 & -1869 \\
    cCHYF-B95 & -10.10 & -441.6 & -16.42 & -1411 & -1879 \\
    \midrule
    Experiment & -10.03 & -432.9 & -14.79 & -1339 & -1670 \\
    \bottomrule
    \end{tabular}
\end{table}

With the explicit solvent shell, the $^1$H and $^{31}$P shifts of (HGe$_3$P)$^{2-}$ are
reproduced well with most functionals and even excellently with cTPSS and cr$^2$SCAN.
The $^1$H shifts deviate by at most 0.7\,ppm from the experimental value, while the $^{31}$P
shifts of most functionals by less than 20\,ppm. For $^{31}$P, TASK and especially cTASK are 
the notable exceptions, deviating considerably more compared to all other functionals, with the
current density dependent cTASK even adding to the deviation, while most other functionals
see an improvement for $^{31}$P upon adding the current density. For (HSn$_3$As)$^{2-}$, 
the $^1$H shifts are overestimated in magnitude by 0.5--1.6\,ppm for most functionals. 
The $^{119}$Sn shifts are generally too negative, by 60--130\,ppm for the \textit{Sn}--H--\textit{Sn}
and by 180--350\,ppm for the Sn--\textit{Sn}--Sn moiety, and are thus less accurate than the $^{31}$P shifts.
Here, the two heavy-element signals are dominated by spin--orbit effects, and the explicit
solvent shell may not yet be sufficient to saturate the environment of the large Sn$_3$As cluster.
Additionally, more than one solvent-solute structure may be needed to represent the system more
realistically, even though the impact of different structures was shown to be
small for (HGe$_3$P)$^{2-}$. \cite{Spang.Beuthert.ea:From.2026}
To sum up, as in the previous example, the best performers are TPSS, TPSSh, and CHYF-B95, 
which are however rather insensitive to the current density due to the reasons 
previously explained.

%%%%%%%%%%%%%%%%%%%%%%%%%%%%%%%%%%%%%%%%%%%%%%%%%%%%%%%%%%%%%%%%%%%%%
%%% Conclusion
%%%%%%%%%%%%%%%%%%%%%%%%%%%%%%%%%%%%%%%%%%%%%%%%%%%%%%%%%%%%%%%%%%%%%
\section{Summary \& Conclusion}
\label{sec:conclusion}
We presented the first GIAO-based approach for NMR shifts with
current-dependent meta-GGAs in the presence of spin--orbit coupling
as well as the first application of local hybrid functionals for
NMR shifts in a relativistic two-component formalism.
This fully unlocks meta-GGA functionals and many local hybrid
functionals for computational NMR spectroscopy of compounds
with heavy elements. For instance, it has been argued in the literature
that Minnesota functionals or SCAN variants should only be used with
the current-dependent generalization.
\cite{Schattenberg.Kaupp:Effect.2021, Franzke.Holzer:Impact.2022}
Similar findings hold for local hybrids which are usually based on
the iso-orbital condition. Additionally, the spin current densities
are already needed to arrive at a covariant formulation for GGAs.
We considered challenging systems in the first NMR evaluation of the
given meta-GGAs and LHFs with spin--orbit coupling. 
Results are generally encouraging so that these functionals will
be beneficial for \textit{in silico} studies.
However, our results also reveal that not only the functional is
of great importance but also the environment and structure sampling
are becoming more and more important due to the improved capabilities
of the functional zoo. Here, heavy elements are still challenging
as many efficient or low-cost computational tools for large-scale
structure sampling and solvent effects rely on parametrization,
\cite{Pracht.Grimme.ea:CREST.2024, Wood.Dzamba.ea:UMA.2026,
Porezag.Frauenheim.ea:Construction.1995, Jha.Heine:DFTB.2022}
which typically focuses more on light elements or organometallic
compounds instead of metal clusters.

%%%%%%%%%%%%%%%%%%%%%%%%%%%%%%%%%%%%%%%%%%%%%%%%%%%%%%%%%%%%%%%%%%%%%
%% Supplementary material
%%%%%%%%%%%%%%%%%%%%%%%%%%%%%%%%%%%%%%%%%%%%%%%%%%%%%%%%%%%%%%%%%%%%%
\section*{Supplementary Material}
Supporting Information is available as a ZIP archive with
\begin{itemize}
   \item Modified x2c-TZVPall-2c/x2c-TZVPall-s basis set (x2c-TZVPall-2c-s)
   for 2c relativistic NMR calculations of H, P, Cd, and Sn (TURBOMOLE format, ASCII txt file).
   \item Structures of the reference compounds optimized in the present work (xyz file).
   \item Spreadsheet for the evaluation of the P and Cd NMR shifts of [Cd$_3$(Ge$_3$P)$_3$]$^{3-}$ (xlsx and ods files).
   \item Spreadsheet for the evaluation of the NMR shifts of (HGe$_3$P)$^{2-}$ and (HSn$_3$As)$^{2-}$ (xlsx and ods files).
\end{itemize}

%%%%%%%%%%%%%%%%%%%%%%%%%%%%%%%%%%%%%%%%%%%%%%%%%%%%%%%%%%%%%%%%%%%%%
%% Acknowledgments
%%%%%%%%%%%%%%%%%%%%%%%%%%%%%%%%%%%%%%%%%%%%%%%%%%%%%%%%%%%%%%%%%%%%%
\begin{acknowledgments}
Applications of Y.J.F.\ were supported through the Emmy-Noether Programme funded by the
Deutsche Forschungsgemeinschaft (DFG, German Research Foundation) -- 574993465. 
\end{acknowledgments}
\medskip

\noindent \textbf{Note}: Claude Opus/Fable (Anthropic) was used to
validate, comment, and leverage/optimize code. The authors reviewed
all modifications of the code and take full responsibility for the
code and the theoretical work.

%%%%%%%%%%%%%%%%%%%%%%%%%%%%%%%%%%%%%%%%%%%%%%%%%%%%%%%%%%%%%%%%%%%%%
%% Author Contributions and Declarations
%%%%%%%%%%%%%%%%%%%%%%%%%%%%%%%%%%%%%%%%%%%%%%%%%%%%%%%%%%%%%%%%%%%%%
\section*{Author Contributions and Declarations}
\noindent \textbf{Yannick J. Franzke}: Conceptualization (equal);
Data curation (equal); Formal analysis (equal); Investigation (equal);
Methodology (equal); Software (equal); Validation (equal);
Visualization (equal); Writing – original draft (equal);
Writing – review \& editing (equal).
\\
\noindent \textbf{Christof Holzer}: Conceptualization (equal);
Data curation (equal); Formal analysis (equal); Investigation (equal);
Methodology (equal); Software (equal); Validation (equal);
Visualization (equal); Writing – original draft (equal);
Writing – review \& editing (equal).

%%%%%%%%%%%%%%%%%%%%%%%%%%%%%%%%%%%%%%%%%%%%%%%%%%%%%%%%%%%%%%%%%%%%%
%% Data Availability Statement
%%%%%%%%%%%%%%%%%%%%%%%%%%%%%%%%%%%%%%%%%%%%%%%%%%%%%%%%%%%%%%%%%%%%%
\section*{Data and Code Availability Statement}
The data that support the findings of this study are available within
the article and its supplementary material.
TURBOMOLE is available from the webpage \url{https://www.turbomole.org/}
as compiled binaries. \cite{TURBOMOLE}
The program suite can be obtained free of charge for students an
educational purposes.

%%%%%%%%%%%%%%%%%%%%%%%%%%%%%%%%%%%%%%%%%%%%%%%%%%%%%%%%%%%%%%%%%%%%%
%%% Appendix
%%%%%%%%%%%%%%%%%%%%%%%%%%%%%%%%%%%%%%%%%%%%%%%%%%%%%%%%%%%%%%%%%%%%%
\appendix
\allowdisplaybreaks

\section{Transformation Laws}
\label{sec:trafo}
We briefly review the transformation laws, which are relevant for the present work.
A local gauge transformation changes every spinor based on $\phi_i\to\exp[\text{i}\theta + \text{i}\boldsymbol{\lambda}\circ\hat{\boldsymbol{\sigma}}]\,\theta_i$
with a real function $\theta$ and a real spin vector $\boldsymbol{\lambda}$ of the position.
\cite{Vignale.Rasolt:Current-.1988, Bencheikh:Spin-orbit.2003}
For the densities, the kinetic energy densities, and
the particle current the infinitesimal changes are consequently obtained by
the insertion $\phi_i + \text{i}(\delta\theta + \delta\boldsymbol{\lambda}\circ\hat{\boldsymbol{\sigma}})\phi_i$
and considering $\hat{\sigma}_u\hat{\sigma}_v = \delta_{uv} \one + \text{i}\epsilon_{uvw}\hat{\sigma_w}$, where
$\epsilon_{uvw}$ denotes the Levi--Civita symbol.
\cite{Pittalis.Vignale.ea:U1.2017, Desmarais.Bencheikh.ea:Physical.2026}
For the phase, this results in
\begin{align}
  \delta\rho_0 &= 0, & \delta\sv m &= 0, \nonumber\\
  \delta j_{0,\alpha} &= (\nabla_{\alpha}\delta\theta)\,\rho_0, &
  \delta\sv Y_{\alpha} &= (\nabla_{\alpha}\delta\theta)\,\sv m, \\
  \delta\tau_0 &= \sum_{\alpha}(\nabla_{\alpha}\delta\theta)\,j_{0,\alpha}, &
  \delta\sv u &= \sum_{\alpha}(\nabla_{\alpha}\delta\theta)\,\sv Y_{\alpha} , \nonumber
  \label{eq:u1}
\end{align}
and for the spin rotation it results in
\begin{align}
  \delta\rho_0 &= 0,   \nonumber\\
  \delta\sv m &= -2\,\delta\boldsymbol{\lambda}\xs\sv m, \nonumber\\
  \delta j_{0,\alpha} &= (\nabla_{\alpha}\delta\boldsymbol{\lambda})\circ\sv m, \\
  \delta\sv Y_{\alpha} &= (\nabla_{\alpha}\delta\boldsymbol{\lambda})\,\rho_0 - 2\,\delta\boldsymbol{\lambda}\xs\sv Y_{\alpha},
  \nonumber\\
  \delta\tau_0 &= \sum_{\alpha}(\nabla_{\alpha}\delta\boldsymbol{\lambda})\circ\sv Y_{\alpha},   \nonumber\\
  \delta\sv u &= \sum_{\alpha}\Big[(\nabla_{\alpha}\delta\boldsymbol{\lambda})\,j_{0,\alpha}
  - \tfrac12\,(\nabla_{\alpha}\delta\boldsymbol{\lambda})\xs\nabla_{\alpha}\sv m\Big]
  - 2\,\delta\boldsymbol{\lambda}\xs\sv u . \nonumber
  \label{eq:su2}
\end{align}
This shows that terms with $\delta\boldsymbol{\lambda}$ itself are a common rotation of all spin vectors.
In contrast, the terms with its gradient are inhomogeneous and have to cancel in an invariant quantity.
Therefore, the GGA closed-shell kernel \cite{Autschbach:role.2013, Kuehn.Weigend:Implementation.2013,
Franzke.Holzer:Exact.2023} already requires an extension as outlined in the present work, as it is not
invariant under a local spin rotation. Of course, this also holds for the Scalmani--Frisch approach for
real open-shell systems in paramagnetic NMR or time-dependent DFT.

\section{Covariant GGA Kernel}
\label{sec:gga-covariant}
The gradient of the spin magnetization consists of two parts. Here, we consider the direction
$\sv e = \sv m/|\sv m|$. Then, the gradient consists of a longitudinal and a transverse contribution
according to
\begin{equation}
  \nabla_{\alpha}\sv m = (\nabla_{\alpha}|\sv m|)\,\sv e + |\sv m|\,\nabla_{\alpha}\sv e ,
  \qquad \sv e\circ \nabla_{\alpha}\sv e = 0 .
  \label{eq:translong}
\end{equation}
The first term, i.e.\ the longitudinal part, is parallel to $\sv m$ and is related to the change of the length.
The transverse gradient, $\nabla_{\alpha}\sv m - \sv e\,(\sv e\circ\nabla_{\alpha}\sv m)$,
is perpendicular to $\sv m$ and related to the change of the direction. The latter vanishes
for a collinear state. For a closed-shell reference, $\sv m$ becomes zero and only the
response $\delta\sv m$ has to be considered. Here, the transverse gradient part breaks
the invariance and a functional that uses the plain gradient of the magnetization component
by component is therefore not invariant. The transverse gradient was motivated by
Scalmani and Frisch to yield a non-vanishing local torque,
\cite{Capelle.Vignale.ea:Spin.2001, Sharma.Dewhurst.ea:First-Principles.2007}
which is important for spin dynamics. \cite{Scalmani.Frisch:New.2012, Egidi.Sun.ea:Two-Component.2017}
However, this transverse gradient necessitates the inclusion of the spin current densities
to meet the invariance under local spin rotations. \cite{Vignale.Rasolt:Current-.1988,
Bencheikh:Spin-orbit.2003, Pittalis.Vignale.ea:U1.2017, Desmarais.Bencheikh.ea:Physical.2026}

To ensure invariance again, we apply the covariant derivative,
\begin{equation}
  \mathcal D_{\alpha}\sv m = \nabla_{\alpha}\sv m + \frac{2}{\rho_0}\,\sv Y_{\alpha}\xs\sv m ,
  \label{eq:covder}
\end{equation}
which remains unchanged by a local phase transformation. Its part along $\sv m$ is
$\nabla_{\alpha}|\sv m|$ and the spin current densities are parallel to $\sv m$.
Consequently, collinear states are not affected by this generalization.
The simple substitution of $\nabla_{\alpha} \sv m$ with the covariant derivative
$\mathcal D_{\alpha}\sv m$ allows us to generalize the conventional Scalmani--Frisch approach.
\cite{Scalmani.Frisch:New.2012, Bulik.Scalmani.ea:Noncollinear.2013, Egidi.Sun.ea:Two-Component.2017,
Komorovsky.Cherry.ea:Four-component.2019}
The new non-collinear variables read
\begin{align}
  \rho_{\uparrow,\downarrow} &= \tfrac12(\rho_0\pm|\sv m|), \qquad
  \gamma_{\uparrow\downarrow} = \tfrac14\big(|\nabla\rho_0|^2 - A\big), \nonumber\\
  \gamma_{\uparrow\uparrow,\downarrow\downarrow} &= \tfrac14\big(|\nabla\rho_0|^2 + A\big)
  \pm \tfrac12\operatorname{sgn}(\bm\Gamma\circ\sv m)\,|\bm\Gamma| ,
\end{align}
with $A = \sum_{\alpha}\mathcal D_{\alpha}\sv m\circ\mathcal D_{\alpha}\sv m$ and
$\bm\Gamma = \sum_{\alpha} g_{\alpha}\,\mathcal D_{\alpha}\sv m$, where $g_{\alpha} = \nabla_{\alpha}\rho_0$.
This meets the requirements and ideas of the literature, i.e.\ the covariant derivative
keeps the torque and adds the missing dependence.
\cite{Vignale.Rasolt:Current-.1988, Bencheikh:Spin-orbit.2003, Pittalis.Vignale.ea:U1.2017}
The closed-shell ground-state limit is correctly recovered and for a collinear state
the variables coincide with those of the spin-unrestricted functional.
A few comments on the weights of the gradients of the spin magnetization along and perpendicular
to its direction are in order. In the Scalmani--Frisch formulation, these use equal weights
and this is only legitimate for weak spin polarization or strong inhomogeneity.
\cite{Eich.Pittalis.ea:Transverse.2013, Komorovsky.Cherry.ea:Four-component.2019}
In the present work, we consider closed-shell ground states and thus the equal weight is
well justified or even required by the isotropy in spin space. Different weights are
further possible and still meet the invariance requirements, as the contribution
along $\sv m$, $\nabla_{\alpha}|\sv m|$, and the part perpendicular to $\sv m$ are
both invariant itself.

The kernel requires the second-order energy expression. For NMR properties, the second-order energy
expression is greatly simplified, as spin channels and the charge channel decouple in second order.
The spin contribution is of a quadratic form in $\delta\sv m$ and in $X_{\alpha,u}$, which is 
the first-order covariant derivative of Eq.~\eqref{eq:covder} according to
\begin{equation}
  X_{\alpha,u} = \nabla_{\alpha}\delta m_u + \frac{2}{\rho_0}\,(\sv Y_{\alpha}\xs\delta\sv m)_u .
  \label{eq:scX}
\end{equation}
The response of the particle density $\rho_0$ and of the spin current densities
$\sv Y_{\alpha}$ are not needed for $X_{\alpha,u}$, as they are multiplied with $\sv m = 0$.
The second-order part of $\mathcal D_{\alpha}\sv m$, i.e.\ $(2/\rho_0)\,\delta\sv Y_{\alpha}\xs\delta\sv m - (2\,\delta\rho_0/\rho_0^2)\,\sv Y_{\alpha}\xs\delta\sv m$,
multiplies a vanishing first derivative and is also not needed for the kernel but the
hyperkernel. Therefore, the second-order XC energy becomes
\begin{equation}
\begin{split}
  & E^{\text{XC}, (2)}_{\text{spin}} = \int\sum_u \frac{f_1}{2}\,\delta m_u^2 \,\text{d} \mathbf{r}  \\
  + & \int\sum_u \Big[
  + f_2\,\delta m_u\,(\sv g\cdot\sv X_u)
  + \frac{f_3}{2}\,(\sv g\cdot\sv X_u)^2
  + \frac{f_4}{2}\sum_{\alpha} X_{\alpha,u}^2\Big]\,\text{d} \mathbf{r} ,
  \label{eq:scE2-gga}  
\end{split}
\end{equation}
with $\sv g\cdot\sv X_u = \sum_{\alpha} g_{\alpha} X_{\alpha,u}$. Only the triplet combinations of the
derivatives of $e^{\text{XC}}$ at $\rho_{\uparrow} = \rho_{\downarrow} = \rho_0/2$ and
$\gamma_{\uparrow\uparrow} = \gamma_{\uparrow\downarrow} = \gamma_{\downarrow\downarrow}
= |\boldsymbol{\nabla}\rho_0|^2/4$, are needed
\begin{align}
  f_1 &= \frac12\Big(\frac{\partial^2 e^{\text{XC}}}{\partial\rho_{\uparrow}\partial\rho_{\uparrow}}
  - \frac{\partial^2 e^{\text{XC}}}{\partial\rho_{\uparrow}\partial\rho_{\downarrow}}\Big), \\
  f_2 &= \frac12\Big(\frac{\partial^2 e^{\text{XC}}}{\partial\rho_{\uparrow}\partial\gamma_{\uparrow\uparrow}}
  - \frac{\partial^2 e^{\text{XC}}}{\partial\rho_{\uparrow}\partial\gamma_{\downarrow\downarrow}}\Big),
 \\
  f_3 &= \frac12\Big(\frac{\partial^2 e^{\text{XC}}}{\partial\gamma_{\uparrow\uparrow}\partial\gamma_{\uparrow\uparrow}}
  - \frac{\partial^2 e^{\text{XC}}}{\partial\gamma_{\uparrow\uparrow}\partial\gamma_{\downarrow\downarrow}}\Big), \\
  f_4 &= \frac{\partial e^{\text{XC}}}{\partial\gamma_{\uparrow\uparrow}}
  - \frac12\,\frac{\partial e^{\text{XC}}}{\partial\gamma_{\uparrow\downarrow}} .
\end{align}
The derivatives remain unchanged upon exchange of the two spin labels and one representative of
each pair is written. Note that the same-spin and the opposite-spin derivatives differ and
consequently the differences in $f_1$ to $f_4$ do not vanish.
The weights $f$ are available from the triplet kernel of a one-component closed-shell calculation
and the spin components use equal weights to ensure isotropy in spin space at $\sv m = 0$.
Further, the spin current densities couple the three channels.

The first-order XC potential follows with the derivatives of the integrand of Eq.~\eqref{eq:scE2-gga}
with respect to $\delta m_u$ at fixed $X_{\alpha,u}$ and with respect to $X_{\alpha,u}$. These read
\begin{align}
  h_u = & f_u\,\delta m_u + f_2\,(\sv g\cdot\sv X_u), \\
  v_{\alpha,u} = & \big(f_2\,\delta m_u + f_3\,(\sv g\cdot\sv X_u)\big)\,g_{\alpha} + f_4\,X_{\alpha,u} .
\end{align}
Note that the spin current densities add a term to $h_u$.
For an implementation purpose, we define so-called rows. Here, a row is the term at a grid point
that multiplies one type of basis function product in the integration of the XC potential matrix. 
That is, the density row (LDA row) multiplies the product $\chi_\mu\chi_\nu$ and the gradient row (GGA row)
multiplies its derivative $\nabla_{\alpha}(\chi_\mu\chi_\nu)$.
Finally, this yields the rows of the first-order potential of spin channel $u$ as
\begin{align}
  \text{density row, spin:} \quad & d_u = h_u
  + \frac{2}{\rho_0}\sum_{\alpha}(\sv v_{\alpha}\xs\sv Y_{\alpha})_u , \label{eq:scrows}\\
  \text{gradient row, spin:} \quad & v_{\alpha,u} .
\end{align}
The first-order XC potential matrix in a basis of real functions $\chi_\mu$ becomes
\begin{equation}
  \delta V^{\text{XC}}_{\mu\nu, u} = \int\Big[d_u\,\chi_\mu\chi_\nu
  + \sum_{\alpha} v_{\alpha,u}\,\nabla_{\alpha}(\chi_\mu\chi_\nu)\Big]\,\text{d} \mathbf{r},
\end{equation}
which multiplies $\hat{\sigma}_u$ in the two-component Fock matrix. The charge channel is related
to the singlet kernel of the closed-shell GGA with the plain gradient of $\delta\rho_0$.
A numerical test with random spinors and a random local transformation confirmed the invariance,
while the plain variables change. We note that valence excitation energies and NMR shieldings
only change slightly but larger changes are expected for systems with an open-shell
ground state.

\section{Details on the Covariant meta-GGA Formulation}
\label{sec:detailed-derivation}
We start by considering the effect of a transformation for the $2 \times 2$ matrices
of the density $\hat{R}$, the kinetic energy density $\hat{Q}$, and their cross term $\hat{K}$ as defined
in Sec.~\ref{subsec:variation-tau}. We note that the first and second terms of $K_{\alpha} = \tfrac12\nabla_{\alpha} \hat{R} + \text{i}\hat{J}_{\alpha}$ are Hermitian and anti-Hermitian, respectively.
Further, it may also be viewed as the matrix counterpart of $\tfrac12\nabla\rho + \text{i}\vc j$.
Rewriting the matrices in spinors yields $\hat{R} = 2\sum_i\phi_i\phi_i^\dagger$,
$\hat{Q} = \sum_{i,\alpha}(\nabla_{\alpha}\phi_i)(\nabla_{\alpha}\phi_i)^\dagger$ and
$\hat{K}_{\alpha} = 2\sum_i(\nabla_{\alpha}\phi_i)\phi_i^\dagger$.
Here, a transformation $\phi_i\to \hat{U}\phi_i$ with $\hat{A}_{\alpha} = -\text{i}\hat{U}^\dagger\nabla_{\alpha} \hat{U}$ results in
$\nabla_{\alpha}\phi_i\to \hat{U}(\nabla_{\alpha} + \text{i}\hat{A}_{\alpha})\phi_i$ and
\begin{align}
  \hat{R}\to & \hat{U}\hat{R}\hat{U}^\dagger, \\
  \hat{K}_{\alpha} \to & \hat{U}(\hat{K}_{\alpha} + \text{i}\hat{A}_{\alpha} \hat{R})\hat{U}^\dagger, \\
  \hat{Q} \to & \hat{U}\Big[\hat{Q}
  + \tfrac{\text{i}}{2}\sum_{\alpha}\big(\hat{A}_{\alpha} \hat{K}_{\alpha}^\dagger - \hat{K}_{\alpha} \hat{A}_{\alpha}\big)
  + \tfrac12\sum_{\alpha} \hat{A}_{\alpha} \hat{R}\hat{A}_{\alpha}\Big]\hat{U}^\dagger .
\end{align}
Now consider $\tfrac12\sum_{\alpha} \hat{K}_{\alpha} \hat{R}^{-1}\hat{K}_{\alpha}^\dagger$ which corresponds
to the lower bound of the kinetic energy density in a matrix sense that a state given above
has. Without spin terms, the bound is given as
\begin{equation}
  \frac{|\nabla\rho|^2}{8\rho} + \frac{|\vc j|^2}{2\rho} .
\end{equation}
The first term is the von-Weizs{\"a}cker part and the second term is the current density part.
Notably, $\tfrac12\sum_{\alpha} \hat{K}_{\alpha} \hat{R}^{-1}\hat{K}_{\alpha}^\dagger$ gives rise to the same
inhomogeneous terms as above. Therefore, we find
\begin{equation}
\begin{split}
  \pk = & \hat{Q} - \frac12\sum_{\alpha} \hat{K}_{\alpha} \hat{R}^{-1}\hat{K}_{\alpha}^\dagger \\
  = & \hat{\bar{Q}} - \frac18\sum_{\alpha}\nabla_{\alpha} \hat{R}\,\hat{R}^{-1}\,\nabla_{\alpha} \hat{R}
  + \frac{\text{i}}{4}\sum_{\alpha}\big(\nabla_{\alpha} \hat{R}\,\hat{R}^{-1}\hat{J}_{\alpha} - \hat{J}_{\alpha} \hat{R}^{-1}\nabla_{\alpha} \hat{R}\big)
\end{split}
\end{equation}
which transforms as $\pk\to \hat{U}\pk \hat{U}^\dagger$ with $\hat{\bar{Q}} = \hat{Q} - \tfrac12\sum_{\alpha} \hat{J}_{\alpha} \hat{R}^{-1}\hat{J}_{\alpha}$.
This matrix corresponds to the matrix form of the kinetic energy density after subtracting
the von-Weizs{\"a}cker and the current density parts. This is the crucial quantity for the
iso-orbital indicator and it becomes zero for a single spinor. We refer to this matrix as
the Pauli kinetic energy matrix. We note in passing that the $\tfrac18\sum_{\alpha} \nabla_{\alpha}  \hat{R}\,\hat{R}^{-1}\,\nabla_{\alpha}  \hat{R}$ term in $\pk$ is not covariant itself but only
its combination with the current density terms is.

For the complete kinetic energy density $\hat{T}$, the von-Weizs{\"acker} term $\hat{W}$ has to be added
and the gauge transformation requirements need to be considered.
Thus, the von-Weizs{\"acker} term  $\hat{W}$ has to use the covariant gradient of $\hat{R}$, which
introduces a dependence on the current densities, so that
\begin{align}
  \hat{W} = & \frac18\sum_{\alpha}\mathcal D_{\alpha} \hat{R}\,\hat{R}^{-1}\,\mathcal D_{\alpha} \hat{R} , \\
  \mathcal D_{\alpha} \hat{R} = & \nabla_{\alpha}\rho_0\,\one + \mathcal D_{\alpha}\sv m\circ\hat{\bm\sigma} .
\end{align}
The gradient of the density is invariant and $\mathcal D_{\alpha} \sv m$ rotates with $\sv m$.
Hence, the transformation results in $\mathcal D_{\alpha}  \hat{R}\to \hat{U}\,\mathcal D_{\alpha}  \hat{R}\,\hat{U}^\dagger$ and $\hat{W}\to \hat{U}\hat{W}\hat{U}^\dagger$.
Rearranging leads to the form of $\hat{W}$ given in Eq.~\eqref{eq:W} with the charge contribution $S$ and the spin contribution $\sv V$.
The sum $\hat{T} = \pk + \hat{W}$ transforms as $\hat{T}\to \hat{U}\hat{T}\hat{U}^\dagger$ and
its eigenvalues are invariant.

For a closed-shell reference, the kinetic energy variables reduce to
\begin{align}
  \hat{W} &= \frac{|\nabla\rho_0|^2}{8\rho_0}\,\one, \qquad
  \pk = \Big[\tau_0 - \frac{|\nabla\rho_0|^2}{8\rho_0}
  - \frac{1}{2\rho_0}\sum_{\alpha}\sv Y_{\alpha}\circ\sv Y_{\alpha}\Big]\one, \nonumber\\
  \hat{T} &= \Big[\tau_0 - \frac{1}{2\rho_0}\sum_{\alpha}\sv Y_{\alpha}\circ\sv Y_{\alpha}\Big]\one .
  \label{eq:Tref}
\end{align}
Generally, the charge and spin channels are available with the partition scheme
\begin{equation}
  \hat{T} = t_0\,\one + \sv t\circ \hat{\boldsymbol{\sigma}}, \qquad
  t_0 = \tfrac12\operatorname{tr}\hat{T}, \qquad t_i = \tfrac12\operatorname{tr}(\hat{\sigma}_u\hat{T}).
\end{equation}
Therefore, $t_0$ is the decisive quantity for the evaluation of the energy density
$e^{\text{XC}}$ of a given functional. In this sense, it replaces $\tau_0$ of the
standard meta-GGA ansatz. Based on the separation of the charge and spin channels,
the open-shell non-collinear canonical and Scalmani--Frisch approach for the XC
potential can also be generalized to this covariant formulation.

The given formulation also meets the von-Weizs{\"a}cker bound and thus generalizes
the findings of Ref.~\citenum{Bates.Furche:Harnessing.2012}, which shows the proof
for the gauge-invariant kinetic energy without SOC based on the Cauchy--Schwarz
inequality for $\tfrac12\nabla\rho + \text{i}\vc j$. With the $2\times N$ matrix
$\Phi$ of the occupied spinors, we get the Pauli forms $\hat{R} = 2\Phi\Phi^\dagger$,
$\hat{K}_{\alpha} = 2(\nabla_{\alpha}\Phi)\Phi^\dagger$, and
$\hat{Q} = \sum_{\alpha}(\nabla_{\alpha}\Phi)(\nabla_{\alpha}\Phi)^\dagger$ to generalize
the proof so that
\begin{equation}
  \pk = \hat{Q} - \frac12\sum_{\alpha} \hat{K}_{\alpha} \hat{R}^{-1}\hat{K}_{\alpha}^\dagger \ge 0, \quad
  \hat{T} = \pk + \hat{W} \ge \hat{W} \ge 0 .
  \label{eq:vwmat}
\end{equation}
Furthermore, the NMR calculations should meet the bound. To do so, the kinetic energy
density variable of the reference state has to include the spin current densities,
the kernel is the second derivative of a form for that the bound holds in every state,
and at grid points where the slack of the bound in a spin channel $\sigma\in\{\uparrow,\downarrow\}$,
$\tilde\tau_\sigma - \gamma_{\sigma\sigma}/(8\rho_\sigma)$, is zero for the reference, it does
not respond to the perturbation. This was verified numerically with random spinors.
Finally, we verified the formulation with the generalized test based on Ref.~\citenum{Bates.Heiche.ea:Erratum.2022}.
Here, $\pk$ becomes zero for every state of two spinors, not only for the ground-state reference.
This means that $\hat{T} = \hat{W}$ is true for a perturbation of a two-electron closed-shell system
and $\hat{W}$ includes variables of a covariant GGA only.

\section{Comparison to the Minimal Substitution}
\label{bec:comparison-sub}
The functionals proposed in Refs.~\citenum{Pittalis.Vignale.ea:U1.2017, Desmarais.Erba:Meta-Generalized.2025}
are also invariant, however, these are built in a different way. For an open-shell system, the energy
density $e^{\text{XC}}$ for the functional form depends on $\rho_0$, its gradient, the spin polarization
$\zeta = m_{\text{s}} / \rho_0$ and one kinetic variable for both spin channels,
\begin{equation}
  \tilde\tau = \tau_0 - \frac{\mathbf{j}_0\cdot \mathbf{j}_0}{2\rho_0} + \frac{\sv m\circ\sv u}{\rho_0}
  - \frac{1}{2\rho_0}\sum_{\alpha}\sv Y_{\alpha}\circ\sv Y_{\alpha}
  + \frac{1}{8\rho_0}\sum_{\alpha}\nabla_{\alpha}\sv m\circ\nabla_{\alpha}\sv m .
  \label{eq:pve}
\end{equation}
For a closed-shell Kramers-restricted reference, it becomes $\tilde\tau = \tau_0 - \sum_{\alpha}\sv Y_{\alpha}\circ\sv Y_{\alpha}/(2\rho_0)$,
which again corresponds to the value of $t_0$ herein. For a time-odd perturbation such as magnetic perturbation,
the first-order variation vanishes and we obtain
\begin{equation}
  \tfrac12\,\delta^2\tilde\tau = -\frac{1}{2\rho_0}\sum_{\alpha}(\delta j_{0,\alpha})^2
  + \frac{\delta\sv m\circ\delta\sv u}{\rho_0}
  + \frac{1}{8\rho_0}\sum_{\alpha}\nabla_{\alpha}\delta\sv m\circ\nabla_{\alpha}\delta\sv m ,
\end{equation}
as the spin current densities do not respond to first order. With the definitions
$f_\tau = \partial e^{\text{XC}}/\partial\tilde\tau$ and
$e^{\text{XC}}_{\zeta\zeta} = \partial^2e^{\text{XC}}/\partial\zeta^2$, the rows of the first-order
XC potential follow as
\begin{align}
  \text{density row, spin:} \quad & \frac{e^{\text{XC}}_{\zeta\zeta}}{\rho_0^2}\,\delta\sv m
  + \frac{f_\tau}{\rho_0}\,\delta\sv u ,  \\
  \text{gradient row, spin:} \quad & \frac{f_\tau}{4\rho_0}\,\nabla_{\alpha}\delta\sv m , \\
  \text{kinetic row, spin:} \quad & \frac{f_\tau}{\rho_0}\,\delta\sv m ,  \\
  \text{current row, particle:} \quad & -\frac{f_\tau}{\rho_0}\,\delta j_{0,\alpha} .
\end{align}
Notably, this is a much simpler formulation at first glance, as the ground-state spin currents
enter through the value of the generalized kinetic energy density $\tilde\tau$ only. The kernel is
invariant and thus both kernels are correct in terms of transformations. However, they belong to
different XC functionals. 

The singlet and triplet structure can be formed also for this kernel.
In a collinear state, the variable of Eq.~\eqref{eq:pve} becomes a function of spin channel variables,
\begin{equation}
  \tilde\tau = \frac{2}{\rho_0}\big(\rho_{\uparrow}\tau_{\uparrow} + \rho_{\downarrow}\tau_{\downarrow}\big)
  - \frac{|\vc j_{\uparrow}|^2 + |\vc j_{\downarrow}|^2}{\rho_0}
  + \frac{\gamma_{\uparrow\uparrow} - 2\gamma_{\uparrow\downarrow} + \gamma_{\downarrow\downarrow}}{8\rho_0}.
\end{equation}
In the collinear limit along the axis $u$, only 
\begin{align}
  \rho_0 = & \rho_{\uparrow} + \rho_{\downarrow}, & \quad m_u = & \rho_{\uparrow} - \rho_{\downarrow}, \nonumber \\
  \tau_0 = & \tau_{\uparrow} + \tau_{\downarrow}, & \quad u_u = & \tau_{\uparrow} - \tau_{\downarrow}, \\
  \vc j_0 = & \vc j_{\uparrow} + \vc j_{\downarrow}, & \quad Y_{\alpha,u} = & j_{\uparrow,\alpha} - j_{\downarrow,\alpha} \nonumber
\end{align}
are generally non-zero and the other spin components vanish. So, the functional has a collinear limit~\cite{Desmarais.Erba:Meta-Generalized.2025}
and the triplet combinations similar to those of the sections~\ref{sec:gga-covariant} and \ref{subsec:variation-tau}
are available. Yet, the weights read differently,
\begin{equation}
\begin{split}
  f_1 = & \frac{\varepsilon_{\zeta\zeta}}{\rho_0^2}, \qquad
  f_4 = \frac{f_\tau}{4\rho_0}, \qquad
  f_5 = \frac{f_\tau}{\rho_0}, \\
  f_2 = & f_3 = f_6 = f_7 = 0 ,
\end{split}
\end{equation}
which was confirmed by numerical differentiation of a test function.
This becomes important for the non-relativistic limit without spin--orbit coupling. The covariant formalism
presented herein in Sec.~\ref{subsec:variation-tau} and \ref{sec:detailed-derivation} reduces to the triplet
kernel of the one-component calculation, while the kernel of this section does not.
The triplet kernel is used for the Fermi-contact and the spin-dipole terms of NMR coupling
constants. \cite{Helgaker.Jaszunski.ea:Ab.1999, Bates.Furche:Harnessing.2012, Holzer.Franzke.ea:Assessing.2021,
Mack.Schattenberg.ea:Nuclear.2020}
Therefore, the covariant formalism herein can be used directly on top of the existing non-relativistic framework.

To sum up, the covariant formulation reduces to the triplet kernel of a one-component
calculation with the existing XC functional in the absence of SOC. In contrast, the kernel of this section
is the triplet kernel of the collinear limit of the new XC functional. Therefore, NMR spin--spin coupling
constants would change even in the non-relativistic limit. NMR shielding constants would agree in the
non-relativistic limit, as only the term with $(\delta j_{0,\alpha})^2$ is required. However, shieldings
would differ with SOC.

\section{Time-Even Response and Relation to TDDFT}
\label{sec:tddft}
So far, the kernel is applicable to NMR properties from a time-odd perturbation such as
shieldings or chemical shifts and coupling constants.
Excitation energies in linear-response TDDFT also necessitate the time-even contribution
of the response $\delta\rho_0$, $x_{\alpha} = \nabla_{\alpha} \delta \rho_0$, $\delta\tau_0$, and $\delta\sv Y_{\alpha}$.
This corresponds to the charge channel of the second-order XC energy.
The complete second-order energy is the sum of the charge and the spin contribution
in Eq.~\eqref{eq:scE2m} according to
\begin{equation}
  E^{\text{XC}, (2)} = E^{\text{XC}, (2)}_{\text{charge}} + E^{\text{XC}, (2)}_{\text{spin}}
  \label{eq:E2sum}
\end{equation}
without mixed terms.
For a closed-shell reference, the generalized kinetic energy variables for the charge part follow as
\begin{align}
	\delta t_0 &= \delta\tau_0 + \kappa\,\delta\rho_0
	- \frac{1}{\rho_0}\sum_{\alpha}\sv Y_{\alpha}\circ\delta\sv Y_{\alpha} , \\
	\tfrac12\,\delta^2t_0 &= -\frac{1}{2\rho_0}\sum_{\alpha}\delta\sv Y_{\alpha}\circ\delta\sv Y_{\alpha}
	+ \frac{\delta\rho_0}{\rho_0^2}\sum_{\alpha}\sv Y_{\alpha}\circ\delta\sv Y_{\alpha}
	- \frac{\kappa}{\rho_0}\,\delta\rho_0^2 .
\end{align}
This means that the singlet kernel does not induce a particle current or a spin
magnetization. The second-order XC energy thus becomes
\begin{equation}
\begin{split}
  E^{\text{XC},(2)}_{\text{charge}} = \int\Big\{&\tfrac12 s_1\,\delta\rho_0^2 \\
  & + s_2\,\delta\rho_0\,(\vc g\cdot\vc x)
  + \tfrac12 s_3\,(\vc g\cdot\vc x)^2
  + \tfrac12 s_4\sum_{\alpha} x_{\alpha}^2 \\
  &+ s_5\,\delta\rho_0\,\delta t_0 + \tfrac12 s_6\,\delta t_0^2
  + s_7\,(\vc g\cdot\vc x)\,\delta t_0 \\
  & + f_\tau\,\tfrac12\,\delta^2t_0\Big\}\,\text{d} \mathbf{r}
\end{split}
\end{equation}
with the singlet weights evaluated at the closed-shell ground-state reference reading
\begin{align}
  s_1 &= \frac12\Big(\frac{\partial^2 e^{\text{XC}}}{\partial\rho_{\uparrow}\partial\rho_{\uparrow}}
  + \frac{\partial^2 e^{\text{XC}}}{\partial\rho_{\uparrow}\partial\rho_{\downarrow}}\Big), \\
  s_2 &= \frac12\Big(\frac{\partial^2 e^{\text{XC}}}{\partial\rho_{\uparrow}\partial\gamma_{\uparrow\uparrow}}
  + \frac{\partial^2 e^{\text{XC}}}{\partial\rho_{\uparrow}\partial\gamma_{\uparrow\downarrow}}
  + \frac{\partial^2 e^{\text{XC}}}{\partial\rho_{\uparrow}\partial\gamma_{\downarrow\downarrow}}\Big),
  \\
  s_3 &= \frac12\Big(\frac{\partial^2 e^{\text{XC}}}{\partial\gamma_{\uparrow\uparrow}\partial\gamma_{\uparrow\uparrow}}
  + 2\,\frac{\partial^2 e^{\text{XC}}}{\partial\gamma_{\uparrow\uparrow}\partial\gamma_{\uparrow\downarrow}}
  + \frac{\partial^2 e^{\text{XC}}}{\partial\gamma_{\uparrow\uparrow}\partial\gamma_{\downarrow\downarrow}}
  + \frac12\,\frac{\partial^2 e^{\text{XC}}}{\partial\gamma_{\uparrow\downarrow}\partial\gamma_{\uparrow\downarrow}}\Big),
  \\
  s_4 & = \frac{\partial e^{\text{XC}}}{\partial\gamma_{\uparrow\uparrow}}
  + \frac12\,\frac{\partial e^{\text{XC}}}{\partial\gamma_{\uparrow\downarrow}} , \\
  s_5 &= \frac12\Big(\frac{\partial^2 e^{\text{XC}}}{\partial\rho_{\uparrow}\partial\tilde\tau_{\uparrow}}
  + \frac{\partial^2 e^{\text{XC}}}{\partial\rho_{\uparrow}\partial\tilde\tau_{\downarrow}}\Big), \\
  s_6 & = \frac12\Big(\frac{\partial^2 e^{\text{XC}}}{\partial\tilde\tau_{\uparrow}\partial\tilde\tau_{\uparrow}}
  + \frac{\partial^2 e^{\text{XC}}}{\partial\tilde\tau_{\uparrow}\partial\tilde\tau_{\downarrow}}\Big),
  \\
  s_7 &= \frac12\Big(\frac{\partial^2 e^{\text{XC}}}{\partial\gamma_{\uparrow\uparrow}\partial\tilde\tau_{\uparrow}}
  + \frac{\partial^2 e^{\text{XC}}}{\partial\gamma_{\uparrow\uparrow}\partial\tilde\tau_{\downarrow}} 
  + \frac{\partial^2 e^{\text{XC}}}{\partial\gamma_{\uparrow\downarrow}\partial\tilde\tau_{\uparrow}}\Big).
\end{align}
The derivatives for the first-order XC potential contribution of the charge part can be evaluated as done for the triplet kernel.
This results in
\begin{align}
  h_0 &= s_1\,\delta\rho_0 + s_2\,(\vc g\cdot\vc x) + s_5\,\delta t_0 , \\
  v_{\alpha,0} &= \big(s_2\,\delta\rho_0 + s_3\,(\vc g\cdot\vc x) + s_7\,\delta t_0\big)\,g_{\alpha}
  + s_4\,x_{\alpha} , \\
  \eta_0 &= s_5\,\delta\rho_0 + s_7\,(\vc g\cdot\vc x) + s_6\,\delta t_0 .
\end{align}
This finally yields the rows for the kernel evaluation according to
\begin{align}
	\begin{split}
	\text{density row, particle:} \quad & h_0 + \kappa\,\eta_0 \\
& + f_\tau\Big[\frac{1}{\rho_0^2}\sum_{\alpha}\sv Y_{\alpha}\circ\delta\sv Y_{\alpha}
- \frac{2\kappa}{\rho_0}\,\delta\rho_0\Big] ,
    \end{split} \\
	\text{gradient row, particle:} \quad & v_{\alpha,0} ,\\
	\text{kinetic row, particle:} \quad & \eta_0 ,  \\
	\begin{split}
	\text{current row, spin:} \quad & -\frac{1}{\rho_0}\,\eta_0\,\sv Y_{\alpha}
	+ \frac{f_\tau}{\rho_0^2}\,\delta\rho_0\,\sv Y_{\alpha} \\
	& - \frac{f_\tau}{\rho_0}\,\delta\sv Y_{\alpha} .
    \end{split}
\end{align}
The dependence of the density row on the spin current densities originates from
$\kappa$ and through the variation of $t_0$. Notably, the covariant generalization
of the gradient in Eq.~\eqref{eq:covder} does not contribute to the charge contribution
of the kernel but the plain particle gradient derivative $x_{\alpha}$ does. The corresponding
GGA kernel for a time-even perturbation does not add the kinetic contributions
and spin current densities at all. Moreover, mixed second-order derivatives of
a time-even and a time-odd variable are zero due to time-reversal symmetry.
Therefore, the closed-shell GGA limit of the kernels presented in the 
literature~\cite{Komorovsky.Cherry.ea:Four-component.2019,
Egidi.Sun.ea:Two-Component.2017, Kuehn.Weigend:Implementation.2013,
Peng.Zou.ea:Time-dependent.2005, Franzke.Holzer:Application.2026}
holds for such a perturbation. TDDFT excitation energies are, however, computed
as a combination of a time-even and a time-odd perturbation. In other words,
the kernel of the excitation energies is the combination of two kernels, i.e.\ the
singlet and triplet kernel, due to the breakdown of spin symmetry with SOC.
The two parts are only coupled through the spinor response to the perturbation but
not the kernels itself.
We note that a non-Hermitian CPKS solver~\cite{Kehry.Franzke.ea:Quasirelativistic.2020}
is applied for the time-dependent DFT excitation energies when using the covariant
generalization to improve the convergence.

%%%%%%%%%%%%%%%%%%%%%%%%%%%%%%%%%%%%%%%%%%%%%%%%%%%%%%%%%%%%%%%%%%%%%
% References
%%%%%%%%%%%%%%%%%%%%%%%%%%%%%%%%%%%%%%%%%%%%%%%%%%%%%%%%%%%%%%%%%%%%%
\section*{References}
\vspace{-0.5cm}
\bibliography{literature}% Produces the bibliography via BibTeX.

@STRING{jcp="J. Chem. Phys."}

@Article{	  Maier.Arbuznikov.ea:Local.2019,
  author	= {Maier, Toni M. and Arbuznikov, Alexei V. and Kaupp,
		  Martin},
  journal	= {Wiley Interdiscip. Rev.: Comput. Mol. Sci.},
  number	= {1},
  pages		= {e1378},
  title		= {Local hybrid functionals: Theory, implementation, and
		  performance of an emerging new tool in quantum chemistry
		  and beyond},
  volume	= {9},
  year		= {2019},
  doi       = {10.1002/wcms.1378},
  url       = {https://doi.org/10.1002/wcms.1378}
}

@Article{	  Kehry.Franzke.ea:Quasirelativistic.2020,
  author	= {Kehry, Max and Franzke, Yannick J. and Holzer, Christof and Klopper, Wim},
  title		= {Quasirelativistic two-component core excitations and polarisabilities from a
               damped-response formulation of the {B}ethe--{S}alpeter equation},
  journal	= {Mol. Phys.},
  volume	= {118},
  number	= {21-22},
  pages		= {e1755064},
  year		= {2020},
  publisher	= {Taylor & Francis},
  doi		= {10.1080/00268976.2020.1755064},
  url		= { https://doi.org/10.1080/00268976.2020.1755064},
}

@Article{	  Plessow.Weigend:Seminumerical.2012,
  author	= {Plessow, Philipp and Weigend, Florian},
  title		= {Seminumerical calculation of the {H}artree--{F}ock exchange
		  matrix: Application to two-component procedures and
		  efficient evaluation of local hybrid density functionals},
  journal	= {J. Comput. Chem.},
  volume	= {33},
  number	= {7},
  pages		= {810--816},
  doi		= {10.1002/jcc.22901},
  url		= {https://onlinelibrary.wiley.com/doi/abs/10.1002/jcc.22901},
  year		= {2012}
}

@Article{	  Franzke.Weigend:NMR.2019,
  author	= {Franzke, Yannick J. and Weigend, Florian},
  title		= {{NMR} Shielding Tensors and Chemical Shifts in
		  Scalar-Relativistic Local Exact Two-Component Theory},
  journal	= {J. Chem. Theory Comput.},
  volume	= {15},
  number	= {2},
  pages		= {1028--1043},
  year		= {2019},
  doi		= {10.1021/acs.jctc.8b01084}
}

@Article{	  Franzke.Tress.ea:Error-consistent.2019,
  author	= {Franzke, Yannick J. and Tre\ss{}, Robert and Pazdera,
		  Tobias M. and Weigend, Florian},
  title		= {Error-consistent segmented contracted all-electron
		  relativistic basis sets of double- and triple-zeta quality
		  for {NMR} shielding constants},
  journal	= {Phys. Chem. Chem. Phys.},
  year		= {2019},
  volume	= {21},
  issue		= {30},
  pages		= {16658--16664},
  publisher	= {Royal Society of Chemistry},
  doi		= {10.1039/C9CP02382H}
}

@Article{	  Maximoff.Scuseria:Nuclear.2004,
  title		= "Nuclear magnetic resonance shielding tensors calculated
		  with kinetic energy density-dependent exchange-correlation
		  functionals",
  journal	= "Chem. Phys. Lett.",
  volume	= "390",
  number	= "4",
  pages		= "408--412",
  year		= "2004",
  issn		= "0009-2614",
  doi		= "10.1016/j.cplett.2004.04.049",
  url		= "https://doi.org/10.1016/j.cplett.2004.04.049",
  author	= "Sergey N. Maximoff and Gustavo E. Scuseria",
}

@Article{	  Autschbach:role.2013,
  author	= { Jochen Autschbach },
  title		= {The role of the exchange-correlation response kernel and
		  scaling corrections in relativistic density functional
		  nuclear magnetic shielding calculations with the
		  zeroth-order regular approximation},
  journal	= {Mol. Phys.},
  volume	= {111},
  number	= {16--17},
  pages		= {2544-2554},
  year		= {2013},
  publisher	= {Taylor & Francis},
  doi		= {10.1080/00268976.2013.796415}
}

@Article{	  Dyall:Relativistic.2011,
  author	= "Dyall, Kenneth G.",
  year		= "2011",
  title		= "Relativistic double-zeta, triple-zeta, and quadruple-zeta
		  basis sets for the 6d elements Rf--Cn",
  journal	= "Theor. Chem. Acc.",
  pages		= "603--613",
  volume	= "129",
  issue		= "3",
  url		= "https://doi.org/10.1007/s00214-011-0906-z",
  doi		= "10.1007/s00214-011-0906-z"
}

@article{Dyall:Relativistic.2007A,
  title={Relativistic double-zeta, triple-zeta, and quadruple-zeta basis sets for the actinides Ac--Lr},
  author={Dyall, Kenneth G},
  journal	= {Theor. Chem. Acc.},
  volume={117},
  number={4},
  pages={491--500},
  year={2007},
  doi = {10.1007/s00214-006-0175-4},
  url= {https://doi.org/10.1007/s00214-006-0175-4}
}

@Article{	  Dyall:Relativistic.2004,
  author	= "Dyall, Kenneth G.",
  title		= "Relativistic double-zeta, triple-zeta, and quadruple-zeta
		  basis sets for the 5d elements Hf--Hg",
  journal	= "Theor. Chem. Acc.",
  year		= "2004",
  volume	= "112",
  number	= "5",
  pages		= "403--409",
  doi		= "10.1007/s00214-004-0607-y"
}

@Article{	  Gomes.Dyall.ea:Relativistic.2010,
  author	= "Gomes, A. S. P. and Dyall, K. G. and Visscher, L.",
  year		= "2010",
  title		= "Relativistic double-zeta, triple-zeta, and quadruple-zeta
		  basis sets for the lanthanides La--Lu",
  journal	= "Theor. Chem. Acc.",
  pages		= "369--381",
  volume	= "127",
  issue		= "4",
  url		= " https://doi.org/10.1007/s00214-009-0725-7",
  doi		= "10.1007/s00214-009-0725-7"
}

@article{Dyall:Relativistic.2012,
  title={Relativistic double-zeta, triple-zeta, and quadruple-zeta basis sets for the 7p elements, with atomic and molecular applications},
  author={Dyall, Kenneth G},
  journal={Theor. Chem. Acc.},
  volume={131},
  number={3},
  pages={1172},
  year={2012},
  doi={10.1007/s00214-012-1172-4},
  url={https://doi.org/10.1007/s00214-012-1172-4}
}

@Article{	  Tao.Perdew.ea:Climbing.2003,
  title		= {Climbing the Density Functional Ladder: Nonempirical
		  Meta--Generalized Gradient Approximation Designed for
		  Molecules and Solids},
  author	= {Tao, Jianmin and Perdew, John P. and Staroverov, Viktor N.
		  and Scuseria, Gustavo E.},
  journal	= {Phys. Rev. Lett.},
  volume	= {91},
  issue		= {14},
  pages		= {146401},
  numpages	= {4},
  year		= {2003},
  doi		= {10.1103/PhysRevLett.91.146401}
}

@Article{	  Staroverov.Scuseria.ea:Comparative.2003,
  author	= {Staroverov,Viktor N. and Scuseria,Gustavo E. and
		  Tao,Jianmin and Perdew,John P. },
  title		= {Comparative assessment of a new nonempirical density
		  functional: Molecules and hydrogen-bonded complexes},
  journal	= {J. Chem. Phys.},
  volume	= {119},
  number	= {23},
  pages		= {12129--12137},
  year		= {2003},
  doi		= {10.1063/1.1626543}
}

@Article{	  Dyall:Relativistic.2006,
  author	= "Dyall, Kenneth G.",
  title		= "Relativistic Quadruple-Zeta and Revised Triple-Zeta and
		  Double-Zeta Basis Sets for the 4p, 5p, and 6p Elements",
  journal	= "Theor. Chem. Acc.",
  year		= "2006",
  volume	= "115",
  number	= "5",
  pages		= "441--447",
  issn		= "1432-2234",
  doi		= "10.1007/s00214-006-0126-0"
}

@misc{       Dyall-Repo,
  author	= {Kenneth G Dyall},
  title		= {{Dyall dz, tz, and qz basis sets for relativistic
		  electronic structure calculations}},
  month		= jan,
  year		= 2023,
  publisher	= {Zenodo},
  version	= 1,
  doi		= {10.5281/zenodo.7574629},
  url		= {https://doi.org/10.5281/zenodo.7574629},
  Note    = {{DOI}: 10.5281/zenodo.7574629}
}

@Misc{		  TURBOMOLE-manual,
  author = {{TURBOMOLE GmbH}},
  year = {2026},
  note		= {Manual of {TURBOMOLE V8.0.1}, a
		  development of {University of Karlsruhe} and
		  {Forschungszentrum Karlsruhe GmbH}, 1989-2007, {TURBOMOLE
		  GmbH}, since 2007; available from
		  \url{https://www.turbomole.org/turbomole/turbomole-documentation/}
		  (retrieved November 6, 2026).}
}

@Misc{		  TURBOMOLE,
  author = {{TURBOMOLE GmbH}},
  year = {2026},
  note		= {Developers' version of {TURBOMOLE V8.1}, a
		  development of {University of Karlsruhe} and
		  {Forschungszentrum Karlsruhe GmbH}, 1989-2007, {TURBOMOLE
		  GmbH}, since 2007; available from
		  \url{https://www.turbomole.org} (retrieved June 23, 2026).}
}

@Article{	  Dunning:Gaussian.1989,
  author	= {Dunning,Thom H. },
  title		= {Gaussian basis sets for use in correlated molecular
		  calculations. {I}. {T}he atoms boron through neon and
		  hydrogen},
  journal	= {J. Chem. Phys.},
  volume	= {90},
  number	= {2},
  pages		= {1007--1023},
  year		= {1989},
  doi		= {10.1063/1.456153}
}

@Article{	  Woon.Dunning:Gaussian.1993,
  author	= {Woon,David E. and Dunning,Thom H. },
  title		= {Gaussian basis sets for use in correlated molecular
		  calculations. {III.} {T}he atoms aluminum through argon},
  journal	= {J. Chem. Phys.},
  volume	= {98},
  number	= {2},
  pages		= {1358--1371},
  year		= {1993},
  doi		= {10.1063/1.464303}
}

@Article{	  Mitzinger.Bandemehr.ea:Ge2P22-.2018,
  author	= "Mitzinger, Stefan and Bandemehr, Jascha and Reiter, Kevin
		  and Scott McIndoe, J. and Xie, Xiulan and Weigend, Florian
		  and Corrigan, John F. and Dehnen, Stefanie",
  title		= "(Ge2P2)2-: a binary analogue of P4 as a precursor to the
		  ternary cluster anion [Cd3(Ge3P)3]3-",
  journal	= "Chem. Commun.",
  year		= "2018",
  volume	= "54",
  issue		= "12",
  pages		= "1421--1424",
  publisher	= "Royal Society of Chemistry",
  doi		= "10.1039/C7CC08348C"
}

@Article{	  Reiter.Mack.ea:Calculation.2018,
  author	= {Reiter, Kevin and Mack, Fabian and Weigend, Florian},
  title		= {Calculation of Magnetic Shielding Constants with meta-{GGA}
		  Functionals Employing the Multipole-Accelerated Resolution
		  of the Identity: Implementation and Assessment of Accuracy
		  and Efficiency},
  journal	= {J. Chem.Theory Comput.},
  volume	= {14},
  number	= {1},
  pages		= {191--197},
  year		= {2018},
  doi		= {10.1021/acs.jctc.7b01115}
}

@Article{	  Ditchfield:Self-consistent.1974,
  author	= { Robert Ditchfield },
  title		= {Self-consistent perturbation theory of diamagnetism},
  journal	= {Mol. Phys.},
  volume	= {27},
  number	= {4},
  pages		= {789--807},
  year		= {1974},
  publisher	= {Taylor & Francis},
  doi		= {10.1080/00268977400100711}
}

@Article{	  Olejniczak.Bast.ea:simple.2012,
  author	= {Olejniczak,Małgorzata and Bast,Radovan and Saue,Trond and
		  Pecul,Magdalena },
  title		= {A simple scheme for magnetic balance in four-component
		  relativistic Kohn-Sham calculations of nuclear magnetic
		  resonance shielding constants in a Gaussian basis},
  journal	= {J. Chem. Phys.},
  volume	= {136},
  number	= {1},
  pages		= {014108},
  year		= {2012},
  doi		= {10.1063/1.3671390}
}

@Article{	  Komorovsky.Repisky.ea:Fully.2010,
  author	= {Komorovsk\'y,Stanislav and Repisk\'y,Michal and
		  Malkina,Olga L. and Malkin,Vladimir G. },
  title		= {Fully relativistic calculations of {NMR} shielding tensors
		  using restricted magnetically balanced basis and gauge
		  including atomic orbitals},
  journal	= {J. Chem. Phys.},
  volume	= {132},
  number	= {15},
  pages		= {154101},
  year		= {2010},
  doi		= {10.1063/1.3359849}
}

@Article{	  Malkin.Malkina.ea:Calculations.1993,
  title		= "Calculations of {NMR} shielding constants by uncoupled
		  density functional theory",
  journal	= "Chem. Phys. Lett.",
  volume	= "204",
  number	= "1",
  pages		= "80--86",
  year		= "1993",
  issn		= "0009-2614",
  doi		= "https://doi.org/10.1016/0009-2614(93)85608-Q",
  url		= "http://www.sciencedirect.com/science/article/pii/000926149385608Q",
  author	= "V.G. Malkin and O.L. Malkina and D.R. Salahub"
}

@Article{	  Peng.Middendorf.ea:efficient.2013,
  author	= {Daoling Peng and Nils Middendorf and Florian Weigend and
		  Markus Reiher},
  title		= {An efficient implementation of two-component relativistic
		  exact-decoupling methods for large molecules},
  journal	= {J. Chem. Phys.},
  volume	= {138},
  number	= {18},
  pages		= {184105},
  year		= {2013},
  doi		= {10.1063/1.4803693}
}

@Article{	  Franzke.Middendorf.ea:Efficient.2018,
  author	= {Yannick J. Franzke and Nils Middendorf and Florian
		  Weigend},
  title		= {Efficient implementation of one- and two-component
		  analytical energy gradients in exact two-component theory},
  journal	= {J. Chem. Phys.},
  volume	= {148},
  number	= {10},
  pages		= {104410},
  year		= {2018},
  doi		= {10.1063/1.5022153}
}

@Article{	  Dyall:exact.1994,
  author	= {Kenneth G. Dyall},
  title		= {An exact separation of the spin‐free and
		  spin‐dependent terms of the {D}irac--{C}oulomb--Breit
		  Hamiltonian},
  journal	= {J. Chem. Phys.},
  volume	= {100},
  number	= {3},
  pages		= {2118--2127},
  year		= {1994},
  doi		= {10.1063/1.466508}
}

@Article{	  Ilias.Saue:infinite-order.2007,
  author	= {Miroslav Ilia\v{s} and Trond Saue},
  title		= {An infinite-order two-component relativistic {H}amiltonian
		  by a simple one-step transformation},
  journal	= {J. Chem. Phys.},
  volume	= {126},
  number	= {6},
  pages		= {064102},
  year		= {2007},
  doi		= {10.1063/1.2436882}
}

@Article{	  Kutzelnigg.Liu:Quasirelativistic.2005,
  author	= {Werner Kutzelnigg and Wenjian Liu},
  title		= {Quasirelativistic theory equivalent to fully relativistic
		  theory},
  journal	= {J. Chem. Phys.},
  volume	= {123},
  number	= {24},
  pages		= {241102},
  year		= {2005},
  doi		= {10.1063/1.2137315}
}

@Article{	  Visscher.Dyall:Dirac-Fock.1997,
  title		= "Dirac-{F}ock atomic electronic structure calculations using
		  different nuclear charge distributions",
  journal	= "At. Data Nucl. Data Tables",
  volume	= "67",
  number	= "2",
  pages		= "207--224",
  year		= "1997",
  issn		= "0092-640X",
  doi		= "10.1006/adnd.1997.0751",
  author	= "L. Visscher and K. G. Dyall"
}

@Article{	  Armbruster.Weigend.ea:Self-consistent.2008,
  author	= "Armbruster, Markus K. and Weigend, Florian and van
		  W{\"u}llen, Christoph and Klopper, Wim",
  title		= "Self-consistent treatment of spin-orbit interactions with
		  efficient {H}artree-{F}ock and density functional methods",
  journal	= "Phys. Chem. Chem. Phys.",
  year		= "2008",
  volume	= "10",
  issue		= "13",
  pages		= "1748--1756",
  publisher	= "Royal Society of Chemistry",
  doi		= "10.1039/B717719D",
}

@Article{	  Sierka.Hogekamp.ea:Fast.2003,
  author	= {Sierka,Marek and Hogekamp,Annika and Ahlrichs,Reinhart },
  title		= {Fast evaluation of the Coulomb potential for electron
		  densities using multipole accelerated resolution of
		  identity approximation},
  journal	= {J. Chem. Phys.},
  volume	= {118},
  number	= {20},
  pages		= {9136--9148},
  year		= {2003},
  doi		= {10.1063/1.1567253}
}

@Article{	  Treutler.Ahlrichs:Efficient.1995,
  author	= {Oliver Treutler and Reinhart Ahlrichs},
  title		= {Efficient molecular numerical integration schemes},
  journal	= {J. Chem. Phys.},
  volume	= {102},
  number	= {1},
  pages		= {346--354},
  year		= {1995},
  doi		= {10.1063/1.469408}
}

@Article{	  Pollak.Weigend:Segmented.2017,
  author	= {Pollak, Patrik and Weigend, Florian},
  title		= {Segmented Contracted Error-Consistent Basis Sets of
		  Double- and Triple-{\textzeta} Valence Quality for One- and
		  Two-Component Relativistic All-Electron Calculations},
  journal	= {J. Chem. Theory Comput.},
  volume	= {13},
  number	= {8},
  pages		= {3696--3705},
  year		= {2017},
  doi		= {10.1021/acs.jctc.7b00593}
}

@Article{	  Schafer.Klamt.ea:COSMO.2000,
  author	= "Sch{\"a}fer, Ansgar and Klamt, Andreas and Sattel, Diana
		  and Lohrenz, John C. W. and Eckert, Frank",
  title		= "{COSMO} Implementation in {TURBOMOLE}: {E}xtension of an
		  efficient quantum chemical code towards liquid systems",
  journal	= "Phys. Chem. Chem. Phys.",
  year		= "2000",
  volume	= "2",
  issue		= "10",
  pages		= "2187--2193",
  publisher	= "Royal Society of Chemistry",
  doi		= "10.1039/B000184H",
}

@Article{	  Cheeseman.Trucks.ea:comparison.1996,
  author	= {Cheeseman, J. R. and Trucks, G. W. and Keith, T. A. and
		  Frisch, M. J.},
  title		= {A comparison of models for calculating nuclear magnetic
		  resonance shielding tensors},
  journal	= {J. Chem. Phys.},
  volume	= {104},
  number	= {14},
  pages		= {5497--5509},
  year		= {1996},
  doi		= {10.1063/1.471789},
  url		= { https://doi.org/10.1063/1.471789}
}

@Article{	  Furness.Verbeke.ea:Current.2015,
  author	= {Furness, J. W. and Verbeke, J. and Tellgren, E. I. and
		  Stopkowicz, S. and Ekstr\"om, U. and Helgaker, T. and
		  Teale, A. M.},
  title		= {Current density functional theory using meta-generalized
		  gradient exchange-correlation functionals},
  journal	= {J. Chem. Theory Comput.},
  volume	= {11},
  number	= {9},
  pages		= {4169--4181},
  year		= {2015},
  doi		= {10.1021/acs.jctc.5b00535},
  url		= { https://doi.org/10.1021/acs.jctc.5b00535},
}

@Article{	  Helgaker.Jaszunski.ea:Ab.1999,
  author	= {Helgaker, T. and Jaszu\'nski, M. and Ruud, K.},
  title		= {Ab initio methods for the calculation of {NMR} shielding and
		  indirect spin-spin coupling constants},
  journal	= {Chem Rev.},
  volume	= {99},
  number	= {1},
  pages		= {293--352},
  url   = {https://doi.org/10.1021/cr960017t},
  doi  = {10.1021/cr960017t},
  year		= 1999
}

@Article{	  Tellgren.Kvaal.ea:Choice.2012,
  author	= {Tellgren, E. I. and Kvaal, S. and Sagvolden, E. and
		  Ekstr\"om, U. and Teale, A. M. and Helgaker, T.},
  title		= {Choice of basic variables in current-density-functional
		  theory},
  journal	= {Phys. Rev. A},
  volume	= {86},
  pages		= {062506},
  doi		= {10.1103/PhysRevA.86.062506},
  issue		= {6},
  numpages	= {13},
  publisher	= {American Physical Society},
  url		= {http://link.aps.org/doi/10.1103/PhysRevA.86.062506},
  year		= 2012
}

@Article{	  Tellgren.Teale.ea:Non-perturbative.2014,
  author	= {Tellgren, E. I. and Teale, A. M. and Furness, J. W. and
		  Lange, K. K. and Ekstr\"om, U. and Helgaker, T.},
  title		= {Non-perturbative calculation of molecular magnetic
		  properties within current-density functional theory},
  journal	= {J. Chem. Phys.},
  volume	= {140},
  pages		= {034101},
  year		= {2014},
  doi		= {10.1063/1.4861427},
  url		= { https://doi.org/10.1063/1.4861427},
}

@Article{	  Reimann.Ekstrom.ea:importance.2015,
  author	= {Reimann, S. and Ekstr\"om, U. and Stopkowicz, S. and
		  Teale, A. M. and Borgoo, A. and Helgaker, T.},
  title		= {The importance of current contributions to shielding
		  constants in density-functional theory},
  journal	= {Phys. Chem. Chem. Phys.},
  volume	= {17},
  pages		= {18834--18842},
  year		= {2015},
  doi		= {10.1039/C5CP02682B},
  url		= {http://dx.doi.org/10.1039/C5CP02682B},
}

@Article{	  Vignale.Rasolt:Current-.1988,
  author	= {Vignale, G. and Rasolt, M.},
  title		= {Current- and spin-density functional theory for
		  inhomogenous electronic systems in strong magnetic fields},
  journal	= {Phys. Rev. B},
  volume	= {37},
  number	= {18},
  pages		= {10685--10696},
  doi = {10.1103/PhysRevB.37.10685},
  url = {https://link.aps.org/doi/10.1103/PhysRevB.37.10685},
  year		= 1988
}

@Article{	  Tellgren.Kvaal.ea:Fermion.2014,
  author	= {Tellgren, E. I. and Kvaal, S. and Helgaker, T.},
  title		= {Fermion $N$-representability for prescribed density and
		  paramagnetic current density},
  journal	= {Phys. Rev. A},
  volume	= {89},
  number	= {1},
  pages		= {012515},
  doi       = {10.1103/PhysRevA.89.012515},
  url       = {https://link.aps.org/doi/10.1103/PhysRevA.89.012515},
  year		= 2014
}

@Article{	  Vaara:Theory.2007,
  author	= {Vaara, J.},
  title		= {Theory and computation of nuclear magnetic resonance
		  parameters},
  journal	= {Phys. Chem. Chem. Phys.},
  volume	= {9},
  number	= {40},
  pages		= {5399--5148},
  url = {https://doi.org/10.1039/B706135H},
  doi = {10.1039/B706135H},
  year		= 2007
}

@Article{	  Lieb.Schrader:Current.2013,
  author	= {Lieb, E. H. and Schrader, R.},
  title		= {Current densities in density-functional theory},
  journal	= {Phys. Rev. A},
  volume	= {88},
  number	= {3},
  pages		= {032516},
  doi       = {10.1103/PhysRevA.88.032516},
  url       = {https://link.aps.org/doi/10.1103/PhysRevA.88.032516},
  year		= 2013
}

@Article{	  Franzke.Spiske.ea:Segmented.2020,
  author	= {Franzke, Yannick J. and Spiske, Lucas and Pollak, Patrik
		  and Weigend, Florian},
  title		= {Segmented Contracted Error-Consistent Basis Sets of
		  Quadruple-${\zeta}$ Valence Quality for One- and Two-Component
		  Relativistic All-Electron Calculations},
  journal	= {J. Chem. Theory Comput.},
  volume	= {16},
  number	= {9},
  pages		= {5658--5674},
  year		= {2020},
  doi		= {10.1021/acs.jctc.0c00546},
  url		= { https://doi.org/10.1021/acs.jctc.0c00546}
}

@Book{	       Kaupp.Buhl.ea:Calculation.2004,
  publisher	= {Wiley-VCH},
  address   = {Weinheim, Germany},
  title	    = {Calculation of {NMR} and {EPR} Parameters. Theory and Applications},
  editor    = {Kaupp, Martin and B\"uhl, Michael and Malkin, Vladimir G.},
  doi		= {10.1002/3527601678},
  url		= {https://onlinelibrary.wiley.com/doi/abs/10.1002/3527601678},
  year		= {2004}
}

@Article{	  Franzke.Mack.ea:NMR.2021,
  author	= {Franzke, Yannick J. and Mack, Fabian and Weigend, Florian },
  title		= {{NMR} Indirect Spin--Spin Coupling Constants in a Modern
                   Quasirelativistic Density Functional Framework},
  journal	= {J. Chem. Theory Comput.},
  volume	= {17},
  number	= {7},
  pages		= {3974--3994},
  year		= {2021},
  doi		= {10.1021/acs.jctc.1c00167},
  url		= {https://doi.org/10.1021/acs.jctc.1c00167},
}

@Article{	  Mack.Schattenberg.ea:Nuclear.2020,
	author	= {Mack, Fabian and Schattenberg, Caspar J. and Kaupp, Martin
		and Weigend, Florian},
	title		= {Nuclear Spin--Spin Couplings: Efficient Evaluation of
		Exact Exchange and Extension to Local Hybrid Functionals},
	journal	= {J. Phys. Chem. A},
	volume	= {124},
	number	= {41},
	pages		= {8529-8539},
	year		= {2020},
	doi		= {10.1021/acs.jpca.0c06897},
	url		= { https://doi.org/10.1021/acs.jpca.0c06897},
}

@Article{	  Holzer:improved.2020,
	author	= {Holzer,Christof },
	title		= {An improved seminumerical {C}oulomb and exchange algorithm
		for properties and excited states in modern density
			functional theory},
	journal	= {J. Chem. Phys.},
	volume	= {153},
	number	= {18},
	pages		= {184115},
	year		= {2020},
	doi		= {10.1063/5.0022755},
	url		= { https://doi.org/10.1063/5.0022755}
}

@Article{	  Holzer.Franzke.ea:Assessing.2021,
  author	= {Holzer, Christof and Franzke, Yannick J. and Kehry, Max},
  title		= {Assessing the Accuracy of Local Hybrid Density Functional
		  Approximations for Molecular Response Properties},
  journal	= {J. Chem. Theory Comput.},
  volume	= {17},
  number	= {5},
  pages		= {2928--2947},
  year		= {2021},
  doi		= {10.1021/acs.jctc.1c00203},
  url		= { https://doi.org/10.1021/acs.jctc.1c00203}
}

@Article{	  Mardirossian.Head-Gordon:Thirty.2017,
  author	= {Narbe Mardirossian and Martin Head-Gordon},
  title		= {Thirty years of density functional theory in computational
		  chemistry: an overview and extensive assessment of 200
		  density functionals},
  journal	= {Mol. Phys.},
  volume	= {115},
  number	= {19},
  pages		= {2315-2372},
  year		= {2017},
  doi		= {10.1080/00268976.2017.1333644},
  url		= { https://doi.org/10.1080/00268976.2017.1333644}
}

@Article{	  Furness.Kaplan.ea:Accurate.2020,
  author	= {Furness, James W. and Kaplan, Aaron D. and Ning, Jinliang
		  and Perdew, John P. and Sun, Jianwei},
  title		= {Accurate and Numerically Efficient {r$^2$SCAN} Meta-Generalized
		  Gradient Approximation},
  journal	= {J. Phys. Chem. Lett.},
  volume	= {11},
  number	= {19},
  pages		= {8208--8215},
  year		= {2020},
  doi		= {10.1021/acs.jpclett.0c02405},
  url		= { https://doi.org/10.1021/acs.jpclett.0c02405}
}

@article{Autschbach:Relativistic.2014,
author = {Autschbach, Jochen },
title = {Relativistic calculations of magnetic resonance parameters: background and some recent developments},
journal = {Philos. Trans. A Math. Phys. Eng. Sci.},
volume = {372},
number = {2011},
pages = {20120489},
year = {2014},
doi = {10.1098/rsta.2012.0489},
URL = {https://royalsocietypublishing.org/doi/abs/10.1098/rsta.2012.0489},
}

@Article{	  Jaramillo.Scuseria.ea:Local.2003,
  author	= {Jaramillo,Juanita and Scuseria,Gustavo E. and
		  Ernzerhof,Matthias },
  title		= {Local hybrid functionals},
  journal	= {J. Chem. Phys.},
  volume	= {118},
  number	= {3},
  pages		= {1068--1073},
  year		= {2003},
  doi		= {10.1063/1.1528936},
  url		= { https://doi.org/10.1063/1.1528936}
}

@Article{	  Bates.Furche:Harnessing.2012,
  title		= {Harnessing the meta-generalized gradient approximation for
		  time-dependent density functional theory},
  author	= {Bates, Jefferson E and Furche, Filipp},
  journal	= {J. Chem. Phys.},
  volume	= {137},
  number	= {16},
  pages		= {164105},
  year		= {2012},
  publisher	= {American Institute of Physics},
  doi		= {10.1063/1.4759080},
  url		= { https://doi.org/10.1063/1.4759080}
}

@Article{	  Bahmann.Kaupp:Efficient.2015,
  author	= {Bahmann, Hilke and Kaupp, Martin},
  journal	= {J. Chem. Theory Comput.},
  number	= {4},
  pages		= {1540--1548},
  title		= {Efficient Self-Consistent Implementation of Local Hybrid
		  Functionals},
  volume	= {11},
  year		= {2015},
  doi       = {0.1021/ct501137x},
  url       = {https://doi.org/10.1021/ct501137x}
}

@Article{	  Haasler.Maier.ea:Local.2020,
  author	= {Haasler, Matthias and Maier, Toni M. and Grotjahn, Robin
		  and G\"uckel, Simon and Arbuznikov, Alexei V. and Kaupp,
		  Martin},
  title		= {A Local Hybrid Functional with Wide Applicability and Good
		  Balance between (De)Localization and Left--Right
		  Correlation},
  journal	= {J. Chem. Theory Comput.},
  volume	= {16},
  number	= {9},
  pages		= {5645--5657},
  year		= {2020},
  doi		= {10.1021/acs.jctc.0c00498},
  url		= { https://doi.org/10.1021/acs.jctc.0c00498}
}

@Article{	  Schattenberg.Reiter.ea:Efficient.2020,
  author	= {Schattenberg, Caspar Jonas and Reiter, Kevin and Weigend,
		  Florian and Kaupp, Martin},
  title		= {An Efficient Coupled-Perturbed {Kohn--Sham} Implementation
		  of {NMR} Chemical Shift Computations with Local Hybrid
		  Functionals and Gauge-Including Atomic Orbitals},
  journal	= {J. Chem. Theory Comput.},
  volume	= {16},
  number	= {2},
  pages		= {931--943},
  year		= {2020},
  doi		= {10.1021/acs.jctc.9b00944},
  url		= { https://doi.org/10.1021/acs.jctc.9b00944}
}

@Article{	  Arbuznikov.Kaupp:Importance.2012,
  author	= {Arbuznikov,Alexei V. and Kaupp,Martin },
  title		= {Importance of the correlation contribution for local
		  hybrid functionals: Range separation and self-interaction
		  corrections},
  journal	= {J. Chem. Phys.},
  volume	= {136},
  number	= {1},
  pages		= {014111},
  year		= {2012},
  doi		= {10.1063/1.3672080},
  url		= { https://doi.org/10.1063/1.3672080}
}

@Article{	  Sun.Ruzsinszky.ea:Strongly.2015,
  title		= {Strongly Constrained and Appropriately Normed Semilocal
		  Density Functional},
  author	= {Sun, Jianwei and Ruzsinszky, Adrienn and Perdew, John P.},
  journal	= {Phys. Rev. Lett.},
  volume	= {115},
  issue		= {3},
  pages		= {036402},
  numpages	= {6},
  year		= {2015},
  publisher	= {American Physical Society},
  doi		= {10.1103/PhysRevLett.115.036402},
  url		= {https://link.aps.org/doi/10.1103/PhysRevLett.115.036402}
}

@Article{	  Schattenberg.Kaupp:Implementation.2021,
  author	= {Schattenberg, Caspar Jonas and Kaupp, Martin},
  title		= {Implementation and Validation of Local Hybrid Functionals
		  with Calibrated Exchange-Energy Densities for Nuclear
		  Shielding Constants},
  journal	= {J. Phys. Chem. A},
  volume	= {125},
  number	= {12},
  pages		= {2697--2707},
  year		= {2021},
  doi		= {10.1021/acs.jpca.1c01135},
  url		= { https://doi.org/10.1021/acs.jpca.1c01135}
}

@Article{	  Schattenberg.Kaupp:Effect.2021,
  author	= {Schattenberg, Caspar J. and Kaupp, Martin},
  title		= {Effect of the current dependence of tau-dependent
               exchange-correlation functionals on nuclear shielding
               calculations},
  journal	= {J. Chem. Theory Comput.},
  volume	= {17},
  number	= {3},
  pages		= {1469--1479},
  year		= {2021},
  doi		= {10.1021/acs.jctc.0c01223},
  url		= { https://doi.org/10.1021/acs.jctc.0c01223},
}

@Article{	  Aschebrock.Kummel:Ultranonlocality.2019,
  title		= {Ultranonlocality and accurate band gaps from a
		  meta-generalized gradient approximation},
  author	= {Aschebrock, Thilo and K\"ummel, Stephan},
  journal	= {Phys. Rev. Res.},
  volume	= {1},
  issue		= {3},
  pages		= {033082},
  numpages	= {14},
  year		= {2019},
  month		= {Nov},
  publisher	= {American Physical Society},
  doi		= {10.1103/PhysRevResearch.1.033082},
  url		= {https://link.aps.org/doi/10.1103/PhysRevResearch.1.033082}
}

@Article{	  Yu.He.ea:MN15.2016,
  author	= "Yu, Haoyu S. and He, Xiao and Li, Shaohong L. and Truhlar,
		  Donald G.",
  title		= "{MN15}: A {Kohn--Sham} global-hybrid exchange--correlation
		  density functional with broad accuracy for multi-reference
		  and single-reference systems and noncovalent interactions",
  journal	= "Chem. Sci.",
  year		= "2016",
  volume	= "7",
  issue		= "8",
  pages		= "5032--5051",
  publisher	= "Royal Society of Chemistry",
  doi		= "10.1039/C6SC00705H",
  url		= "http://dx.doi.org/10.1039/C6SC00705H",
}

@Article{	  Sundholm.Dimitrova.ea:Current.2021,
  author	= "Sundholm, Dage and Dimitrova, Maria and Berger, Raphael J.
		  F.",
  title		= "Current density and molecular magnetic properties",
  journal	= "Chem. Commun.",
  year		= "2021",
  volume	= "57",
  issue		= "93",
  pages		= "12362--12378",
  publisher	= "Royal Society of Chemistry",
  doi		= "10.1039/D1CC03350F",
  url		= "http://dx.doi.org/10.1039/D1CC03350F",
}

@Article{	  Pausch.Holzer:Linear.2022,
  author	= {Pausch, Ansgar and Holzer, Christof},
  title		= {Linear Response of Current-Dependent Density Functional
		  Approximations in Magnetic Fields},
  journal	= {J. Phys. Chem. Lett.},
  volume	= {13},
  number	= {19},
  pages		= {4335--4341},
  year		= {2022},
  doi		= {10.1021/acs.jpclett.2c01082},
  url		= { https://doi.org/10.1021/acs.jpclett.2c01082}
}

@Article{	  Zhao.Truhlar:M06.2008,
  title		= {The {M06} suite of density functionals for main group
		  thermochemistry, thermochemical kinetics, noncovalent
		  interactions, excited states, and transition elements: two
		  new functionals and systematic testing of four {M06}-class
		  functionals and 12 other functionals},
  author	= {Zhao, Yan and Truhlar, Donald G},
  journal	= {Theor. Chem. Acc.},
  volume	= {120},
  number	= {1},
  pages		= {215--241},
  year		= {2008},
  doi       = {10.1007/s00214-007-0310-x},
  url       = {https://doi.org/10.1007/s00214-007-0310-x },
}

@Article{	  Yu.He.ea:MN15-L.2016,
  author	= {Yu, Haoyu S. and He, Xiao and Truhlar, Donald G.},
  title		= {{MN15-L}: A New Local Exchange-Correlation Functional for
		  {Kohn--Sham} Density Functional Theory with Broad Accuracy
		  for Atoms, Molecules, and Solids},
  journal	= {J. Chem. Theory Comput.},
  volume	= {12},
  number	= {3},
  pages		= {1280--1293},
  year		= {2016},
  doi		= {10.1021/acs.jctc.5b01082},
  url		= { https://doi.org/10.1021/acs.jctc.5b01082}
}

@Article{	  Bruder.Franzke.ea:Paramagnetic.2022,
  author	= {Bruder, Florian and Franzke, Yannick J. and Weigend,
		  Florian},
  title		= {Paramagnetic {NMR} Shielding Tensors Based on Scalar Exact
		  Two-Component and Spin--Orbit Perturbation Theory},
  doi		= {10.1021/acs.jpca.2c03579},
  url       = { https://doi.org/10.1021/acs.jpca.2c03579},
  journal	= {J. Phys. Chem. A},
  volume    = {126},
  number    = {30},
  pages     = {5050--5069},
  year		= {2022},
}

@Article{	  Holzer.Franzke:Local.2022,
  author	= {Holzer, Christof and Franzke, Yannick J.},
  title		= {A Local Hybrid Exchange Functional Approximation from
		  First Principles},
  journal	= {J. Chem. Phys.},
  volume	= {157},
  number	= {3},
  pages		= {034108},
  year		= {2022},
  doi		= {10.1063/5.0100439},
  url		= { https://aip.scitation.org/doi/10.1063/5.0100439},
}

@Article{	  Dobson:Alternative.1993,
  author	= {Dobson,John F. },
  title		= {Alternative expressions for the Fermi hole curvature},
  journal	= {J. Chem. Phys.},
  volume	= {98},
  number	= {11},
  pages		= {8870-8872},
  year		= {1993},
  doi		= {10.1063/1.464444},
  url		= { https://doi.org/10.1063/1.464444}
}

@Article{	  Becke:Current.2002,
  author	= {Becke,Axel D. },
  title		= {Current density in exchange-correlation functionals:
		  {A}pplication to atomic states},
  journal	= {J. Chem. Phys.},
  volume	= {117},
  number	= {15},
  pages		= {6935-6938},
  year		= {2002},
  doi		= {10.1063/1.1503772},
  url		= { https://doi.org/10.1063/1.1503772}
}

@article{Tao:Explicit.2005,
  title = {Explicit inclusion of paramagnetic current density in the
            exchange-correlation functionals of current-density functional theory},
  author = {Tao, Jianmin},
  journal = {Phys. Rev. B},
  volume = {71},
  issue = {20},
  pages = {205107},
  year = {2005},
  publisher = {American Physical Society},
  doi = {10.1103/PhysRevB.71.205107},
  url = {https://link.aps.org/doi/10.1103/PhysRevB.71.205107}
}

@Article{	  Sen.Tellgren:Benchmarking.2021,
  author	= {Sen, Sangita and Tellgren, Erik I.},
  title		= {Benchmarking Density Functional Approximations for
		  Diamagnetic and Paramagnetic Molecules in Nonuniform
		  Magnetic Fields},
  journal	= {J. Chem. Theory Comput.},
  volume	= {17},
  number	= {3},
  pages		= {1480-1496},
  year		= {2021},
  doi		= {10.1021/acs.jctc.0c01222},
  url		= { https://doi.org/10.1021/acs.jctc.0c01222}
}

@Article{	  Irons.David.ea:Optimizing.2021,
  author	= {Irons, Tom J. P. and David, Gr\'egoire and Teale, Andrew
		  M.},
  title		= {Optimizing Molecular Geometries in Strong Magnetic
		  Fields},
  journal	= {J. Chem. Theory Comput.},
  volume	= {17},
  number	= {4},
  pages		= {2166-2185},
  year		= {2021},
  doi		= {10.1021/acs.jctc.0c01297},
  url		= { https://doi.org/10.1021/acs.jctc.0c01297}
}

@Article{	  Irons.Spence.ea:Analyzing.2020,
  author	= {Irons, Tom J. P. and Spence, Lucy and David, Gr\'egoire and
		  Speake, Benjamin T. and Helgaker, Trygve and Teale, Andrew
		  M.},
  title		= {Analyzing Magnetically Induced Currents in Molecular
		  Systems Using Current-Density-Functional Theory},
  journal	= {J. Phys. Chem. A},
  volume	= {124},
  number	= {7},
  pages		= {1321-1333},
  year		= {2020},
  doi		= {10.1021/acs.jpca.9b10833},
  url		= { https://doi.org/10.1021/acs.jpca.9b10833}
}

@Article{	  Liang.Feng.ea:Revisiting.2022,
  author	= {Liang, Jiashu and Feng, Xintian and Hait, Diptarka and
		  Head-Gordon, Martin},
  title		= {Revisiting the Performance of Time-Dependent Density
		  Functional Theory for Electronic Excitations: Assessment of
		  43 Popular and Recently Developed Functionals from Rungs
		  One to Four},
  journal	= {J. Chem. Theory Comput.},
  volume	= {18},
  number	= {6},
  pages		= {3460-3473},
  year		= {2022},
  doi		= {10.1021/acs.jctc.2c00160},
  url		= { https://doi.org/10.1021/acs.jctc.2c00160}
}

@Article{	  Franzke.Holzer:Impact.2022,
  author	= {Franzke,Yannick J. and Holzer,Christof },
  title		= {Impact of the current density on
		  paramagnetic {NMR} properties},
  journal	= {J. Chem. Phys.},
  volume	= {157},
  number	= {3},
  pages		= {031102},
  year		= {2022},
  doi		= {10.1063/5.0103898},
  url		= { https://doi.org/10.1063/5.0103898},
}

@InCollection{	  InC-Repisky.Komorovsky.ea:2016,
  author	= "Repisky, Michal and Komorovsky, Stanislav and Bast,
		  Radovan and Ruud, Kenneth",
  title		= "Relativistic Calculations of Nuclear Magnetic
		  Resonance Parameters",
  booktitle	= "Gas Phase NMR",
  year		= "2016",
  pages		= "267--303",
  chapter   = "8",
  publisher	= "Royal Society of Chemistry",
  isbn		= "978-1-78262-161-4",
  doi		= "10.1039/9781782623816-00267",
  url		= "http://dx.doi.org/10.1039/9781782623816-00267",
  editor    = "Karol Jackowski and Micha\l{} Jaszu\'{n}ski",
  address   = "Cambridge, United Kingdom"
}

@Article{	  Wang.Ziegler.ea:calculation.2005,
  title		= {The calculation of excitation energies based on the
		  relativistic two-component zeroth-order regular
		  approximation and time-dependent density-functional with
		  full use of symmetry},
  author	= {Wang, Fan and Ziegler, Tom and van Lenthe, Erik and van
		  Gisbergen, Stan and Baerends, Evert Jan},
  journal	= {J. Chem. Phys.},
  volume	= {122},
  number	= {20},
  pages		= {204103},
  year		= {2005},
  publisher	= {AIP},
  url		= { https://doi.org/10.1063/1.1899143},
  doi		= {10.1063/1.1899143}
}

@Article{	  Gao.Zou.ea:Time-dependent.2005,
  title		= {Time-dependent four-component relativistic
		  density-functional theory for excitation energies. II. The
		  exchange-correlation kernel},
  author	= {Gao, Jun and Zou, Wenli and Liu, Wenjian and Xiao, Yunlong
		  and Peng, Daoling and Song, Bo and Liu, Chengbu},
  journal	= {J. Chem. Phys.},
  volume	= {123},
  number	= {5},
  pages		= {054102},
  year		= {2005},
  publisher	= {AIP},
  url		= {https://doi.org/10.1063/1.1940609},
  doi		= {10.1063/1.1940609}
}

@Article{	  Scalmani.Frisch:New.2012,
  author	= {Scalmani, Giovanni and Frisch, Michael J.},
  title		= {A New Approach to Noncollinear Spin Density Functional
		  Theory beyond the Local Density Approximation},
  journal	= {J. Chem. Theory Comput.},
  volume	= {8},
  number	= {7},
  pages		= {2193--2196},
  year		= {2012},
  doi		= {10.1021/ct300441z},
  url		= { https://doi.org/10.1021/ct300441z}
}

@Article{	  Komorovsky.Cherry.ea:Four-component.2019,
  author	= {Komorovsky,Stanislav and Cherry,Peter J. and
		  Repisky,Michal },
  title		= {Four-component relativistic time-dependent
		  density-functional theory using a stable noncollinear DFT
		  ansatz applicable to both closed- and open-shell systems},
  journal	= {J. Chem. Phys.},
  volume	= {151},
  number	= {18},
  pages		= {184111},
  year		= {2019},
  doi		= {10.1063/1.5121713},
  url		= { https://doi.org/10.1063/1.5121713}
}

@Article{	  Desmarais.Komorovsky.ea:Spin-orbit.2021,
  author	= {Desmarais,Jacques K. and Komorovsky,Stanislav and
		  Flament,Jean-Pierre and Erba,Alessandro },
  title		= {Spin--orbit coupling from a two-component self-consistent
		  approach. II. Non-collinear density functional theories},
  journal	= {J. Chem. Phys.},
  volume	= {154},
  number	= {20},
  pages		= {204110},
  year		= {2021},
  doi		= {10.1063/5.0051447},
  url		= { https://doi.org/10.1063/5.0051447}
}

@Article{	  Egidi.Sun.ea:Two-Component.2017,
  author	= {Egidi, Franco and Sun, Shichao and Goings, Joshua J. and
		  Scalmani, Giovanni and Frisch, Michael J. and Li,
		  Xiaosong},
  title		= {Two-Component Noncollinear Time-Dependent Spin Density
		  Functional Theory for Excited State Calculations},
  journal	= {J. Chem. Theory Comput.},
  volume	= {13},
  number	= {6},
  pages		= {2591--2603},
  year		= {2017},
  doi		= {10.1021/acs.jctc.7b00104},
  url		= { https://doi.org/10.1021/acs.jctc.7b00104}
}

@Article{	  Holzer.Franzke.ea:Current.2022,
  author	= {Holzer,Christof and Franzke,Yannick J. and Pausch,Ansgar
		  },
  title		= {Current density functional framework for spin--orbit
		  coupling},
  journal	= {J. Chem. Phys.},
  volume	= {157},
  number	= {20},
  pages		= {204102},
  year		= {2022},
  doi		= {10.1063/5.0122394},
  url		= { https://doi.org/10.1063/5.0122394}
}

@Article{	  Schattenberg.Kaupp:Extended.2021,
  author	= {Schattenberg, Caspar Jonas and Kaupp, Martin},
  title		= {Extended Benchmark Set of Main-Group Nuclear Shielding
		  Constants and {NMR} Chemical Shifts and Its Use to Evaluate
		  Modern DFT Methods},
  journal	= {J. Chem. Theory Comput.},
  volume	= {17},
  number	= {12},
  pages		= {7602-7621},
  year		= {2021},
  doi		= {10.1021/acs.jctc.1c00919},
  url		= { https://doi.org/10.1021/acs.jctc.1c00919}
}

@Article{	  Schattenberg.Lehmann.ea:Systematic.2022,
  author	= {Schattenberg, Caspar Jonas and Lehmann, Morten and B\"uhl,
		  Michael and Kaupp, Martin},
  title		= {Systematic Evaluation of Modern Density Functional Methods
		  for the Computation of {NMR} Shifts of 3d Transition-Metal
		  Nuclei},
  journal	= {J. Chem. Theory Comput.},
  volume	= {18},
  number	= {1},
  pages		= {273-292},
  year		= {2022},
  doi		= {10.1021/acs.jctc.1c00964},
  url		= { https://doi.org/10.1021/acs.jctc.1c00964}
}

@Article{	  Bates.Heiche.ea:Erratum.2022,
  author	= {Jefferson E. Bates and Maximillian C. Heiche and Jiashu
		  Liang and Filipp Furche},
  journal	= {J. Chem. Phys.},
  number	= {15},
  pages		= {159902},
  title		= {Erratum: {\textquotedblleft}Harnessing the
		  meta-generalized gradient approximation for time-dependent
		  density functional theory{\textquotedblright} [J. Chem.
		  Phys. 137, 164105 (2012)]},
  volume	= {156},
  year		= {2022}
}

@Article{	  Grotjahn.Furche.ea:Importance.2022,
  author	= {Grotjahn,Robin and Furche,Filipp and Kaupp,Martin},
  journal	= {J. Chem. Phys.},
  number	= {11},
  pages		= {111102},
  title		= {Importance of imposing gauge invariance in time-dependent
		  density functional theory calculations with
		  meta-generalized gradient approximations},
  volume	= {157},
  year		= {2022},
  doi = {10.1063/5.0113083},
  url = {https://doi.org/10.1063/5.0113083}
}

@Article{	  Kuehn.Weigend:Implementation.2013,
  author	= {K{\"u}hn, Michael and Weigend, Florian},
  title		= {Implementation of Two-Component Time-Dependent Density
		  Functional Theory in TURBOMOLE},
  journal	= {J. Chem. Theory Comput.},
  volume	= {9},
  number	= {12},
  pages		= {5341-5348},
  year		= {2013},
  doi		= {10.1021/ct400743r},
  url		= { https://doi.org/10.1021/ct400743r}
}

@Article{	  Peralta.Scuseria.ea:Noncollinear.2007,
  title		= {Noncollinear magnetism in density functional
		  calculations},
  author	= {Peralta, Juan E. and Scuseria, Gustavo E. and Frisch,
		  Michael J.},
  journal	= {Phys. Rev. B},
  volume	= {75},
  issue		= {12},
  pages		= {125119},
  numpages	= {8},
  year		= {2007},
  publisher	= {American Physical Society},
  doi		= {10.1103/PhysRevB.75.125119},
  url		= {https://link.aps.org/doi/10.1103/PhysRevB.75.125119}
}

@article{     Franzke:Reducing.2023,
  title     = {Reducing Exact Two-Component Theory for {NMR} Couplings to a One-Component Approach:
               Efficiency and Accuracy},
  author    = {Franzke, Yannick J.},
  journal   = {J. Chem. Theory Comput.},
  volume    = {19},
  number    = {7},
  pages     = {2010--2028},
  year      = {2023},
  doi       = {10.1021/acs.jctc.2c01248},
  url       = {https://doi.org/10.1021/acs.jctc.2c01248},
}

@Article{	  Sherrill.Manolopoulos.ea:Electronic.2020,
  author	= {Sherrill,C. David and Manolopoulos,David E. and
		  Martínez,Todd J. and Michaelides,Angelos },
  title		= {Electronic structure software},
  journal	=  jcp,
  volume	= {153},
  number	= {7},
  pages		= {070401},
  year		= {2020},
  doi		= {10.1063/5.0023185},
  url		= { https://doi.org/10.1063/5.0023185}
}

@Article{	  Vicha.Novotny.ea:Relativistic.2020,
  author	= {V\'icha, Jan and Novotn\'y, Jan and Komorovsky, Stanislav
		  and Straka, Michal and Kaupp, Martin and Marek, Radek},
  title		= {Relativistic Heavy-Neighbor-Atom Effects on {NMR} Shifts:
		  Concepts and Trends Across the Periodic Table},
  journal	= {Chem. Rev.},
  volume	= {120},
  number	= {15},
  pages		= {7065--7103},
  year		= {2020},
  doi		= {10.1021/acs.chemrev.9b00785},
  url		= { https://doi.org/10.1021/acs.chemrev.9b00785}
}

@Article{	  Franzke.Holzer.ea:TURBOMOLE.2023,
  author	= {Franzke, Yannick J. and Holzer, Christof and Andersen,
		  Josefine H. and Begu\v{s}i\'c, Tomislav and Bruder, Florian and
		  Coriani, Sonia and Della Sala, Fabio and Fabiano, Eduardo
		  and Fedotov, Daniil A. and Fürst, Susanne and Gillhuber,
		  Sebastian and Grotjahn, Robin and Kaupp, Martin and Kehry,
		  Max and Krsti\'c, Marjan and Mack, Fabian and Majumdar,
		  Sourav and Nguyen, Brian D. and Parker, Shane M. and Pauly,
		  Fabian and Pausch, Ansgar and Perlt, Eva and Phun, Gabriel
		  S. and Rajabi, Ahmadreza and Rappoport, Dmitrij and Samal,
		  Bibek and Schrader, Tim and Sharma, Manas and Tapavicza,
		  Enrico and Treß, Robert S. and Voora, Vamsee and
		  Wody{\'n}ski, Artur and Yu, Jason M. and Zerulla, Benedikt and
		  Furche, Filipp and Hättig, Christof and Sierka, Marek and
		  Tew, David P. and Weigend, Florian},
  title		= {{TURBOMOLE}: {T}oday and {T}omorrow},
  journal	= {J. Chem. Theory Comput.},
  volume	= {19},
  number	= {20},
  pages		= {6859--6890},
  year		= {2023},
  doi		= {10.1021/acs.jctc.3c00347},
  url		= { https://doi.org/10.1021/acs.jctc.3c00347}
}

@Article{	  Caldeweyher.Ehlert.ea:Generally.2019,
  author	= {Caldeweyher, Eike and Ehlert, Sebastian and Hansen,
		  Andreas and Neugebauer, Hagen and Spicher, Sebastian and
		  Bannwarth, Christoph and Grimme, Stefan},
  title		= {A generally applicable atomic-charge dependent London
		  dispersion correction},
  journal	= {J. Chem. Phys.},
  volume	= {150},
  number	= {15},
  pages		= {154122},
  year		= {2019},
  month		= {04},
  doi		= {10.1063/1.5090222},
  url		= {https://doi.org/10.1063/1.5090222},
}

@Article{	  Jameson.De-Dios.ea:Absolute.1990,
  title		= {Absolute shielding scale for $^{31}$P from gas-phase NMR
		  studies},
  journal	= {Chem. Phys. Lett.},
  volume	= {167},
  number	= {6},
  pages		= {575-582},
  year		= {1990},
  issn		= {0009-2614},
  doi		= {https://doi.org/10.1016/0009-2614(90)85472-O},
  url		= {https://www.sciencedirect.com/science/article/pii/000926149085472O},
  author	= {Cynthia J. Jameson and Angel {De Dios} and A. K.
		  Jameson},
}

@Article{	  Franzke.Holzer:Exact.2023,
  author	= {Franzke, Yannick J. and Holzer, Christof},
  title		= {Exact two-component theory becoming an efficient tool for
		  {NMR} shieldings and shifts with spin--orbit coupling},
  journal   = {J. Chem. Phys.},
  volume	= {159},
  number	= {18},
  pages		= {184102},
  year		= {2023},
  issn		= {0021-9606},
  doi		= {10.1063/5.0171509},
  url		= {https://doi.org/10.1063/5.0171509},
}

@article{     Bruder.Franzke.ea:Zero-Field.2023,
  title     = {Zero-Field Splitting Parameters within Exact Two-Component
               Theory and Modern Density Functional Theory Using
               Seminumerical Integration},
  journal   = {J. Chem. Phys.},
  author    = {Bruder, Florian and Franzke, Yannick J. and Holzer, Christof and Weigend, Florian},
  volume    = {159},
  number    = {19},
  pages     = {194117},
  year      = {2023},
  doi       = {10.1063/5.0175758},
  url       = {https://doi.org/10.1063/5.0175758},
}

@Article{	  Franzke.Bruder.ea:Paramagnetic.2024,
  title		= {Paramagnetic Nuclear Magnetic Resonance Shifts for Triplet Systems and Beyond
		  with Modern Relativistic Density Functional Methods},
  doi		= {10.1021/acs.jpca.3c07093},
  url       = {https://doi.org/10.1021/acs.jpca.3c07093},
  journal	= {J. Phys. Chem. A},
  author	= {Franzke, Yannick J. and Bruder, Florian and Gillhuber,
		  Sebastian and Holzer, Christof and Weigend, Florian},
  volume    = {128},
  number    = {3},
  pages     = {670--686},
  year      = {2024},
}

@Article{	  Grotjahn.Furche:Gauge-Invariant.2023,
  author	= {Grotjahn, Robin and Furche, Filipp},
  title		= {Gauge-Invariant Excited-State Linear and Quadratic
		  Response Properties within the Meta-Generalized Gradient
		  Approximation},
  journal	= {J. Chem. Theory Comput.},
  volume	= {19},
  number	= {15},
  pages		= {4897-4911},
  year		= {2023},
  doi		= {10.1021/acs.jctc.3c00259},
  url		= { https://doi.org/10.1021/acs.jctc.3c00259},
}

@Article{	  Franzke.Holzer:Current.2024,
  author	= {Franzke, Yannick J. and Holzer, Christof},
  title		= {Current density functional framework for spin--orbit
		  coupling: Extension to periodic systems},
  journal	= {J. Chem. Phys.},
  volume	= {160},
  number	= {18},
  pages		= {184101},
  year		= {2024},
  issn		= {0021-9606},
  doi		= {10.1063/5.0209704},
  url		= {https://doi.org/10.1063/5.0209704}
}

@Article{	  Brack.Jennings.ea:On.1976,
  title		= {On the extended {T}homas-{F}ermi approximation to the kinetic
		  energy density},
  journal	= {Phys. Lett.},
  volume	= {65B},
  number	= {1},
  pages		= {1-4},
  year		= {1976},
  issn		= {0370-2693},
  doi		= {10.1016/0370-2693(76)90519-0},
  url		= {https://www.sciencedirect.com/science/article/pii/0370269376905190},
  author	= {M. Brack and B.K. Jennings and Y.H. Chu},
}

@Article{         Surjuse.Valeev:SAP-X2C.2026,
  author        = {Surjuse, Kshitijkumar A. and Valeev, Edward F.},
  title         = {SAP-X2C: Optimally-Simple Two-Component Relativistic
                  Hamiltonian with Size-Intensive Picture Change},
  journal       = {J. Chem. Theory Comput.},
  volume        = {22},
  number        = {7},
  pages         = {3443--3452},
  year          = {2026},
  doi           = {10.1021/acs.jctc.6c00032},
  url           = {https://doi.org/10.1021/acs.jctc.6c00032}
}

@article{Dyall:Relativistic.2016,
  title={Relativistic double-zeta, triple-zeta, and quadruple-zeta basis sets for the light elements H--Ar},
  author={Dyall, Kenneth G},
  journal={Theor. Chem. Acc.},
  volume={135},
  number={5},
  pages={128},
  year={2016},
  DOI={10.1007/s00214-016-1884-y},
  url={https://doi.org/10.1007/s00214-016-1884-y}
}

@article{Dyall:Relativistic.2009,
  title={Relativistic double-zeta, triple-zeta, and quadruple-zeta basis sets for the 4s, 5s, 6s, and 7s elements},
  author={Dyall, Kenneth G},
  journal={J. Phys. Chem. A},
  volume={113},
  number={45},
  pages={12638--12644},
  year={2009},
  doi={0.1021/jp905057q},
  url={https://doi.org/0.1021/jp905057q}
}

@Article{	  Holzer.Franzke:General.2025,
  title		= {A General and Transferable Local Hybrid Functional for
		  Electronic Structure Theory and Many-Fermion Approaches},
  author	= {Holzer, Christof and Franzke, Yannick J},
  journal	= {J. Chem. Theory Comput.},
  doi		= {10.1021/acs.jctc.4c01309},
  url		= {https://doi.org/10.1021/acs.jctc.4c01309},
  volume  = {21},
  issue   = {1},
  pages   = {202--217},
  year		= {2025}
}

@InBook{	  InB-Franzke:2025,
  author	= {Franzke, Yannick J.},
  editor	= {Aucar, Gustavo Adolfo},
  isbn		= {978-1-83767-260-8},
  title		= {{NMR} {S}pectroscopic {P}arameters in {L}ocal {E}xact
		  {T}wo-component {T}heory},
  booktitle	= {{NMR} Spectroscopic Parameters: Theories and Models,
		  Computational Codes and Calculations},
  publisher	= {Royal Society of Chemistry},
  address	= {Cambridge, United Kingdom},
  year		= {2025},
  pages   = {98--134},
  chapter = {3},
  doi		= {10.1039/9781837678020-00098},
  url		= {https://doi.org/10.1039/9781837678020-00098}
}

@Book{	  Aucar:2025,
  editor	= {Aucar, Gustavo Adolfo},
  booktitle	= {{NMR} Spectroscopic Parameters: Theories and Models,
		  Computational Codes and Calculations},
  publisher	= {Royal Society of Chemistry},
  address	= {Cambridge, United Kingdom},
  year		= {2025},
  doi		= { 0.1039/9781837678020 },
  url		= { https://doi.org/10.1039/9781837678020}
}

@article{Gagliardi:Sustainable.2023,
author = {Gagliardi, Laura},
title = {Sustainable Electronic Structure Software for Everyone, Everywhere},
journal = {J. Chem. Theory Comput.},
volume = {19},
number = {20},
pages = {6857--6858},
year = {2023},
doi = {10.1021/acs.jctc.3c00968},
URL = {https://doi.org/10.1021/acs.jctc.3c00968}
}

@article{Pausch:Consistent.2024,
author = {Pausch, Ansgar},
title = {Consistent Analytical Second Derivatives of the Kohn–Sham DFT Energy in the Framework of the Conductor-Like Screening Model through Gaussian Charge Distributions},
journal = {J. Chem. Theory Comput.},
volume = {20},
number = {8},
pages = {3169-3183},
year = {2024},
doi = {10.1021/acs.jctc.4c00052},
url= {https://doi.org/10.1021/acs.jctc.4c00052}
}

@article{Franzke.Holzer:Application.2026,
      title={Application of SAP-X2C to Spectroscopy and Comparison to Screened Nuclear Spin--Orbit Approximations}, 
      author={Yannick J. Franzke and Christof Holzer},
      volume={22},
      number={18},
      pages={9608--9625},
      year={2026},
      journal={J. Chem. Theory Comput},
      doi={10.1021/acs.jctc.6c01360},
      url={https://doi.org/10.1021/acs.jctc.6c01360}, 
}

@article{Holzer.Franzke:Relativistic.2026,
  author  = {Holzer, Christof and Franzke, Yannick J.},
  title   = {Relativistic {K}ohn--{S}ham Theory of Paramagnetic Magnetic-Resonance Properties},
  journal = {arXiv},
  year    = {2026},
  note    = {arXiv:2611.03170},
  doi     = {10.48550/arXiv.2611.03170},
  url     = {https://doi.org/10.48550/arXiv.2611.03170}
}

@Article{         Franzke.Sharma.ea:Recent.2026,
  author        = {Franzke, Yannick J. and Sharma, Manas and Bhasin,
                  Priyanshi and Bruder, Florian and Huniar, Uwe and
                  Komonvasee, Poramas and Krsti{\'c}, Marjan and Nejad, Arman
                  and Pausch, Amber and Sorathia, Kesha and S\"ulzner, Niklas
                  and Tiska, \"Omer Faruk Can and Zhu, Andrew and Zielke,
                  Roman and H\"attig, Christof and Holzer, Christof and
                  Sierka, Marek and Tew, David P. and Weigend, Florian},
  title         = {Recent developments in TURBOMOLE},
  journal       = {J. Chem. Phys.},
  volume        = {165},
  number        = {16},
  pages         = {162503},
  year          = {2026},
  doi           = {10.1063/5.0351594},
  url           = {https://doi.org/10.1063/5.0351594},
}

@Article{	  Franzke.Pausch.ea:Application.2025,
  author	= {Franzke, Yannick J. and Pausch, Ansgar and Holzer,
		  Christof},
  title		= {Application of the noncollinear Scalmani–Frisch
		  formalism to current density functional theory},
  journal	= {J. Chem. Phys.},
  volume	= {162},
  number	= {8},
  pages		= {084104},
  year		= {2025},
  doi		= {10.1063/5.0246433},
  url		= {https://doi.org/10.1063/5.0246433}
}

@Article{	  Spang.Beuthert.ea:From.2026,
  author	= {Nils Spang and Katrin Beuthert and Tamina Z. Kirsch and
		  Lukas Guggolz and Yannick J. Franzke and Stefanie Dehnen},
  title		= {From 'KGeP' to the Protonated P$_4$-like Anion
		  (HGe$_3$P)$^{2-}$ and the Endohedral Zintl Cluster
		  [Mn@Ge$_{10}$P$_2$]$^{3-}},
  journal	= {J. Am. Chem. Soc.},
  year		= {2026},
  doi		= {10.1021/jacs.6c17420},
  url		= {https://doi.org/10.1021/jacs.6c17420}
}

@Article{         Caldeweyher.Bannwarth.ea:Extension.2017,
  author        = {Caldeweyher, Eike and Bannwarth, Christoph and Grimme,
                  Stefan},
  title         = {Extension of the D3 dispersion coefficient model},
  journal       = {J. Chem. Phys.},
  volume        = {147},
  number        = {3},
  pages         = {034112},
  year          = {2017},
  issn          = {0021-9606},
  doi           = {10.1063/1.4993215},
  url           = {https://doi.org/10.1063/1.4993215},
}

@Article{         Ellis.Odom.ea:Comments.1993,
  author        = {Ellis, Paul D. and Odom, Jerome D. and Lipton, Andrew S.
                  and Gulick, James M.},
  title         = {Comments concerning the computation of cadmium-113
                  chemical shifts},
  journal       = {J. Am. Chem. Soc.},
  volume        = {115},
  number        = {2},
  pages         = {755-759},
  year          = {1993},
  month         = {01},
  issn          = {0002-7863},
  doi           = {10.1021/ja00055a055},
  url           = {https://doi.org/10.1021/ja00055a055},
}

@Article{         Kubis.Peerless.ea:SnAs,
  author        = {Kubis, Maximilian and Peerless, Benjamin and Franzke, Yannick J. and Dehnen, Stefanie},
  title         = {Sn/As-based P$_4$-Analogous Zintl Anions (Sn$_2$As$_2$)$^{2-}$ and (HSn$_3$As)$^{2-}$},
  journal       = {Inorg. Chem.},
  volume        = {},
  number        = {},
  pages         = {},
  year          = {2026},
  doi           = {10.1021/acs.inorgchem.6c04492},
  url           = {https://doi.org/10.1021/acs.inorgchem.6c04492},
}

@Article{Steffen.Szych.ea:Di.2025,
author ="Steffen, Leo and Szych, Lilian S. and Franzke, Yannick J. and Kopp, Richard O. and Ernst, Moritz J. and Weber, Manuela and Müller, Christian",
title  ="Di-{\pi}-methane rearrangement in 1-phosphabarrelenes: formation and reactivity of an unprecedented 2-phosphasemibullvalene",
journal  ="Chem. Commun.",
year  ="2025",
volume  ="61",
issue  ="77",
pages  ="14907-14910",
publisher  ="The Royal Society of Chemistry",
doi  ="10.1039/D5CC04015A",
url  ="http://dx.doi.org/10.1039/D5CC04015A",
}

@Article{         Ketzel.Li.ea:Benchmark.2025,
  author        = {Ketzel, Anton Florian and Li, Xiaolu and Kaupp, Martin and
                  Sun, Han and Schattenberg, Caspar Jonas},
  title         = {Benchmark of Density Functional Theory in the Prediction
                  of $^{13}$C Chemical Shielding Anisotropies for Anisotropic
                  Nuclear Magnetic Resonance-Based Structural Elucidation},
  journal       = {J. Chem. Theory Comput.},
  volume        = {21},
  number        = {2},
  pages         = {871-885},
  year          = {2025},
  issn          = {1549-9618},
  doi           = {10.1021/acs.jctc.4c01407},
  url           = {https://doi.org/10.1021/acs.jctc.4c01407},
}

@Article{	  Desmarais.Maul.ea:Spin.2024,
  title		= {Spin Currents via the Gauge Principle for Meta-Generalized
		  Gradient Exchange-Correlation Functionals},
  author	= {Desmarais, Jacques K. and Maul, Jefferson and Civalleri,
		  Bartolomeo and Erba, Alessandro and Vignale, Giovanni and
		  Pittalis, Stefano},
  journal	= {Phys. Rev. Lett.},
  volume	= {132},
  issue		= {25},
  pages		= {256401},
  numpages	= {7},
  year		= {2024},
  month		= {Jun},
  publisher	= {American Physical Society},
  doi		= {10.1103/PhysRevLett.132.256401},
  url		= {https://link.aps.org/doi/10.1103/PhysRevLett.132.256401}
}

@Article{	  Desmarais.Vignale.ea:Electron.2024,
  title		= {Electron Localization Function for Noncollinear Spins},
  author	= {Desmarais, Jacques K. and Vignale, Giovanni and Bencheikh,
		  Kamel and Erba, Alessandro and Pittalis, Stefano},
  journal	= {Phys. Rev. Lett.},
  volume	= {133},
  issue		= {13},
  pages		= {136401},
  numpages	= {6},
  year		= {2024},
  month		= {Sep},
  publisher	= {American Physical Society},
  doi		= {10.1103/PhysRevLett.133.136401},
  url		= {https://link.aps.org/doi/10.1103/PhysRevLett.133.136401}
}

@Article{	  Maier.Ikabata.ea:Restoring.2019,
  author	= {Maier, Toni M. and Ikabata, Yasuhiro and Nakai, Hiromi},
  title		= {Restoring the iso-orbital limit of the kinetic energy
		  density in relativistic density functional theory},
  journal	= {J. Chem. Phys.},
  volume	= {151},
  number	= {17},
  pages		= {174114},
  year		= {2019},
  issn		= {0021-9606},
  doi		= {10.1063/1.5125634},
  url		= {https://doi.org/10.1063/1.5125634},
}

@Article{         Teale.Helgaker.ea:DFT.2022,
  author        = {Teale, Andrew M. and Helgaker, Trygve and Savin, Andreas
                  and Adamo, Carlo and Aradi, Bálint and Arbuznikov, Alexei
                  V. and Ayers, Paul W. and Baerends, Evert Jan and Barone,
                  Vincenzo and Calaminici, Patrizia and Cancès, Eric and
                  Carter, Emily A. and Chattaraj, Pratim Kumar and Chermette,
                  Henry and Ciofini, Ilaria and Crawford, T. Daniel and De
                  Proft, Frank and Dobson, John F. and Draxl, Claudia and
                  Frauenheim, Thomas and Fromager, Emmanuel and Fuentealba,
                  Patricio and Gagliardi, Laura and Galli, Giulia and Gao,
                  Jiali and Geerlings, Paul and Gidopoulos, Nikitas and Gill,
                  Peter M. W. and Gori-Giorgi, Paola and Görling, Andreas
                  and Gould, Tim and Grimme, Stefan and Gritsenko, Oleg and
                  Jensen, Hans Jørgen Aagaard and Johnson, Erin R. and
                  Jones, Robert O. and Kaupp, Martin and Köster, Andreas M.
                  and Kronik, Leeor and Krylov, Anna I. and Kvaal, Simen and
                  Laestadius, Andre and Levy, Mel and Lewin, Mathieu and Liu,
                  Shubin and Loos, Pierre-François and Maitra, Neepa T. and
                  Neese, Frank and Perdew, John P. and Pernal, Katarzyna and
                  Pernot, Pascal and Piecuch, Piotr and Rebolini, Elisa and
                  Reining, Lucia and Romaniello, Pina and Ruzsinszky, Adrienn
                  and Salahub, Dennis R. and Scheffler, Matthias and
                  Schwerdtfeger, Peter and Staroverov, Viktor N. and Sun,
                  Jianwei and Tellgren, Erik and Tozer, David J. and Trickey,
                  Samuel B. and Ullrich, Carsten A. and Vela, Alberto and
                  Vignale, Giovanni and Wesolowski, Tomasz A. and Xu, Xin and
                  Yang, Weitao},
  title         = {DFT exchange: sharing perspectives on the workhorse of
                  quantum chemistry and materials science},
  journal       = {Phys. Chem. Chem. Phys.},
  volume        = {24},
  number        = {47},
  pages         = {28700--28781},
  year          = {2022},
  doi           = {10.1039/d2cp02827a},
  url           = {https://doi.org/10.1039/d2cp02827a},
}

@InBook{    InB-Perdew:2024,
  title         = {Modern Density Functionals Derived From First Principles},
  editor        = {Manuel Y{\'a}{\~{n}}ez and Russell J. Boyd},
  booktitle     = {Comprehensive Computational Chemistry},
  publisher     = {Elsevier},
  address       = {Oxford, United Kingdom},
  pages         = {69-77},
  year          = {2024},
  isbn          = {978-0-12-823256-9},
  doi           = {https://doi.org/10.1016/B978-0-12-821978-2.00125-2},
  url           = {https://www.sciencedirect.com/science/article/pii/B9780128219782001252},
  author        = {John P. Perdew},
}

@article{Huebsch.Tran.ea:Capturing.2026,
  title = {Capturing exchange-correlation spin-torque effects with a semilocal functional},
  author = {Huebsch, Marie-Therese and Tran, Fabien and Marsman, Martijn},
  journal = {Phys. Rev. B},
  volume = {114},
  issue = {10},
  pages = {105101},
  numpages = {18},
  year = {2026},
  publisher = {American Physical Society},
  doi = {10.1103/67hn-5s46},
  url = {https://link.aps.org/doi/10.1103/67hn-5s46}
}

@article{Desmarais.Bencheikh.ea:Physical.2026,
  title = {Physical Spin Torques from Exactly Constrained Exchange-Correlation Torques},
  author = {Desmarais, Jacques K. and Bencheikh, Kamel and Vignale, Giovanni and Pittalis, Stefano},
  journal = {Phys. Rev. Lett.},
  volume = {136},
  issue = {1},
  pages = {016403},
  numpages = {7},
  year = {2026},
  publisher = {American Physical Society},
  doi = {10.1103/lt1f-8pz2},
  url = {https://link.aps.org/doi/10.1103/lt1f-8pz2}
}

@Article{	  Lee.Handy.ea:density.1995,
  author	= {Lee, Aaron M. and Handy, Nicholas C. and Colwell, Susan
		  M.},
  title		= {The density functional calculation of nuclear shielding
		  constants using London atomic orbitals},
  journal	= {J. Chem. Phys.},
  volume	= {103},
  number	= {23},
  pages		= {10095--10109},
  year		= {1995},
  issn		= {0021-9606},
  doi		= {10.1063/1.469912},
  url		= {https://doi.org/10.1063/1.469912},
}

@article{Grimme.Bannwarth.ea:Fully.2017,
author = {Grimme, Stefan and Bannwarth, Christoph and Dohm, Sebastian and Hansen, Andreas and Pisarek, Jana and Pracht, Philipp and Seibert, Jakob and Neese, Frank},
title = {Fully Automated Quantum-Chemistry-Based Computation of Spin–Spin-Coupled Nuclear Magnetic Resonance Spectra},
journal = {Angew. Chem. Int Ed.},
volume = {56},
number = {46},
pages = {14763-14769},
doi = {https://doi.org/10.1002/anie.201708266},
url = {https://onlinelibrary.wiley.com/doi/abs/10.1002/anie.201708266},
year = {2017}
}

@Article{	  Pracht.Grimme.ea:CREST.2024,
  author	= {Pracht, Philipp and Grimme, Stefan and Bannwarth,
		  Christoph and Bohle, Fabian and Ehlert, Sebastian and
		  Feldmann, Gereon and Gorges, Johannes and Müller, Marcel
		  and Neudecker, Tim and Plett, Christoph and Spicher,
		  Sebastian and Steinbach, Pit and Wesołowski, Patryk A. and
		  Zeller, Felix},
  title		= {CREST--A program for the exploration of low-energy
		  molecular chemical space},
  journal	= {J. Chem. Phys.},
  volume	= {160},
  number	= {11},
  pages		= {114110},
  year		= {2024},
  issn		= {0021-9606},
  doi		= {10.1063/5.0197592},
  url		= {https://doi.org/10.1063/5.0197592}
}

@article{Wood.Dzamba.ea:UMA.2026,
      title={UMA: A Family of Universal Models for Atoms}, 
      author={Brandon M. Wood and Misko Dzamba and Xiang Fu and Meng Gao and Muhammed Shuaibi and Luis Barroso-Luque and Kareem Abdelmaqsoud and Vahe Gharakhanyan and John R. Kitchin and Daniel S. Levine and Kyle Michel and Anuroop Sriram and Taco Cohen and Abhishek Das and Ammar Rizvi and Sushree Jagriti Sahoo and Zachary W. Ulissi and C. Lawrence Zitnick},
      year={2026},
      journal={arXiv},
      doi={10.48550/arXiv.2506.23971},
      url={https://doi.org/10.48550/arXiv.2506.23971}, 
      note={arXiv:2506.23971},
}

@article{Porezag.Frauenheim.ea:Construction.1995,
  title = {Construction of tight-binding-like potentials on the basis of density-functional theory: Application to carbon},
  author = {Porezag, D. and Frauenheim, Th. and K{\"o}hler, Th. and Seifert, G. and Kaschner, R.},
  journal = {Phys. Rev. B},
  volume = {51},
  issue = {19},
  pages = {12947--12957},
  numpages = {0},
  year = {1995},
  month = {May},
  publisher = {American Physical Society},
  doi = {10.1103/PhysRevB.51.12947},
  url = {https://link.aps.org/doi/10.1103/PhysRevB.51.12947}
}

@Article{         Jha.Heine:DFTB.2022,
  author        = {Jha, Gautam and Heine, Thomas},
  title         = {DFTB Parameters for the Periodic Table: Part III,
                  Spin-Orbit Coupling},
  journal       = {J. Chem. Theory Comput.},
  volume        = {18},
  number        = {7},
  pages         = {4472-4481},
  year          = {2022},
  issn          = {1549-9618},
  doi           = {10.1021/acs.jctc.2c00376},
  url           = {https://doi.org/10.1021/acs.jctc.2c00376},
}

@Article{	  Bulik.Scalmani.ea:Noncollinear.2013,
  title		= {Noncollinear density functional theory having proper
		  invariance and local torque properties},
  author	= {Bulik, Ireneusz W. and Scalmani, Giovanni and Frisch,
		  Michael J. and Scuseria, Gustavo E.},
  journal	= {Phys. Rev. B},
  volume	= {87},
  issue		= {3},
  pages		= {035117},
  numpages	= {8},
  year		= {2013},
  month		= {Jan},
  publisher	= {American Physical Society},
  doi		= {10.1103/PhysRevB.87.035117},
  url		= {https://link.aps.org/doi/10.1103/PhysRevB.87.035117}
}

@article{Pittalis.Vignale.ea:U1.2017,
  title = {$\text{U}(1)\ifmmode\times\else\texttimes\fi{}\mathrm{SU}(2)$ gauge invariance made simple for density functional approximations},
  author = {Pittalis, S. and Vignale, G. and Eich, F. G.},
  journal = {Phys. Rev. B},
  volume = {96},
  issue = {3},
  pages = {035141},
  numpages = {8},
  year = {2017},
  publisher = {American Physical Society},
  doi = {10.1103/PhysRevB.96.035141},
  url = {https://link.aps.org/doi/10.1103/PhysRevB.96.035141}
}

@article{Bencheikh:Spin-orbit.2003,
doi = {10.1088/0305-4470/36/48/002},
url = {https://doi.org/10.1088/0305-4470/36/48/002},
year = {2003},
month = {nov},
publisher = {},
volume = {36},
number = {48},
pages = {11929},
author = {K Bencheikh},
title = {Spin–orbit coupling in the spin-current-density-functional theory},
journal = {J. Phys. A},
}

@article{Desmarais.Erba:Meta-Generalized.2025,
  title = {Meta-Generalized Gradient Approximation Made Magnetic},
  author = {Desmarais, Jacques K. and Erba, Alessandro and Vignale, Giovanni and Pittalis, Stefano},
  journal = {Phys. Rev. Lett.},
  volume = {134},
  issue = {10},
  pages = {106402},
  numpages = {7},
  year = {2025},
  month = {Mar},
  publisher = {American Physical Society},
  doi = {10.1103/PhysRevLett.134.106402},
  url = {https://link.aps.org/doi/10.1103/PhysRevLett.134.106402}
}

@article{Capelle.Vignale.ea:Spin.2001,
  title = {Spin Currents and Spin Dynamics in Time-Dependent Density-Functional Theory},
  author = {Capelle, K. and Vignale, G. and Gy{\"o}rffy, B. L.},
  journal = {Phys. Rev. Lett.},
  volume = {87},
  issue = {20},
  pages = {206403},
  numpages = {4},
  year = {2001},
  publisher = {American Physical Society},
  doi = {10.1103/PhysRevLett.87.206403},
  url = {https://link.aps.org/doi/10.1103/PhysRevLett.87.206403}
}

@article{Sharma.Dewhurst.ea:First-Principles.2007,
  title = {First-Principles Approach to Noncollinear Magnetism: Towards Spin Dynamics},
  author = {Sharma, S. and Dewhurst, J. K. and Ambrosch-Draxl, C. and Kurth, S. and Helbig, N. and Pittalis, S. and Shallcross, S. and Nordstr{\"o}m, L. and Gross, E. K. U.},
  journal = {Phys. Rev. Lett.},
  volume = {98},
  issue = {19},
  pages = {196405},
  numpages = {4},
  year = {2007},
  publisher = {American Physical Society},
  doi = {10.1103/PhysRevLett.98.196405},
  url = {https://link.aps.org/doi/10.1103/PhysRevLett.98.196405}
}

@article{Eich.Pittalis.ea:Transverse.2013,
  title = {Transverse and longitudinal gradients of the spin magnetization in spin-density-functional theory},
  author = {Eich, F. G. and Pittalis, S. and Vignale, G.},
  journal = {Phys. Rev. B},
  volume = {88},
  issue = {24},
  pages = {245102},
  numpages = {8},
  year = {2013},
  publisher = {American Physical Society},
  doi = {10.1103/PhysRevB.88.245102},
  url = {https://link.aps.org/doi/10.1103/PhysRevB.88.245102}
}

@Article{         Peng.Zou.ea:Time-dependent.2005,
  author        = {Peng, Daoling and Zou, Wenli and Liu, Wenjian},
  title         = {Time-dependent quasirelativistic density-functional theory
                  based on the zeroth-order regular approximation},
  journal       = {J. Chem. Phys.},
  volume        = {123},
  number        = {14},
  pages         = {144101},
  year          = {2005},
  issn          = {0021-9606},
  doi           = {10.1063/1.2047554},
  url           = {https://doi.org/10.1063/1.2047554},
}

\end{document}